\documentclass[aps,twocolumn,showpacs,superscriptaddress,nofootinbib,preprintnumbers]{revtex4-1}

\usepackage{graphicx}  % needed for figures
\usepackage{dcolumn}   % needed for some tables
\usepackage{amssymb}   % for math
\usepackage{mathtools}
\usepackage{amsmath}

\usepackage{makecell}

\usepackage{epsfig,amsmath,amsfonts,amssymb,amstext,afterpage,psfrag}
\usepackage{slashed}
\usepackage{multirow}
\usepackage{hyperref}
\usepackage{color}
\usepackage{soul}
\usepackage[utf8x]{inputenc}
\usepackage[caption=false]{subfig}
\newcommand{\ct}{{\tt CosmoTransitions}} 
\newcommand{\gl}{{\tt \href{https://gitlab.com/claudius-krause/ew_nr}{GitLab} }}
\newcommand{\tlambda}{\widetilde{\lambda}}

\begin{document}

% the following line is for submission, including submission to the arXiv!!
\hspace{5.2in} \mbox{\vbox{\noindent
    FERMILAB-PUB-21-146-T\\ \today}}
\preprint{FERMILAB-PUB-21-146-T}
%\preprint{\today}

\title{
A New Approach to Electroweak Symmetry Non-Restoration}
\date{April 2021}


\author{Marcela~Carena}
\email{carena@fnal.gov}
\affiliation{Fermi National Accelerator Laboratory, P.~O.~Box 500, Batavia, IL 60510, USA}
\affiliation{Enrico Fermi Institute and Kavli Institute for Cosmological Physics,\\ University of Chicago, Chicago, IL 60637, USA}
\author{Claudius~Krause}
\email{Claudius.Krause@rutgers.edu}
\affiliation{Fermi National Accelerator Laboratory, P.~O.~Box 500, Batavia, IL 60510, USA}
\affiliation{NHETC, Dept.~of Physics and Astronomy, Rutgers University, Piscataway, NJ 08854, USA}
\author{Zhen~Liu}
\email{zliuphys@umn.edu}
\affiliation{School of Physics and Astronomy, University of Minnesota,~Minneapolis, MN 55455, USA}
\author{Yikun~Wang}
\email{yikwang@uchicago.edu}
\affiliation{Fermi National Accelerator Laboratory, P.~O.~Box 500, Batavia, IL 60510, USA}
\affiliation{Enrico Fermi Institute and Kavli Institute for Cosmological Physics,\\ University of Chicago, Chicago, IL 60637, USA}

\begin{abstract}

Electroweak symmetry non-restoration up to high temperatures well above the electroweak scale offers new alternatives for baryogenesis. We propose a new approach for electroweak symmetry non-restoration via an inert Higgs sector that couples to the~Standard Model Higgs as well as an extended scalar singlet sector. We implement renormalization group improvements and thermal resummation, necessary to evaluate the effective potential spanning over a broad range of energy scales and temperatures. We present examples of benchmark scenarios that allow for electroweak symmetry non-restoration all the way up to hundreds of TeV temperatures, and also feature suppressed sphaleron washout factors down to the electroweak scale. Our method for transmitting the Standard Model broken electroweak symmetry to an inert Higgs sector has several intriguing implications for (electroweak) baryogenesis, early universe thermal histories, and can be scrutinized through~Higgs physics phenomenology and electroweak precision measurements at the HL-LHC.

 \end{abstract}

\maketitle

\section{Introduction}
\label{sec:intro}
The Standard Model (SM) of particle physics accurately describes the behavior of the particles making up the ordinary matter, but it fails to provide an explanation of how they came to be. Under the assumption that particles and anti-particles are produced in equal numbers in the early Universe, the SM predicts that they would have long annihilated each other without leaving any remnant matter today. Sakharov~\cite{Sakharov:1967dj} enunciated that producing a Baryon Asymmetry (BA), i.e., more matter than anti-matter, requires baryon number violation, C and CP violation, and out-of-equilibrium processes to all occur at the same time. Although the SM provides sources of C, CP, and baryon number violation through the electroweak interactions and sphalerons, respectively, it fails to explain the observed BA. Indeed, the SM Electroweak Phase Transition (EWPT) is a smooth crossover and, thus, is not giving rise to sufficient deviations from thermal equilibrium~\cite{Morrissey:2012db}. In addition, the amount of C and CP violation in the SM is insufficient to generate the observed baryon asymmetry~\cite{Gavela:1994dt}. In order to generate the observed baryon asymmetry, sources of CP violation and out-of-equilibrium processes beyond those found in the SM must be realized in nature.

There are many mechanisms proposed in the literature to explain the generation of a net Baryon number B, and in most cases, sphaleron processes that are capable of violating B+L, but conserve B-L, play a relevant role (with L the lepton number). One interesting possibility to achieve sphaleron-induced B number generation is via a Strong First Order Electroweak Phase Transition (SFOEWPT), yielding promising conditions for {\it electroweak baryogenesis}~\cite{Kuzmin:1985mm}. Accommodating a SFOEWPT demands modifications of the Higgs potential. Such modifications may be induced predominantly by thermal effects, as it happens e.g., in the Minimal Supersymmetric extension of the Standard Model (MSSM)~\cite{Carena:1996wj,Delepine:1996vn,Laine:1998qk,Cline:1998hy,Balazs:2004ae,Lee:2004we,Carena:2008vj}, or by zero-temperature effects that have a lasting consequence after thermal effects are taken into account. The latter situation naturally occurs in models of new physics containing additional light scalar particles with sizable couplings to the Higgs.

In this study, we are interested in models in which the electroweak (EW) symmetry is broken at temperatures well above the EW~scale. Taking a bottom-up approach, we called these scenarios: i) delayed restoration, if the electroweak symmetry is restored at very high temperatures, or ii) non-restoration if the electroweak symmetry remains broken all the way up to some high energy scale $\Lambda$ of validity of the theory. Electroweak non-restoration or delayed restoration scenarios have advantages in modeling mechanisms for baryogenesis. For example, in the case of electroweak baryogenesis (EWBG), one important advantage is that the additional, required sources of CP violation will only be effective at high energies and, therefore, will avoid current electric dipole moment experimental bounds.

Symmetry non-restoration at high temperatures has been first studied a long time ago \cite{Weinberg:1974hy,Mohapatra:1979qt,Mohapatra:1979vr,Mohapatra:1979bc,Dvali:1995cc,Dvali:1995cj,Bajc:1999cn} and recently~\cite{Patel:2013zla,Kilic:2015joa,Ramsey-Musolf:2017tgh,Meade:2018saz,Baldes:2018nel,Glioti:2018roy,Carena:2019une,Matsedonskyi:2020mlz,Bai:2021hfb}. In particular, new ideas of electroweak symmetry non-restoration or delayed restoration have been discussed~\cite{Meade:2018saz,Baldes:2018nel,Glioti:2018roy,Matsedonskyi:2020mlz} by extending the SM Higgs sector with additional singlet scalars that couple to the SM Higgs and provide it with a negative thermal mass at very high temperatures. Such models typically require several hundreds of new scalar fields. On top of the new scalar sectors, for models with delayed restoration, the Ultraviolet (UV) completions typically require additional scalar and/or fermion fields that couple with the EW sector and yield electroweak symmetry restoration, as well as a strong first-order phase transition, at very high temperatures \cite{Baldes:2018nel,Glioti:2018roy,Matsedonskyi:2020mlz}.

\begin{figure*}[t]
  \includegraphics[width=\textwidth]{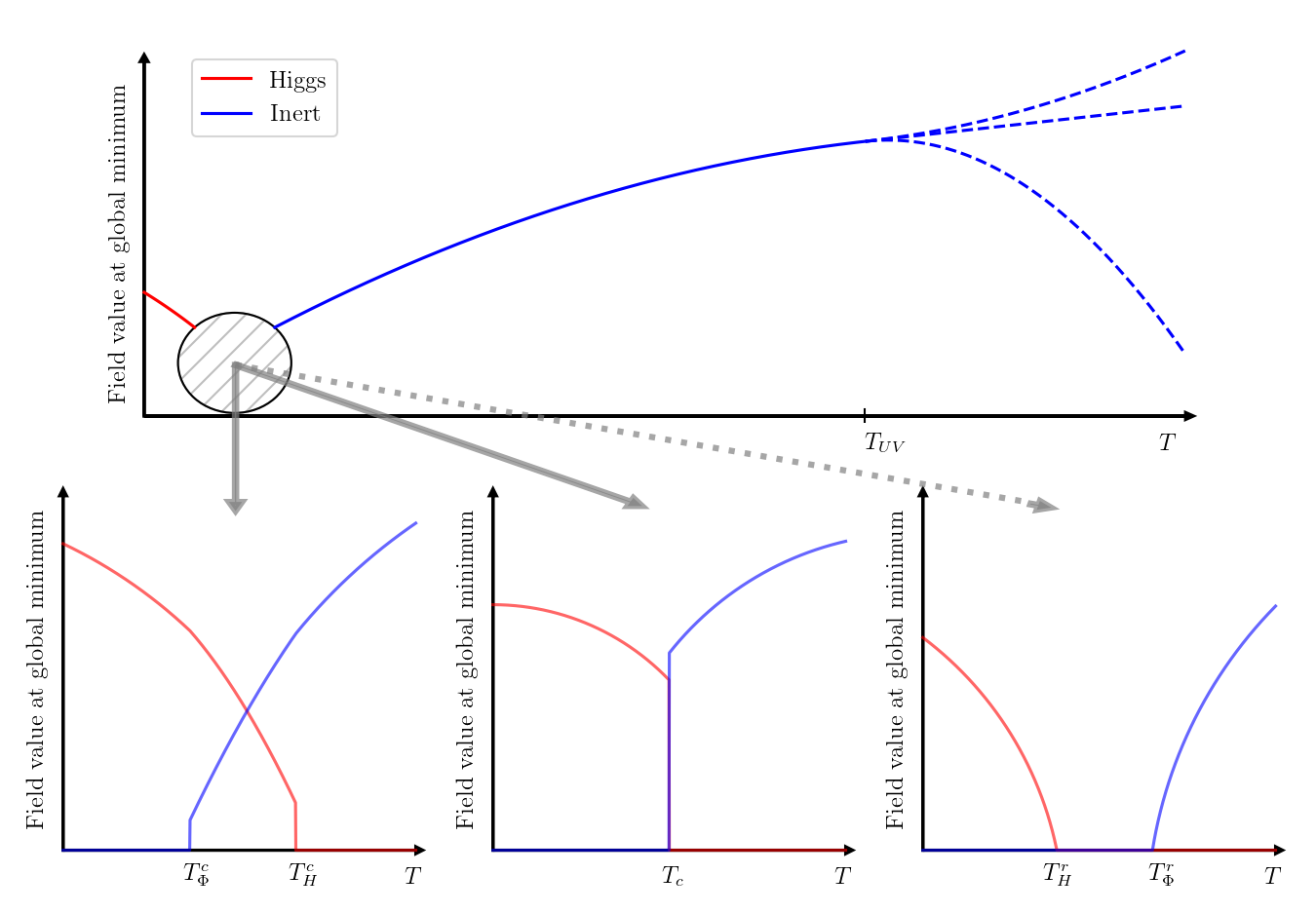}
  \caption{Schematic illustration of the phase values of Higgs and Inert scalar at the global minimum for non-restoration and delayed restoration. The top panel shows the global view up to very high temperatures, where the non-retoration can either persist or the symmetry becomes eventually restored (depicted by dashed lines). The lower panel zooms into the gray region and shows three different scenarios of the transition between the two doublets.}
  \label{fig:NRglobal}
\end{figure*}

In this work, we explore the EW non-restoration or delayed restoration with an extended-Inert Two Higgs Doublet Model (I2HDM) \cite{Branco:2011iw,Gustafsson:2010zz}, where,  instead of the SM Higgs, it is the inert Higgs who acquires a non-zero vacuum expectation value (vev) up to very high temperature by coupling to an additional scalar sector. Such a model requires minimal couplings between the new scalar fields and the SM Higgs boson, and  opens the window to  different  realizations for  baryogenesis at very high energy scales. Due to the lack of large Yukawa couplings to the inert sector, the number of scalars required to achieve negative thermal masses is somewhat reduced. Specific new physics models for high scale EW baryogenesis in the context  of extensions of the I2HDM will be the topic of a forthcoming publication.

The EW non-restoration sets the boundary conditions at high temperatures ($\lesssim T_{\rm{UV}}$), while the observed EW vacuum defines them at zero temperature; see the top row of~\autoref{fig:NRglobal} for a schematic view. For intermediate temperatures, the I2HDM allows different phase histories that we depict in the bottom row of~\autoref{fig:NRglobal}. There could either be a temperature range (between $T_{H}^{c}$ and $T_{\Phi}^{c}$ - to be precisely defined below) for which the global minumum is given by non-vanishing vevs of both the Higgs and the inert fields (left plot), or there could be a discrete jump between the Higgs and the inert phases at a critical Temperature $T_{c}$ (central plot). A third option is given by a scenario in which the Higgs vev goes to zero at a restoration temperature $T_{H}^{r}$ lower than the temperature $T_{\Phi}^{r}$ above which the inert vev starts to grow (right plot). In the temperature range between $T_{H}^{r}$ and $T_{\Phi}^{r}$ the system is in a EW preserving vacuum.

In this work, we utilize the perturbative effective potential (EP) method to calculate the finite-temperature phase structure and quantities relevant to the baryon asymmetry. However, unlike for typical EWPT calculations for which the electroweak symmetry breaking takes place close to the EW scale, here we need to take into account important effects due to the large scale separation between the high temperatures ($\gtrsim \mathcal{O}(1-100)$ TeV)- high field values and the EW scale, which requires careful treatment and improvement of the perturbative calculation. For this purpose, we will implement a Renormalization Group (RG) improvement and daisy resummation of the EP to ameliorate the perturbative convergence.

This paper is organized as follows: in~\autoref{sec:model}, we introduce our model and discuss its zero temperature constraints. In~\autoref{sec:pert}, we investigate the validity of the radiatively corrected, finite temperature effective potential in calculating the phase structures, introducing the RG improvement and daisy resummation, and we set up schemes for the improved perturbative calculation. In~\autoref{sec:pt}, we present an analytical study of the possible thermal histories based on a mean-field approach. In~\autoref{sec:num}, we present the full numerical computation of the finite-temperature phase structure for two benchmark (BM) scenarios. In~\autoref{sec:sph}, we discuss the baryon washout conditions and consider them in light of the thermal history results for the two BM scenarios presented in the previous section. We also discuss the impact of future model building on high-temperature baryogenesis. In~\autoref{sec:pheno}, we discuss phenomenological constraints in this type of model. Finally, we present our conclusion in~\autoref{sec:concl}. We collect various technical aspects in appendices.

\section{The Model} \label{sec:model}
\subsection{The effective potential at tree level}

We consider an extension of the SM Higgs sector that includes an Inert Higgs Doublet with additional singlet scalars. In such case, the most general $\mathbb{Z}_{2}$-symmetric potential reads\footnote{Here $\protect\mathbb{Z}_{2}$ is defined as $H\rightarrow H$, $\chi_i\rightarrow -\chi_i$, and $\Phi\rightarrow -\Phi$. As we shall discuss later, instead of the $\mathbb{Z}_{2}$, we require a continuous global $U(1)$ symmetry on the doublet $\Phi$ to ensure it being inert, which forbids additional terms that we omitted here in the potential.},
\begin{widetext}
\begin{align}
  \label{eq:Model.Lag}
 V_{\mathbb{Z}_{N}+{\rm I2HDM}} = & - \mu^{2}_{H} H^{\dagger}H + \lambda_{H} (H^{\dagger}H)^{2} 
                                      +\mu^{2}_{\Phi}(\Phi^{\dagger}\Phi)+\lambda_{\Phi}(\Phi^{\dagger}\Phi)^{2}  
                                      +\lambda_{H\Phi}(H^{\dagger}H)(\Phi^{\dagger}\Phi)+ \tlambda_{H\Phi}(H^{\dagger}\Phi)(\Phi^{\dagger}H)  \\
                                &      + \frac{\mu^{2}_{\chi}}{2}\chi_{i}^{2}+ \frac{\tlambda_{\chi}}{4}\chi_{i}^{4} + \frac{\lambda_{\chi}}{4}(\chi_{i}\chi_{i})^{2} 
                                      + \frac{\lambda_{\Phi\chi}}{2}\chi_{i}^{2}(\Phi^{\dagger}\Phi) + \frac{\lambda_{H\chi}}{2}\chi_{i}^{2}(H^{\dagger}H) \nonumber ,
\end{align}
\end{widetext}
where the two Higgs doublets are written as
\begin{align}
  \begin{aligned}
    \label{eq:}
   & H= \left(\begin{array}{c}
    G^+\\
        \frac{1}{\sqrt{2}} (h + i G_0)\\
                                        \end{array}\right)
  \end{aligned}
\end{align}
\begin{align}
  \begin{aligned}
    \label{eq:}
   & \Phi= \left(\begin{array}{c}
   \phi^+ \\
      \frac{1}{\sqrt{2}} (\varphi + i \phi_0)    \\
                                        \end{array}\right),
  \end{aligned}
\end{align}
and the fields $\chi_{i}$ represent $N$ real, singlet scalars. Assuming that extra sources of CP violation will come from a new sector, once we study the complete UV theory, we impose CP invariance in the Higgs sector and define all model parameters to be real. The assumed $\mathbb{Z}_{2}$-symmetry forbids couplings of the type $\mu_{12}^{2}(H^{\dagger}\Phi), \lambda_{6}(H^{\dagger}\Phi H^{\dagger}H)$, and $ \lambda_{7}(H^{\dagger}\Phi \Phi^{\dagger}\Phi)$. Portal couplings of the form $(\Phi^{\dagger}H)(\Phi^{\dagger}H)$ and $ (H^{\dagger}\Phi)(H^{\dagger}\Phi)$ are allowed by the $\mathbb{Z}_{2}$ symmetry and are related to the operator $(H^{\dagger}\Phi)(\Phi^{\dagger}H)$ by custodial symmetry~\cite{Pomarol:1993mu}. However, assuming a $U(1)$-symmetry on (one of the) doublets forbids these additional portal couplings and simplifies the potential. Given the custodial symmetry and the additional $U(1)$-symmetry, we can set the coupling $\tlambda_{H\Phi}$ to $0$ as well. However, this is not stable under RG-running, as the hypercharge gauge coupling breaks custodial symmetry. We therefore keep track of the operator with the coefficient $\tlambda_{H\Phi}$ for future RG improvement of the EP; see discussion below in~\autoref{sec:pert}. In addition, to better accommodate phenomenological constraints, we set $\lambda_{H\chi} = 0$, although, similarly to $\tlambda_{H\Phi} $ this coupling will also be induced by the renormalization group evolution (RGE), and we will keep track of its effects. Finally, observe that $\tlambda_{\chi} = 0$ is protected by an $SO(N)$ symmetry of the singlet sector, and we shall impose such symmetry. In the case of a potential with generic values of $\tlambda_{\chi}$, the singlet sector exhibits a discrete $\mathbb{Z}_{N}$ symmetry.

To summarize, parameters in the above potential can be separated as follows:
\begin{itemize}
  \label{eq:Model.para}
\item  fixed\ parameters: $\{\mu_{H}^2, \lambda_{H} \}$, 
\item  free\ parameters: $\{\mu_{\Phi}^2, \mu_{\chi}^2, \lambda_{\Phi}, \lambda_{\chi}, \lambda_{\Phi\chi}, \lambda_{H\Phi}, N\}$,
\item free parameters   set  to  zero:  $\{\tlambda_{H\Phi}, \lambda_{H\chi}, \tlambda_{\chi}\}$,  
\item RGE\ induced\ parameters: $\{\tlambda_{H\Phi}, \lambda_{H\chi}\}$, 
\end{itemize}
where the two fixed parameters are given by the current observation of the EW~vacuum expectation value (vev) $v_0=246$ GeV and the SM Higgs mass $m_{h}=125$ GeV.

In general, there could be charge breaking and CP breaking minima in two Higgs doublet models. However, \cite{Ferreira:2004yd,Barroso:2005sm} showed that at tree level, if an EW breaking minimum exists, any possibly existing charge breaking or CP breaking extremum is necessarily a saddle point above the EW breaking minimum. Although the validity of this result may not hold after the inclusion of radiative corrections, and its validation requires a more detailed analysis beyond the scope of this work, we shall only allow for the neutral CP even components to develop non-zero vacuum expectation values at any temperature. Therefore, from now on, we focus on analyzing the effective potential of the CP-even components of the two Higgs doublets and the singlet sector. The tree-level CP even potential reads,
\begin{widetext}
\begin{align}
  \label{eq:Model.V0}
  V_{0,{\rm CP~even}}^{\mathbb{Z}_{N}+{\rm I2HDM}} = 
                                      &- \frac{\mu_{H}^{2}}{2} h^{2} + \frac{\lambda_{H}}{4} h^{4} \nonumber
                                      +\frac{\mu^{2}_{\Phi}}{2}\varphi^{2} +\frac{\lambda_{\Phi}}{4}\varphi^{4}  \nonumber 
                                      +\frac{\lambda_{H\Phi} + \tlambda_{H\Phi}}{4}h^{2}\varphi^{2}\nonumber \\
                                      & + \frac{\mu^{2}_{\chi}}{2}\chi_{i}^{2}+ \frac{\tlambda_{\chi}}{4}\chi_{i}^{4} + \frac{\lambda_{\chi}}{4}(\chi_{i}\chi_{i})^{2}\nonumber
                                      + \frac{\lambda_{\Phi\chi}}{4}\chi_{i}^{2}\varphi^{2}+\frac{\lambda_{H\chi}}{4}\chi_{i}^{2}h^2. \nonumber \\
\end{align}
\end{widetext}

The particles in the plasma include bosons $\{h, G_0, G^{\pm}, \varphi, \phi_0, \phi^{\pm}, \chi , \gamma, W^{\pm}, Z\}$ with corresponding particle degrees of freedom (d.o.f.) $n_{bos} = \{1, 1, 2, 1, 1, 2, N , 3, 6, 3\}$, and fermions, $\{ t\}$ with corresponding particle d.o.f. $n_{ferm} = \{12\}$ that couple (self-couple) to the dynamical fields. Notice that we work in the Landau gauge so there are no ghost d.o.f. We collect the effective, field-dependent masses of these particles in~appendix~\autoref{app:mass}.

\subsection{Zero temperature constraints} \label{sec:t0}

In this section, we present the tree-level, zero temperature constraints on our model, including the bounded from below (BFB) conditions, and the correct vacuum structure of the tree-level potential. This study provides guidance, later on, in defining the viable parameter space for which we shall perform numerical calculations to constrain the model after the inclusion of radiative corrections.

\subsubsection{Bounded From Below Conditions}

The bounded from below (BFB) conditions, which need to be satisfied simultaneously, for the generic tree level potential given in eq.~\eqref{eq:Model.Lag} are
\begin{widetext}
 \begin{align}
  \label{eq:BFB1}
&\lambda_{H} >0, \quad \quad \quad \lambda_{\Phi} >0,\quad\quad \quad \Lambda_{\chi, n} > 0, \nonumber\\
  \Lambda_{H\Phi} > - \sqrt{4 \lambda_{H} \lambda_{\Phi}},  &
  \quad \quad \lambda_{\Phi\chi} > - \sqrt{4 \lambda_{\Phi} \Lambda_{\chi,n}}, 
   \quad \quad \lambda_{H\chi} > - \sqrt{ 4 \lambda_{H} \Lambda_{\chi, n}},   \\
    \sqrt{4 \lambda_{H} \lambda_{\Phi} \Lambda_{\chi,n}} + \Lambda_{H\Phi} \sqrt{\Lambda_{\chi,n}} + &\lambda_{\Phi\chi} \sqrt{\lambda_{H}} + \lambda_{H\chi} \sqrt{\lambda_{\Phi}} 
    + \sqrt{\left( \Lambda_{H\Phi} + \sqrt{4 \lambda_{H} \lambda_{\Phi}}\right) \left( \lambda_{\Phi\chi} + \sqrt{4 \lambda_{\Phi}  \Lambda_{\chi,n}}\right)\left( \lambda_{H\chi} + \sqrt{ 4 \lambda_{H} \Lambda_{\chi,n} }\right) } > 0, \nonumber
 \end{align}
 \end{widetext}
where for simplicity we define the effective couplings
  \begin{align}
  \label{eq:BFBaux}
    \Lambda_{\chi, n}\equiv \frac{1}{n}\tlambda_{\chi} +  \lambda_{\chi} \quad \text{and}\quad \Lambda_{H\Phi}\equiv \lambda_{H\Phi}+\tlambda_{H\Phi}\rho^2.
\end{align}
There are two variables  in these conditions, $n\in \{1, \dots, N\}$ and $\rho^2\in [0,1]$, see appendix~\autoref{app:BFB} for details. The conditions \eqref{eq:BFB1} have to hold for all values of $n$ and $\rho$. Notice that they only enter the conditions through $\Lambda_{\chi, n}$ and $\Lambda_{H\Phi}$. If $\tlambda_{\chi} >0$, $\Lambda_{\chi, n}$ is the smallest when $n = N$, while if $\tlambda_{\chi} <0$, the smallest $\Lambda_{\chi, n}$ is found for $n= 1$. Similar considerations apply to $\Lambda_{H\Phi}$ and $\rho$. A detailed derivation of these conditions can be found in appendix~\autoref{app:BFB}.

\subsubsection{Vacuum Structure}

In order to be consistent with the current Higgs  and EW precision measurements, as the inert doublet is charged under the EW gauge group, we consider the case that at zero temperature, both the inert Higgs and the singlets have zero vev, say the physical vacuum is 
\begin{align}
  \label{eq:ew.vac}
\langle \{ h, \varphi, \chi_1, \cdots , \chi_N \} \rangle =\{ v_0, 0, 0, \cdots , 0 \},
\end{align}
where $v_0 = 246$ GeV, and we require such vacuum state to be the global minimum of the zero temperature potential. Firstly, for the physical vacuum to be a minimum, one needs to avoid tachyonic solutions, which give constraints on the bare mass parameters of the potential (at tree level)
\begin{align}
  \label{eq:tach}
\mu_{\Phi}^2 + \frac{\lambda_{H\Phi}}{2} v_{\rm EW}^2 \ge 0,\quad \mu_{\chi}^2 \ge 0. 
\end{align}
Equation \eqref{eq:tach} does not involve the RG-generated parameters, $\widetilde{\lambda}_{H\Phi}$ and $\lambda_{H\chi}$  since it refers to the couplings at the physical minimum.

As stated above, at tree level, any possibly existing CP or charge breaking extrema are saddle points lying above the EW vacuum, which, therefore, do not put any further constraints on the viable parameter space. To secure that the EW vacuum is the global minimum of the tree-level potential in the subfield space of the two CP even components and the singlet degrees of freedom,  we find all possible extrema of the polynomial potential (see all possible extrema in appendix~\autoref{app:BFB} at tree level) and we numerically impose the necessary conditions to establish that for each extremum either it cannot exist, or it is above the physical one. 

\section{Radiative Corrected, Finite Temperature Potential }
\label{sec:pert}

In the perturbative effective potential calculation, two types of radiative corrections to the tree-level potential need to be considered, i.e.~the zero temperature loop corrections and the finite temperature radiative corrections.

At one-loop order, the zero temperature loop correction can be taken into account through the Coleman-Weinberg (CW) potential \cite{Coleman:1973jx,Coleman:1985rnk}
\begin{widetext}
\begin{equation}
\label{eq:T.QFT.1}
V_{CW} \left( \{M^2_i (\hat{\Phi} ) \}; \mu_{\rm R}\right)=\frac{1}{64 \pi^{2}} \sum_{i = {\rm B,F}}{ (-1)^{2S_i} n_i M_i^{4}( \hat{\Phi})\left[\log{\frac{M_i^{2}( \hat{\Phi})}{\mu_{R}^{2}}}-a_i \right]},
\end{equation}
\end{widetext}
under the $\overline{\rm MS}$-renormalization scheme, and where $S_i =$ 1 or 1/2 for i =B or F, respectively. The short-handed notation has been introduced for the dynamical fields $\hat{\Phi} \equiv \{h, \varphi, \chi_1, \chi_2,\cdots, \chi_N\}$. The specie $i$ is summed over all degrees of freedoms in the plasma. The constant $a_i$ has a value of $\frac{3}{2}$ for scalars, longitudinal gauge bosons and fermions while $\frac{1}{2}$ for transverse gauge bosons. $\mu_{R}$ is the renormalization scale, and finally $M_i^{2}(\hat{\Phi})$ is the field-dependent mass eigenvalue. The field dependent masses of all degrees of freedom in the plasma for our model are given in appendix~\autoref{app:mass}.
We work in the Landau gauge \cite{Coleman:1985rnk}, which introduces a gauge-dependence of the EP \cite{Jackiw:1974cv,Kang:1974yj,Dolan:1974gu,Fukuda:1975di,Aitchison:1983ns,Patel:2011th,Garny:2012cg,Andreassen:2014eha,Andreassen:2014gha} \footnote{Given that we observe that the high-temperature expansion approximation is in good qualitative agreement with the full treatment of the temperature effects when considering the electroweak symmetry non-restoration analysis, we argue that the main results of this work will not be qualitatively changed by effects of gauge dependence. Indeed, the EW non-restoration at high temperatures relies on a negative thermal mass for the inert Higgs that is governed by the leading order term in the high-temperature expansion, which in turn does not exhibit gauge dependence. Indeed Refs. \cite{Patel:2011th,Garny:2012cg} show that the gauge dependence appears only in the sub-leading temperature-dependent terms in the high-temperature expansion. A dedicated study of the gauge dependence considering a numerical analysis of the full temperature-dependent EP would be necessary to fully understand the relevance of gauge-dependent effects in the analysis of EW non-restoration, which is beyond the scope of this work.}.

The CW potential changes the shape of the zero temperature potential, introducing deviations from the tree level constraints at zero temperature that we discussed in the last section. Specifically, to accommodate the Higgs vev of $246$ GeV and a $125$ GeV mass eigenstate the parameters $\mu^{2}_{H}$ and $\lambda_{H}$ have to be adjusted to recover the two physical conditions at $T=0$. Other zero temperature constraints, including the BFB and correct vacuum structure, also need to be adjusted numerically, as necessary, so that they remain robust after the inclusion of loop corrections.

The leading temperature dependence is given by the thermal one-loop effective potential (see reviews e.g.~\cite{Quiros:1999jp})
\begin{widetext}
\begin{equation}
\label{eq:T.QFT.8}
V^T_{\rm 1-loop} (\{M^2_k (\hat{\Phi} ) \}, T)
= \frac{T^4}{2\pi^2} \left[ \sum_{i = B} n_i J_B \left(\frac{M_i^2( \hat{\Phi})} {T^2}\right) - \sum_{i = F} n_i J_F \left(\frac{M_i^2( \hat{\Phi})} {T^2}\right) \right],
\end{equation}
\end{widetext}
where relevant notation has been introduced above, and
\begin{equation}
\label{eq:T.QFT.9}
J_{B/F}(y) = \int_{0}^{\infty}\;dx\; x^{2} \log{\left(1\mp e^{-\sqrt{x^{2}+y}}\right)},
\end{equation}
where in the thermal potential the argument $y = M_i^2( \hat{\Phi})/T^2$. The $J_{B/F}$ functions can be evaluated numerically, see appendix~\autoref{app:numerics} for the details of our implementation. To gain analytical understanding of the thermal history, a high-temperature expansion can be used to obtain an analytical expression for the thermal potential:
\begin{align}
\begin{aligned}
\label{eq:JhighT}
J^{{\rm high-}T}_{B}(y) &= - \frac{\pi^4}{45} + \frac{\pi^2}{12} y - \frac{\pi}{6} y^{\frac{3}{2}} - \frac{1}{32} y^2 \log\left( \frac{y}{a_b} \right) + \cdots,\\
J^{{\rm high-}T}_{F}(y) &= \frac{7\pi^4}{360} - \frac{\pi^2}{24} y - \frac{1}{32} y^2 \log\left( \frac{y}{a_f} \right)+ \cdots,\\
\end{aligned}
\end{align}
where $a_b= 16\pi^2 \exp(3/2 - 2 \gamma_E)$, $a_f= \pi^2 \exp(3/2 - 2 \gamma_E)$ and $\gamma_E$ is the Euler constant. The high-temperature expansion in eq.~\eqref{eq:JhighT} guarantees a good convergence for values of the argument of the $J_{B/F}$ functions up to $2 -5$, while values are constrained to be below/about $1$ without inclusion of the logarithmic terms.

At very high temperatures and very large field values, which are the relevant scales for the electroweak symmetry non-restoration or delayed restoration scenarios, perturbative convergence of the fixed order calculation becomes compromised, for both the CW and the one-loop thermal potential. In the following, we discuss the improvements that we will implement to deal with both shortcomings.

As it is well understood in the literature, at finite temperature, the self energy of a particle receives higher loop corrections from daisy diagrams, e.g.~see \cite[Fig.~3a]{FENDLEY1987175}. Such corrections at $\mathcal{N}$-loop order contain powers of a field- and temperature-dependent parameter $\alpha$ (up to a normalization factor)\cite{Weinberg:1974hy,PhysRevD.9.3320,Kirzhnits:1974as,FENDLEY1987175,PhysRevD.45.4695},
\begin{align} \label{eq:highTd}
\alpha^{\mathcal{N}} = \lambda_i^{\mathcal{N}} \frac{T^{2\mathcal{N}}}{M_i^{2\mathcal{N}} (\hat{\Phi})},
\end{align}
where $\lambda_i$ is the coupling corresponding to $M_i (\hat{\Phi})$ in the theory. At large temperatures, such contributions exhibit severe IR divergence for some field values such that $M_i (\hat{\Phi}) \ll T$, for example around the origin, where higher loop contributions dominate and the fixed order calculation becomes problematic.
Various treatments have been proposed to resum higher loop thermal contributions and solve the associated IR problem \cite{Weinberg:1974hy,PhysRevD.9.3320,Kirzhnits:1974as,FENDLEY1987175,PhysRevD.45.4695,Nakkagawa:1998xc,Arnold:1992rz,Curtin:2016urg,Croon:2020cgk}.
A full dressing daisy resummation involves adding thermal corrections to the tree level effective masses in the effective potential. For the one loop EP it follows,
\begin{align} \label{eq:fd}
& V_{\rm CW} (\{M^2_i (\hat{\Phi})\}; \mu_R^2)+V^T_{\rm 1-loop} (\{M^2_i (\hat{\Phi})\}; T)\rightarrow \nonumber \\
&V_{\rm CW} \left(\{M^2_i (\hat{\Phi} ) + \Pi^2_i \}; \mu_R^2\right)+V^T_{\rm 1-loop} \left(\{M^2_i (\hat{\Phi})+ \Pi^2_i \}; T\right),
\end{align}
where $\Pi_i ^2$ is the squared thermal mass for the specie $``i"$. Such a procedure effectively resums higher order corrections from daisy diagrams\footnote{There are several relevant discussions in the literature, e.g.~\cite{Arnold:1992rz,Curtin:2016urg,Laine:2017hdk}, pointing out different types of finite temperature contributions due to the different implementation of thermal mass effects, including full vs partial daisy resummation, as well as higher order loop corrections from finite temperature resummations such as those coming from superdaisy, lollipop and sunset diagrams. In this study we restrict ourselves to the full daisy resummation approach and leave further investigation for future work.}.

\begin{figure}[t]
\includegraphics[width=0.5\textwidth]{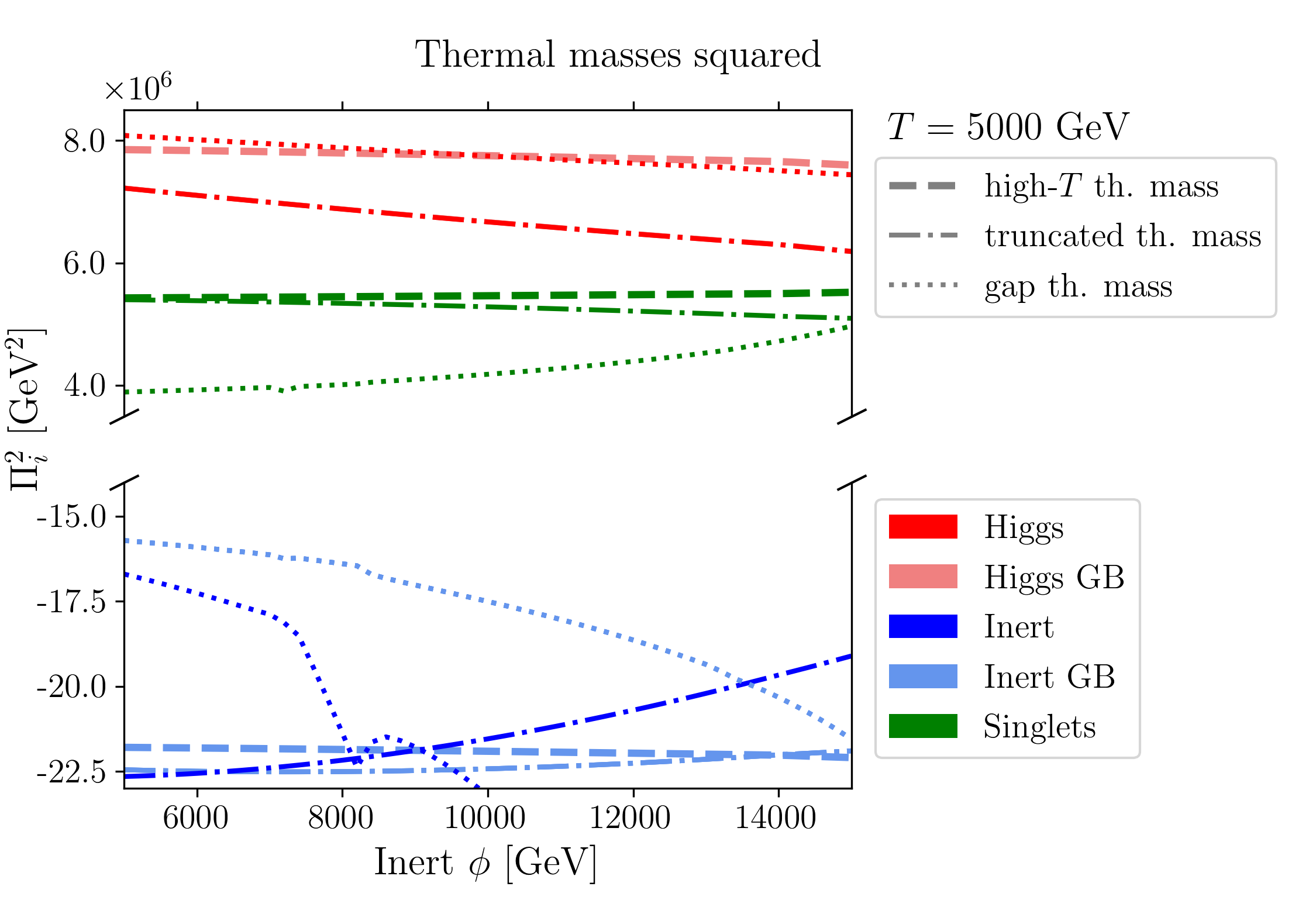}
\caption{Squared thermal mass of the scalars for the BM scenario B as a function of the inert field values at $T=5000$~GeV for different thermal mass implementations. Several lines are overlapping: Higgs and Higgs Goldstone boson (GB) thermal masses are degenerate when scanning the Inert direction; the high-T thermal mass for Inert and Inert GB are always degenerate.}
\label{fig:thermal.mass}
\end{figure}

The squared thermal masses $\Pi_i ^2$ are in general field- and temperature-dependent and can be solved by gap equations. At one-loop level the gap equations read
\begin{align}
\label{eq:daisygap}
\Pi_{i,{\rm gap}}^2 = \frac{\partial^2 }{\partial \hat{\Phi}_i^2} \sum_kV^T_{\rm 1-loop} \left(\{M^2_k (\hat{\Phi})+ \Pi^2_{k,\rm gap} \}; T\right) ,
\end{align}
where the degree of freedom $i$ appears as a background field in the EP. A truncated treatment involves doing an expansion of the right hand side of the gap equation with respect to $\Pi^2_k$ and truncate to a given order. To the leading order, the truncated squared thermal mass reads
\begin{align}
\label{eq:daisytrunc}
\Pi_{i,{\rm trunc}}^2 = \frac{\partial^2 }{\partial \hat{\Phi}_i^2} \sum_kV^T_{\rm 1-loop} \left(\{M^2_k (\hat{\Phi}) \}; T\right).
\end{align}
If the thermal potential is evaluated to leading order in high-temperature expansion, one obtains the well known field-independent form of the squared thermal masses
\begin{align}
\label{eq:daisyhT}
\Pi_{i,{\rm hT}}^2 = c_i T^2.
\end{align}
The $c_i$ are constant coefficients dependent on couplings determined by the theory, and we collect the thermal mass coefficients $c_i$ for all degrees of freedom in our model in appendix~\autoref{sec:odaisy}. We implement the high-T~thermal masses in eq.~\eqref{eq:daisyhT}, the truncated thermal masses in eq.~\eqref{eq:daisytrunc}, and the gap thermal masses in eq.~\eqref{eq:daisygap} in comparison, to effectively resum higher-order daisy diagrams. In~\autoref{fig:thermal.mass}, we show the squared thermal mass of the scalars as a function of the inert field values at a temperature of $T=5000$~GeV, computed with the different levels of accuracy described above, for the BM scenario B to be defined in~\autoref{table:BM}. The high-T thermal masses should be independent of the inert field value $\phi$, however,~\autoref{fig:thermal.mass} shows a small variation with respect to the field value due to the RG improvement implementation to be discussed below. The truncated thermal masses have an enhanced dependence of the inert field value, especially for the inert thermal mass itself, but a more sizable variation occurs for the gap thermal masses. The differences among thermal masses for different implementations as shown in~\autoref{fig:thermal.mass} will end up, however, having a very small impact on the results relevant for the phase structure of the EW non-restoration BMs.

The fixed order EP at finite temperature, including both the zero temperature and thermal contributions, depends on the scale $\mu_R$ at which the theory is renormalized. For example, at one loop order, using the high-temperature expansion in eq.~\eqref{eq:JhighT}, the potential has a logarithmic dependence on the renormalization scale as
\begin{align}
\label{eq:cancel}
\log \left( \frac{T^2}{\mu_{\rm R}^2} \right)^{},
\end{align}
where the $\log (M^2_i (\hat{\Phi})+ \Pi^2_i)$ piece is cancelled between the CW and logarithmic term in the high-temperature expansion of the thermal potential contribution. By implementing RG improvement, where the parameters, fields and vacuum energy of the potential are evaluated at the scale $\mu_R$, one would cancel the scale dependence to the order of the calculation. As we only calculate the effective potential and the RG improvement at one-loop order, the scale $\mu_R$ needs to be chosen wisely to avoid un-resummed large logarithms from higher-order loop effects. Formally, at $T=0$ with a convenient choice of the renormalization scale, the L-loop effective potential with an RG improvement at (L + 1)-loop order, is exact up to L-th-to-leading log order \cite{Bando:1992np,Bando:1992wy,Casas:1998cf}.
At finite temperature, the choice of the renormalization scale, should vanish or minimize the un-resummed logarithms such as $\log^{\mathcal{N}} \left( \frac{M^2_i (\hat{\Phi})+ \Pi^2_i }{\mu_{\rm R}^2} \right)^{}$ for ${\mathcal{N}} \ge 2$ \cite{Arnold:1992rz,Laine:2017hdk}. Our model at hand involves multiple degrees of freedoms, therefore, there is no single choice of the scale to make all the logarithms negligible. In this work, we choose
\begin{align}
\label{eq:ren.scale}
\mu^2_{R} = {\rm Max} \left\{ M^2_i (\hat{\Phi})+ c_i T^2 ; (246\ {\rm GeV})^2\right\},
\end{align}
where $i$ runs over all degrees of freedom (mass eigenstates) in the plasma. This is a convenient choice as long as there is no large separation between scales of the particles' masses, including the thermal mass contribution, as well as between the particle masses and the temperature, as it is the case in our study. The CW potential further includes polynomial contributions of the radiative corrections. It also partially accounts for multi-scale particle threshold effects beyond the one single scale threshold taken into account through the RG improvement.
We collect the one-loop beta functions and wave function renormalization factors for our model in appendix~\autoref{app:rge} and implement the RG improvement for all numerical calculations \footnote{Notice that here the RG improvement we perform does not involve temperature flow as has been proposed, for example in \cite{Liao:1995gt,Nakkagawa:1996ju,Nakkagawa:1997hg}, where they treat temperature as an independent scale that participates in the RG flow and thermal diagrams, like daisy and super-daisy, would have been resumed as a result.}.

\section{Mean field analysis for the thermal history}
\label{sec:pt}

This section provides an analytical understanding of the model parameter space compatible with the desired thermal history - the electroweak symmetry stays non-restored in the inert sector up to temperatures much higher than the EW scale, whereas the agent of the electroweak symmetry breaking changes at temperatures around the EW scale from the inert Higgs sector to the SM one.
In this work, we do not explicitly discuss the UV scale physics completion that may lead to electroweak symmetry restoration at even higher energies and hence would allow for the possibility of EWBG. However, we will study the conditions necessary for the suppression of the sphaleron rate as a function of the model parameter space through the whole temperature regime for which the electroweak symmetry is broken. More specifically, we will explore
the constraints on the ratio between the electroweak symmetry breaking vevs to the temperature that may allow for such a suppressed sphaleron rate.
This will provide a framework for future EWBG~model building.
If, instead, the new physics UV completion would directly provide a source of baryon asymmetry at the high scale, such as, for example, in the case of Leptogenesis, GUT-baryogenesis or Affleck-Dine baryogenesis~\cite{Affleck:1984fy}, then the requirement on the sphaleron rate could be ignored. A discussion of possibilities for baryogenesis as well as specific details on the sphaleron rate relevant for our model will be presented in~\autoref{sec:sph}.

We summarize the above desired thermal history with three conditions as follows
\begin{itemize}

\item{\bf{C1: Non-restoration of the electroweak symmetry}}

  This is realized up to very high temperatures by having a non-trivial inert phase: $\langle \varphi \rangle_{\rm highT} \ne 0$;

\item{\bf{C2: Phase transitions from the inert Higgs phase to the SM Higgs phase}}

  This condition secures that the Universe is at the SM vacuum at zero temperature, while being compatible with C1.

\item{\bf{C3\footnote{As discussed above, this condition is optional.}:
      $\!$Sufficiently suppressed sphaleron rate after EWSB}}

  This would allow preserving any baryon number density that may be generated through an EWBG~mechanism at the ultraviolet.

\end{itemize}

To gain an analytical understanding of the model parameter space compatible with the above conditions, we use a mean-field approximation of the finite temperature effective potential, where the thermal potential is evaluated up to leading order of the high-temperature expansion
\begin{widetext}
\begin{align}
\begin{aligned}
\label{eq:SN+2HDM.V}
V_{\mathbb{Z}_{N}+{\rm I2HDM}}^{\rm MF} =
&-\frac{1}{2} \left( \mu_H^{2} - c_h T^2\right) h^{2}+ \frac{1}{2}\left( \mu_{\Phi}^2 + c_{\varphi} T^2\right)\varphi^{2} +\frac{1}{2}\left( \mu_{\chi}^2+c_{\chi} T^2\right)\chi_{i}^{2}\\
&+ \frac{\lambda_H}{4}h^{4} + \frac{\lambda_{\Phi}}{4}\varphi^{4} +\frac{\widetilde{\lambda}_{\chi}}{4}\chi_{i}^{4} +\frac{\lambda_{\chi}}{4}(\chi_i\chi_{i})^{2}
+\frac{\Lambda_{H\Phi}}{4}\varphi^{2}h^{2}+\frac{\lambda_{\Phi\chi}}{4}\varphi^{2}\chi_{i}^{2} +\frac{\lambda_{H\chi}}{4} h^2 \chi_i^2,
\end{aligned}
\end{align}
\end{widetext}
where $c_i$ for $i = h,\varphi,\chi$ are given in eqs.~\eqref{eq:thermal.scalar.1}--\eqref{eq:thermal.scalar.3}. Such a mean-field potential is a reliable approximation before considering RG improvement and daisy resummation, especially at high temperatures. We shall include resummations in the next section for a full numerical study at high field values and temperatures. Here we provide an analytical study based on the mean-field potential to obtain a coarse understanding of how the desired thermal history is achieved within our model.

Let us first study the SM and inert Higgs sector phases of the potential in eq.~\eqref{eq:SN+2HDM.V} . An inert phase P$_{\Phi}$, where only the inert Higgs field has a non-zero field value, reads
\begin{align}
{\rm P}_{\Phi}: \langle \left( h, \varphi, \chi_1,\cdots, \chi_N\right) \rangle = \left(0, w(T), 0, \cdots, 0\right)
\end{align}
with
\begin{align}
w(T) &= \sqrt{-\frac{\mu_{\Phi}^2 + c_{\varphi} T^2}{\lambda_{\Phi} }}.\label{eq:wIDME}
\end{align}
At very high temperatures, $T^2 \gg \mu_{\Phi}^2$, one can approximate
\begin{align}
w(T)\approx \sqrt{-\frac{ c_{\varphi}}{\lambda_{\Phi} }} T.\label{eq:wIDMEh}
\end{align}
Given the BFB condition that $\lambda_{\Phi} >0$, a negative thermal mass coefficient $c_{\varphi}$,
\begin{align}
\label{eq:cphi}
c_{\varphi} = \frac{\lambda_{\Phi}}{2} + \frac{\lambda_{H\Phi} + \tlambda_{H\Phi}/2}{6}+ \frac{3g^{2}+g^{\prime 2}}{16} + N \frac{\lambda_{\Phi\chi}}{24},
\end{align}
generates a non-zero inert phase at very high temperatures, which is the key to achieve electroweak symmetry non-restoration
(or delayed restoration) in the inert sector in our model. This provides for condition C1 in the mean field approximation as
\begin{align}
{\rm C1}_{\rm MF} \quad \rightarrow \quad c_{\varphi} < 0.
\end{align}
The main driver of a negative $c_{\varphi}$ is a negative cross quartic between the inert and the singlet sector $\lambda_{\Phi\chi}$, whose negative contribution is magnified by the number of singlets $N$. If the inert mass parameter $\mu^2_{\Phi} \ge 0$, such a phase where only the inert field has a non-zero vev would disappear at a temperature $T_{ \Phi}^r$ (either as a global or local minimum), where
\begin{align}
T_{ \Phi}^r &= \sqrt{\frac{\mu_{\Phi}^2}{-c_{\varphi} }}.\label{eq:trphi}
\end{align}
A low restoration temperature $T_{\Phi}^r$ facilitates the existence of phase transitions between the inert and SM Higgs phases as well as the associated condition for a suppressed sphaleron rate, which will be discussed in more detail below. Instead, if $\mu^2_{\Phi} < 0$, this inert phase exists at zero temperature, which puts a constraint
\begin{align} \label{eq:const1}
\mu^2_{\Phi} > - \sqrt{\frac{\lambda_{\Phi}}{\lambda_H}} \mu_{H}^2
\end{align}
for it to be above the EW vacuum at $T=0$, i.e.~$V_0 (0, w(0), 0, \cdots 0) > V_0 (v_0, 0, 0, \cdots 0)$, in addition to condition in eq.~\eqref{eq:tach}.

A SM Higgs phase P$_H$ of the potential, where only the SM Higgs has a non-zero field value, reads
\begin{align}
{\rm P}_H: \langle \left( h, \varphi, \chi_1,\cdots, \chi_N\right) \rangle =(v(T), 0, 0, \cdots, 0)\label{eq:}
\end{align}
with
\begin{align}
v(T) &= \sqrt{\frac{\mu_H^2 - c_{h} T^2}{\lambda_H }},\label{eq:vIDME}
\end{align}
where at zero temperature it becomes the EW vacuum with $ v(0) = v_0$. Such a phase appears at a temperature
\begin{align}
T_{H}^r &= \sqrt{\frac{\mu_H^2}{c_h }} .\label{eq:}
\end{align}

Another phase that possibly exists during the thermal history is when both the SM Higgs and the inert Higgs fields acquire simultaneously non-zero values
\begin{align}
{\rm P}_{H\Phi}: \langle \left( h, \varphi, \chi_1,\cdots, \chi_N\right) \rangle =(\widetilde{v}(T), \widetilde{w}(T), 0, \cdots, 0)\label{eq:}
\end{align}
where
\begin{align}
\widetilde{v} (T) &= \sqrt{\frac{\widetilde{\mu}_H^2 - \widetilde{c}_{h} T^2}{\widetilde{\lambda}_H }},\quad
\widetilde{w} (T)= \sqrt{- \frac{\widetilde{\mu}_{\Phi}^2 + \widetilde{c}_{\varphi} T^2}{\widetilde{\lambda}_{\Phi} }}\label{eq:vtwt}
\end{align}
with
\begin{equation}
\begin{split}
\widetilde{\mu}_H^2 \equiv \mu_H^2 + \frac{\Lambda_{H\Phi}}{2 \lambda_{\Phi}} \mu_{\Phi}^2 ,&\quad
\widetilde{\mu}_{\Phi}^2 \equiv \mu_{\Phi}^2 + \frac{\Lambda_{H\Phi}}{2 \lambda_H} \mu_H^2 , \\
\widetilde{c}_{h} \equiv c_h - \frac{\Lambda_{H\Phi}}{2 \lambda_{\Phi}} c_{\varphi},&\quad
\widetilde{c}_{\varphi} \equiv c_{\varphi} - \frac{\Lambda_{H\Phi}}{2 \lambda_H} c_h, \\
\widetilde{\lambda}_H\equiv \lambda_H - \frac{\Lambda_{H\Phi}^2}{4 \lambda_{\Phi}} ,&\quad
\widetilde{\lambda}_{\Phi} \equiv \lambda_{\Phi} - \frac{\Lambda_{H\Phi}^2}{4 \lambda_H} ,
\end{split}
\end{equation}
implying that this phase is governed by the Higgs-Inert mixing coupling $\Lambda_{H\Phi}$ defined in eq.~\eqref{eq:BFBaux}. An important feature of this phase is that given the potential in eq.~\eqref{eq:SN+2HDM.V}, the potential difference reads
\begin{equation}
\begin{split}
&V({\rm P}_{H\Phi}; T) -V ({\rm P}_H; T) \propto - (4\lambda_{\Phi} \lambda_H - \lambda_{H\Phi}^2)^{-1} \\
&V ({\rm P}_{H\Phi}; T) - V ({\rm P}_{\Phi}; T) \propto- (4\lambda_{\Phi} \lambda_H - \lambda_{H\Phi}^2)^{-1} ,
\end{split}
\end{equation}
where the proportionality coefficients are always positive independent of the temperature. Thus,
if $4\lambda_{\Phi} \lambda_H - \lambda_{H\Phi}^2 \le 0$, the Higgs-inert phase P$_{H\Phi}$ is irrelevant as it is always shallower than either the SM or inert Higgs phases. On the contrary, if $4\lambda_{\Phi} \lambda_H - \lambda_{H\Phi}^2 \ge 0$, as long as such a Higgs-inert phase exits, it is deeper than both the SM or inert Higgs phases, thus becoming the global minimum.

Concentrating on the case where P$_{H\Phi}$ is the global minimum at a given temperature, notice that the situation $4\lambda_{\Phi} \lambda_H - \lambda_{H\Phi}^2 \ge 0$ coincides with the BFB condition if $\lambda_{H\Phi} \le 0$, hence for negative/zero cross quartic, the Higgs-inert phase will be the global minimum at finite temperature.
Moreover, at zero temperature, the non-tachyonic condition enforced in eq.~\eqref{eq:tach} implies $\widetilde{\mu}_{\Phi}^2 = m_{\phi}^2 \ge 0$. This yields that whenever $4\lambda_{\Phi} \lambda_H - \lambda_{H\Phi}^2 \ge 0 \rightarrow \widetilde{\lambda}_\Phi \ge 0$, there is no real solution for $\widetilde{w}(0)$ in eq.~\eqref{eq:vtwt}, as expected since the non-tachyonic solution was derived under the assumption that the P$_H$ at $T=0$ being the physical vacuum.
In addition, let's recall that at very high temperatures we have restricted our case to the inert phase P$_\Phi$ being the global minimum (no electroweak symmetry breaking in the SM Higgs sector), hence eq.~\eqref{eq:vtwt} implies that we voluntarily enforced
\begin{align}
\widetilde{c}_h \ge 0 
\lor
\widetilde{c}_{\varphi} \ge 0
\end{align}
whenever $T^2 \gg \widetilde{\mu}_{H(\Phi)}^2$.
Given the above constraints (P$_{H}$ and P$_{\Phi}$ are the global minimum at $T=0$ and high temperatures, respectively), if the phase P$_{H\Phi}$ ever appears, in a temperature regime $T^2 \sim \widetilde{\mu}_{H(\Phi)}^2$, it develops at a temperature $ {\rm Max} \{ \widetilde{T}_{H}^r , \widetilde{T}_{\Phi}^r \}$ and must disappear at a lower temperature $ {\rm Min} \{ \widetilde{T}_{H}^r , \widetilde{T}_{\Phi}^r \}$. These two characteristic restoration temperatures are defined from eq.~\eqref{eq:vtwt} demanding that either $\widetilde{v} (\widetilde{T}_{H}^r )= 0$ or $\widetilde{w} (\widetilde{T}_{\Phi}^r )= 0$, respectively. Observe that, within the mean field approximation we are considering, from eqs.~\eqref{eq:wIDME}, \eqref{eq:vIDME} and \eqref{eq:vtwt}, it follows $w(T_{H}^c) = \widetilde{w} (T_{H}^c)$ and $v (T_{\Phi}^c) = \widetilde{v} (T_{\Phi}^c )$. As a consequence, $V({\rm P}_{H\Phi}; \widetilde{T}_{H}^r) = V({\rm P}_{\Phi}; \widetilde{T}_{H}^r)$ and $V({\rm P}_{H\Phi}; \widetilde{T}_{\Phi}^r) = V({\rm P}_{H}; \widetilde{T}_{\Phi}^r)$, which implies that, the critical temperature defining the transition between the P$_{H\Phi}$ and P$_{\Phi}$(P$_{H}$) phases is given by $T_{H}^c = \widetilde{T}_{H}^r $ ($ T_{\Phi}^c = \widetilde{T}_{\Phi}^r$), indicating that these transitions are second order within the mean field approximation. These equalities imply that,
\begin{align}
T_{H}^c = \sqrt{\frac{\widetilde{\mu}_H^2}{\widetilde{c}_h}},\quad T_{\Phi}^c = \sqrt{\frac{\widetilde{\mu}_{\Phi}^2}{-\widetilde{c}_{\varphi}}} ,
\end{align}
and their existence demands
\begin{align} \label{eq:cond1}
\frac{\widetilde{\mu}_H^2}{\widetilde{c}_h} \ge 0 
\land
\frac{\widetilde{\mu}_{\Phi}^2}{-\widetilde{c}_{\varphi}} \ge 0.
\end{align}
In the numerical study where we consider the full thermal potential as well as daisy contributions, such phase transitions could be affected and become first order. However, they would hardly be strongly first order in the absence of large thermal or tree level barriers.

Other possible phases associated with the finite temperature potential \eqref{eq:SN+2HDM.V} include the trivial point, which, as long as any of the above phases exist, yields a shallower value of the potential, as well as phases involving singlets with non-zero field values. The latter will not be further considered in this section as they are unlikely to participate in the thermal history. When evaluating the thermal history in the numerical section, however, all possible phases will be taken into account.

After having considered the existence of all possible phases and some of their properties, let us now concentrate on the specifics of the phase transitions from the inert sector to the SM Higgs sector.

First, we discuss the simpler case where the phase ${\rm P}_{H\Phi}$ either never appears or is irrelevant. In such a case, there should be a phase transition from ${\rm P}_{\Phi}$ to ${\rm P}_{H}$, as illustrated on the middle penal of the second row in~\autoref{fig:NRglobal}.
Given the potential in eq.~\eqref{eq:SN+2HDM.V}, such a transition happens at a critical temperature
\begin{equation} \label{eq:tc}
\begin{split}
T_c = \sqrt{\frac{\mu_H^2+ \sqrt{\lambda_H/\lambda_{\Phi}} \mu_{\Phi}^2 }{ c_h- \sqrt{\lambda_H/\lambda_{\Phi}} c_{\varphi}} },
\end{split}
\end{equation}
at which the potential becomes degenerate $V({\rm P}_{H};T_c) =V({\rm P}_{\Phi};T_c) $. The condition for such a $T_c$ to exist reads (with help from the zero temperature constraint eq.~\eqref{eq:tach})
\begin{align}
\label{eq:c2MF}
- \sqrt{\frac{\lambda_{\Phi}}{\lambda_H}} \mu_H^2 < \mu_{\Phi}^2 <- \frac{c_{\varphi}}{ c_h }\mu_H^2,
\end{align}
and this will have a relevant impact on the allowed values of the inert Higgs boson mass, as will be discussed later on.

As mentioned in C3, to allow for the possibility of a EWBG~after UV completion, we will look at the conditions on the sphaleron rate at finite temperatures. The dilution of the baryon number density after the
onset of a UV induced EWPT responsible for the EWBG
will be double exponentially suppressed by the ratio of the sphaleron energy to temperature, see discussion in~\autoref{sec:sph}. Hence successful EWBG~in the complete model will require (see e.g.~\cite{Quiros:1999jp})
\begin{align}
\label{eq:c2MFdef}
\xi(T) = \frac{v_{\rm EW}(T)}{T} \equiv \frac{\sqrt{ \langle \varphi (T)\rangle^2 + \langle h (T) \rangle^2}}{T} \gtrsim 1,
\end{align}
where $\varphi$ and $h$ are the inert and SM Higgs fields charged under the EW gauge group. This condition should be satisfied at any temperatures throughout the thermal history from the creation of baryon asymmetry up to present times. It can be shown that such a condition can be satisfied if the phase transition ${\rm P}_{\Phi}$ to ${\rm P}_{H}$ fulfills
\begin{align}
\label{eq:c2MFtc}
{\rm Min} \left\{ \frac{ w(T_c)}{T_c} ,\frac{v(T_c)}{T_c} \right\}
\gtrsim 1.
\end{align}
This follows from the fact that as long as $\mu_{\Phi}^2 \ge 0$, as will be implemented in our BM scenarios,
\begin{align}
\label{eq:whytc}
&
\xi (T) = \frac{w(T)}{T} \ge \frac{w_c}{T_c} \quad \rm{for} \quad T \ge T_c,\\
&
\xi (T)= \frac{v(T)}{T} \ge \frac{v_c}{T_c} \quad \rm{for} \quad T \le T_c.
\end{align}

\begin{figure}[t]
\begin{center}
\includegraphics[width=1.\columnwidth]{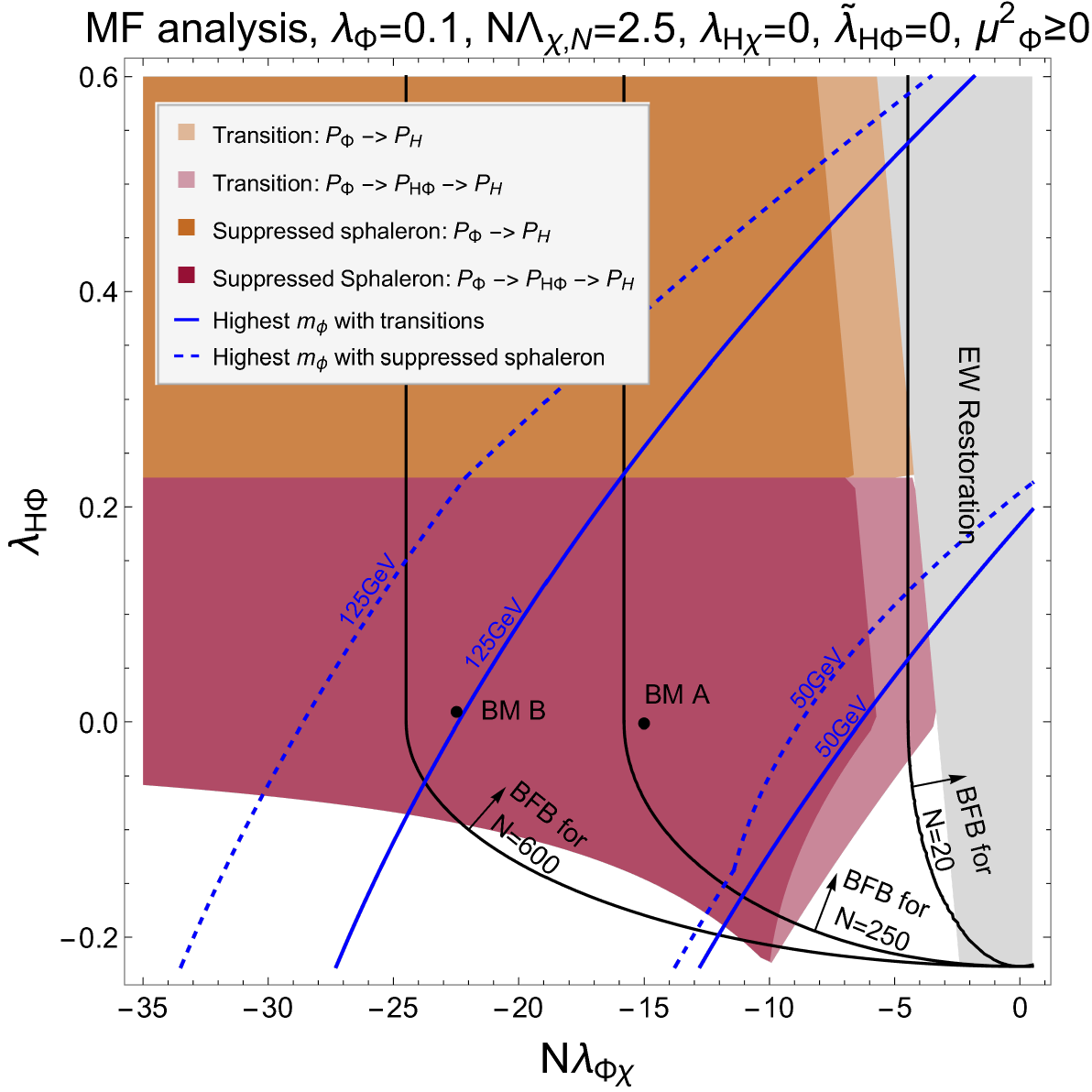}
\caption{Parameter space on the $N\lambda_{\Phi \chi} - \lambda_{H \Phi}$ plane compatible with desired thermal histories based on a mean field analysis. Relevant zero temperature constraints are also shown. Other parameters are fixed: $\lambda_{\Phi} = 0.1$, $N\Lambda_{\chi,n} = 2.5$, $\lambda_{H\chi} =0$, $\tlambda_{H\Phi} =0$, and $\mu_{\Phi}^2 \ge 0$ is imposed.}
\label{fig:MF}
\end{center}
\end{figure}

Next, we discuss the case where the phase ${\rm P}_{H\Phi}$ is relevant and appears as a global minimum in the thermal history, as illustrated on the left panel of the second row in~\autoref{fig:NRglobal}. To have a two step phase transition near the EW~scale
\begin{align}
{\rm P}_{\Phi} \xrightarrow[]{T_{H}^c} {\rm P}_{H\Phi} \xrightarrow[]{T_{\Phi}^c} {\rm P}_{H},
\end{align}
one needs
\begin{align} \label{eq:cond2}
T_{H}^c \ge T_{\Phi}^c,
\end{align}
with
\begin{align} \label{eq:}
\widetilde{\mu}_H^2\ge 0,\quad \widetilde{c}_h\ge 0,\quad \widetilde{\mu}_{\Phi}^2 \ge 0,\quad \widetilde{c}_{\varphi} \le 0,
\end{align}
which corresponds to the condition for these two temperatures to exist given by eq.~\eqref{eq:cond1}.

Analogous to the previous case, the condition to avoid baryon asymmetry washout in the context of a EWBG in an UV completed theory, would require
\begin{align} \label{eq:cond3}
{\rm Min} \left\{ \frac{w(T_{H}^c)}{T_{H}^c}, \frac{v(T_{\Phi}^c)}{T_{\Phi}^c} \right\}
\gtrsim 1.
\end{align}
Another thing to notice in this case is the role played by the mixing quartic $\lambda_{H\Phi}$, which controls the deviation from $T_{H}^c$ to $T_{H}^r$ and from $T_{\Phi}^c$ to $T_{\Phi}^r$. The smaller the mixing quartic, which is the region that we are mainly interested in, the smaller the deviations are. Moreover, in the region of small $\lambda_{H\Phi}$, the phase transition pattern ${\rm P}_{\Phi} \to {\rm P}_{H\Phi} \to {\rm P}_{H}$ is most likely to happen due to the decoupled contributions from the inert and SM Higgs minima to render the ${\rm P}_{H\Phi}$ minimum in the intermediate temperature range. This is apparent in~\autoref{fig:MF} to be discussed below.

It is also possible to have a temporary electroweak symmetry restoration at temperatures between those supporting the two EW breaking phase structures ${\rm P}_{\Phi}$ and ${\rm P}_{H}$. This is the case when $T_{\Phi}^{r}$ is higher than $T_{H}^{r}$, as illustrated on the right penal of the second row in~\autoref{fig:NRglobal}. Since in the temperature range between $T_{H}^{r}$ and $T_{\Phi}^{r}$ the system is in a EW restoring phase, this scenario
would allow for the EW sphaleron to be active in this regime. The sphaleron will wash out any baryon asymmetry that could have been generated by high scale EWBG. At this moment, we will mainly focus on the previous cases that are compatible with an UV EWBG mechanism.

Another possible case is a more fine-tuned four-step phase transition when $T_{H}^c \le T_{\Phi}^c$ and $T_{H}^r \ge T_{\Phi}^r$. This case will require large mixing quartic and significant fine-tuning of the parameter space. We do not further concentrate on this case.

In~\autoref{fig:MF}, we show the parameter space spanned by $N\lambda_{\Phi \chi} - \lambda_{H \Phi}$ considering the zero temperature constraints discussed in~\autoref{sec:model} and the different thermal history possibilities discussed above. The region violating condition C1$_{\rm MF}$ is shaded gray, while the regions satisfying the thermal history patterns and the non-washout conditions are highlighted with light and dark orange (light and dark maroon) for the transition pattern ${\rm P}_{\Phi} \to {\rm P}_{H}$ (${\rm P}_{\Phi} \to {\rm P}_{H\Phi} \to {\rm P}_{H}$), respectively.
There is no region that satisfies the rare four-step phase transition. The conditions for the correct zero temperature vacuum structure are satisfied on the whole parameter space if we impose $\mu_{\Phi}^2, \mu_{\chi}^2 \ge 0$.
The region giving a tree-level BFB potential, calculated from conditions \eqref{eq:BFB1}, is at the right side of the black solid lines for different number of singlet scalars $N$.
Notice that, within the mean-field approximation, the thermal history patterns, as well as the non-washout requirements, are independent on $N$ as long as the value of $N\lambda_{\Phi \chi} $ is kept a constant. Since both the thermal histories and non-washout conditions are strongly correlated to the inert mass parameter, the mass of the inert Higgs boson is in turn also constrained. In~\autoref{fig:MF}, we show solid blue lines that determine the maximal value of the inert Higgs boson mass compatible with the corresponding phase transition patterns for a given value of $N\lambda_{\Phi \chi}$ and $\lambda_{H \Phi}$. Higher values of the inert Higgs boson mass can be achieved to the left of the lines. Similar lines for the suppressed sphaleron rate conditions are shown by the dotted blue lines. Other parameters have been fixed in~\autoref{fig:MF} to be $\lambda_{\Phi} = 0.1$, $N\Lambda_{\chi,n} = 2.5$ and $\lambda_{H\chi} =0$, $\tlambda_{H\Phi} =0$. The SM Higgs sector parameters are fixed to satisfy the Higgs vev and mass at the tree level. We constrain the discussion to the case $\mu_{\Phi}^2 \ge 0$, which makes conditions \eqref{eq:c2MFtc} and \eqref{eq:cond3} sufficient to secure a suppressed sphaleron rate within the mean field approximation as discussed above. In addition, in~\autoref{fig:MF}, we also show the two benchmark points A and B \footnote{BM point B has a slightly different value of $N\Lambda_{\chi,n} $ than the one used in~\autoref{fig:MF}. However, the error of this point's position in the $N\lambda_{\Phi \chi} - \lambda_{H \Phi}$ plane is within the thickness of the point drawn in the plot.}, which will be discussed in the full numerical study in the next section.

From~\autoref{fig:MF}, one notices that the region where the cross quartic coupling between the inert and the SM Higgs sectors almost vanishes, i.e.~$\lambda_{H\Phi} \sim 0$ and hence the SM Higgs sector is minimally perturbed, can be compatible with the desired thermal history. Main constraints on the parameter space come from the tension between the BFB and desired thermal history: the more negative the cross quartic $N\lambda_{\Phi \chi}$, the easier the non-restoration and the lower the critical temperatures which yield larger EW vev to temperature ratios $\xi(T)$. A more negative cross quartic coupling $N \lambda_{\Phi \chi}$ makes it harder for the potential to be BFB, as shown in eq.~\eqref{eq:BFB1}. Moreover, a larger number of singlets in turn helps to relax the BFB condition on $N \lambda_{\Phi \chi}$ by relaxing its lower bound while increasing the singlet effective quartic coupling $N\lambda_{\chi}$. As mentioned above, another constraint is on the mass of the inert Higgs boson. The restriction on the parameter space is alleviated for a lighter inert Higgs boson mass, especially in the region where the cross quartic $\lambda_{H\Phi}$ is small. This can be easily understood, for example in the ${\rm P}_{\Phi} \to {\rm P}_{H}$ phase transition pattern, since a smaller inert mass parameter $\mu_{\Phi}^2$ yields a lower critical temperature $T_c$, as is shown in eq.~\eqref{eq:tc}. A similar argument, although more involved, applies to the two-step phase transition.
The direct correlation between a smaller inert Higgs boson mass and a smaller inert mass parameter $\mu_{\Phi}^2$ especially holds in the region of small $\lambda_{H\Phi}$, as the one considered here. Observe however, an inert Higgs boson mass above half of the SM Higgs mass can be achieved, even with $\lambda_{H\Phi} \sim 0$, as long as the number of singlets is sufficient to be in the BFB allowed region.

The analysis in this section is based on the mean-field approach, where we consider the leading order high-temperature expansion of the thermal potential. For temperatures well above the EW scale however, including the RG improvement and the daisy resummation becomes necessary. In the next section, we perform a full numerical study for two benchmark points and present the results for different approximations.

\section{Numerical results on benchmark points} \label{sec:num}

In this section, we explore the thermal histories of two model benchmark points based on numerical calculation of the finite temperature effective potential prescriptions as described in~\autoref{sec:pert}. Appendix~\autoref{app:numerics} contains the details of our {\tt python} implementation. The main result of the algorithm is the value of the global minimum at a given temperature. The set of all global minima at a given set of temperatures defines the phase history we consider. A phase transition is observed when there is a change of phase pattern (e.g. from an inert-only-phase to an inert-SM Higgs phase) at a certain temperature\footnote{ We leave a detailed scan of the transition using nucleation temperatures instead of critical temperatures to later work.}. In this section, we also explore the value of the EW vev to temperature ratio $\xi(T)$, which is relevant for obtaining information on the sphaleron rate.

We define two characteristic benchmark points for our model - benchmark A that has inert Higgs eigenstates with masses slightly above half of the $Z$ boson mass, and a benchmark B that has inert Higgs eigenstates with masses slightly above half of the SM Higgs boson mass.
Inert mass eigenstates with masses above $100$ GeV can be achieved, but they would either lead to restoration of the electroweak symmetry at intermediate temperature scales or would require a number of singlet scalars of order $\mathcal{O} (1000)$ or more. The specific values of the model parameters and masses are given in~\autoref{table:BM}.

\begin{table}[t]
\caption{Parameter choices for the BMs A and B. The dimensionful quantities are in units of GeV.}
\label{table:BM}
\begin{tabular}{c|c|c|c|c|c|c|c|c}
& $\mu_{H}^{2}$ & $\lambda_{H}$ & $\mu_{\Phi}^{2}$ & $\lambda_{\Phi}$& $\mu_{\chi}^{2}$ & $\lambda_{\chi}$&$\lambda_{H\Phi}$&$\tlambda_{H\Phi}$\\
\hline
BM A& 8994.45& 0.119& 2500& 0.1 & 100 & 0.01&-0.001&0\\
\hline
BM B& 8991.84 & 0.119 & 5800& 0.1 & 5000 & 0.004&0.01&0\\
\multicolumn{8}{c}{}\\
 & $\lambda_{\Phi\chi}$&$\tlambda_{\chi}$&$\lambda_{H\chi}$ & \multicolumn{1}{c}{$N$}&\multicolumn{1}{||c|}{$m_{h}$} &$m_{\phi}$&$m_{\chi}$\\
  \hline
  BM A& -0.06& 0&0&\multicolumn{1}{c}{250}&  \multicolumn{1}{||c|}{125} &48.47&9.8\\
  \hline
  BM B&-0.0375 & 0&0&\multicolumn{1}{c}{600}& \multicolumn{1}{||c|}{125} & 84.58 & 68.87\end{tabular}
\end{table}  

\begin{figure}[t]
\begin{center}
\includegraphics[width=1.\columnwidth]{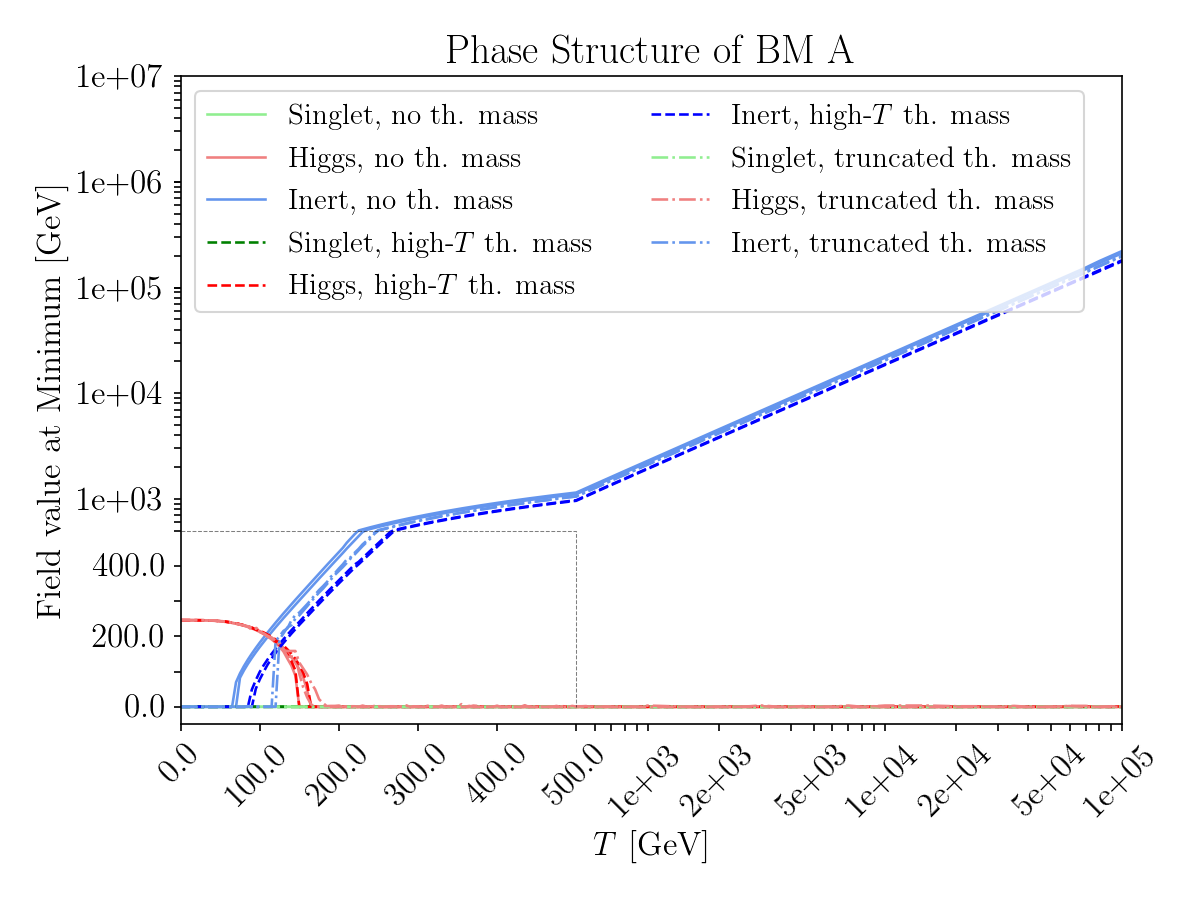}
\includegraphics[width=1.\columnwidth]{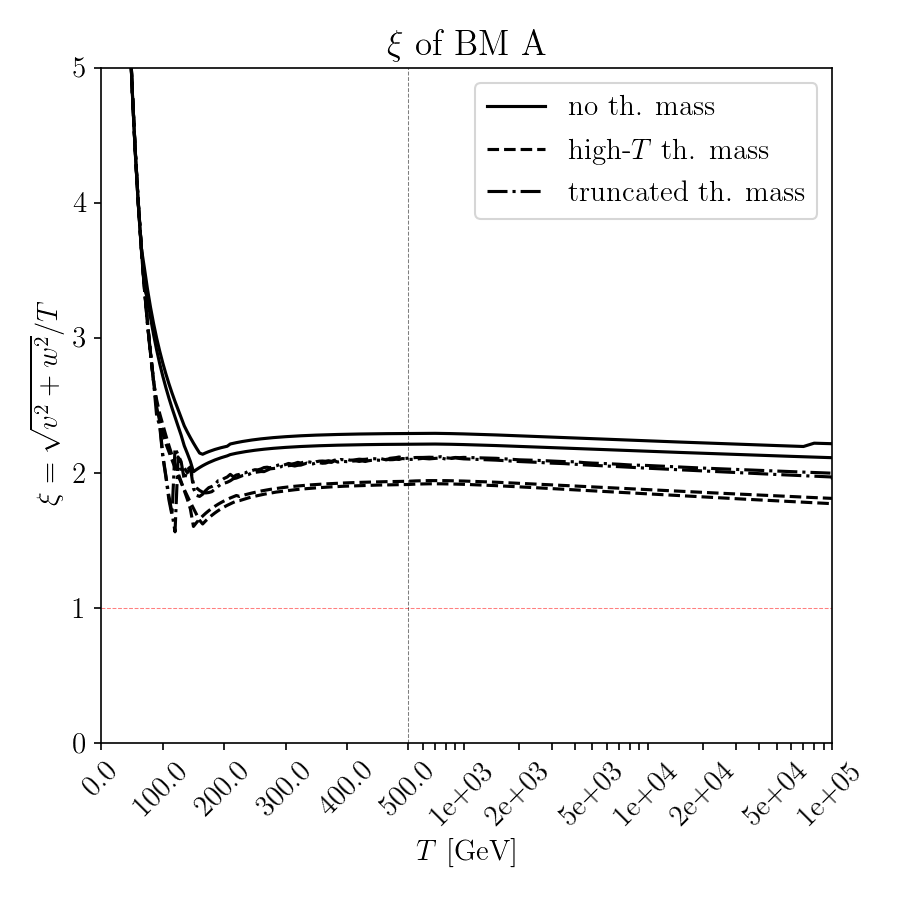}
\caption{Phase structure (upper panel) and EW vev-temperature ratio (lower panel) as a function of temperature, for different finite temperature implementations, for BM point A as defined in~\autoref{table:BM}. The singlets never obtain a non-zero vev in the entire temperature range. The black dotted line separates linear from logarithmic axes scales. The red dotted line in the lower panel marks $\xi=1$.}
\label{fig:BMlight}
\end{center}
\end{figure}
  
We implement the RG improvement on the BFB conditions of eq.~\eqref{eq:BFB1} and find that, at scales of the order of $10^5$ GeV, these conditions are violated for both BMs \footnote{There is a small dependence on the CW treatment that somewhat perturbs the SM Higgs quartic coupling as is explained in~\autoref{sec:pert}.}. Such an energy scale is of the order of the scale at which the SM Higgs quartic coupling becomes negative through its SM one loop RGE. This is expected since we consider that the SM Higgs only interacts with the extended scalar sector through a tiny inert-Higgs coupling, and therefore its quartic coupling evolution should be minimally perturbed compared to its SM behavior. The scale above gives a rough estimate of the energy scales up to which our results can be trusted.
By minimizing the finite temperature potential numerically, we have checked that the potential remains stable up to high energy scales shown below for both BMs. We also ran the RGE of the model (see eqs.\eqref{eq:RGEs}) for BMs A and B and found that Landau poles appear at energies around $2.5 \cdot 10^{14}$~GeV and $10^{15}$~GeV, respectively - well above the scale of validity of the theory at the one-loop RGE level.

\begin{figure}[t]
\begin{center}
\includegraphics[width=1.\columnwidth]{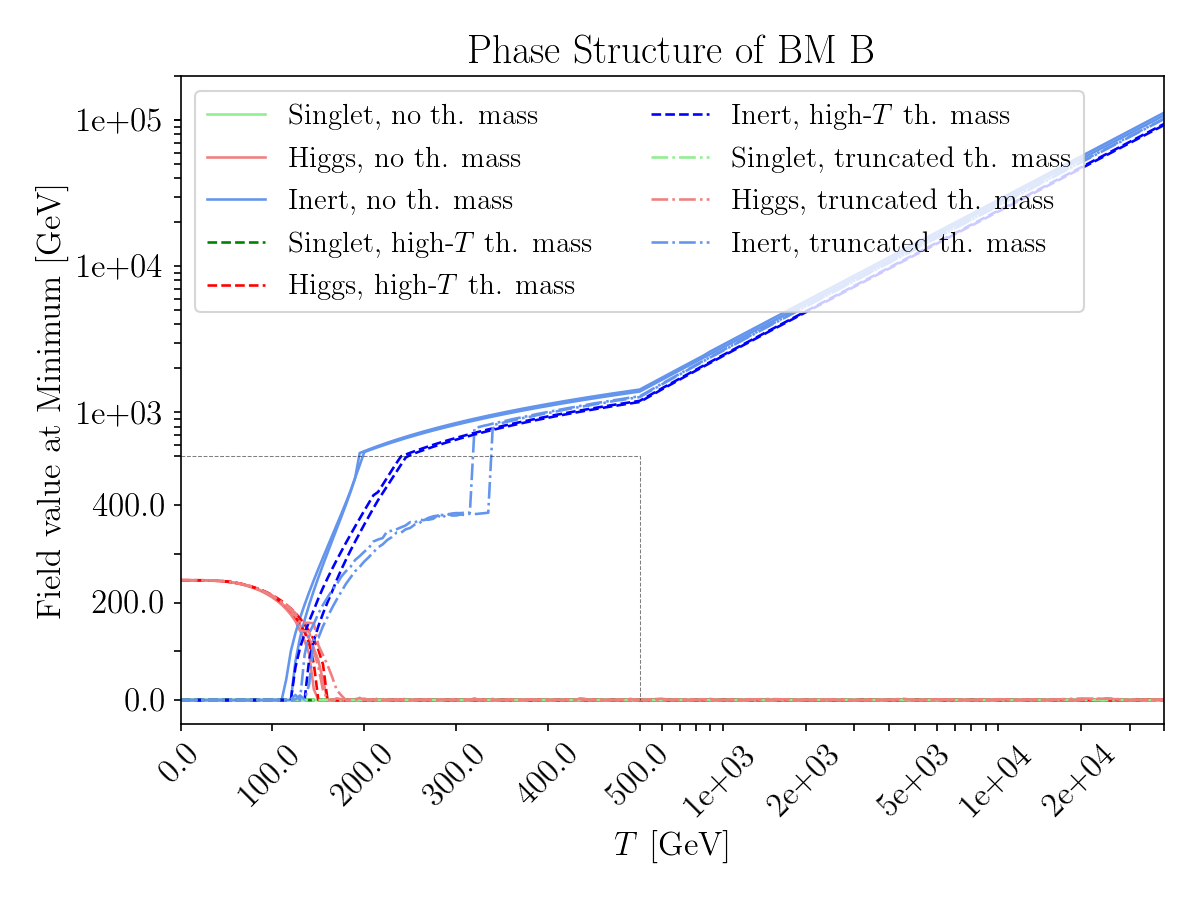}
\includegraphics[width=1.\columnwidth]{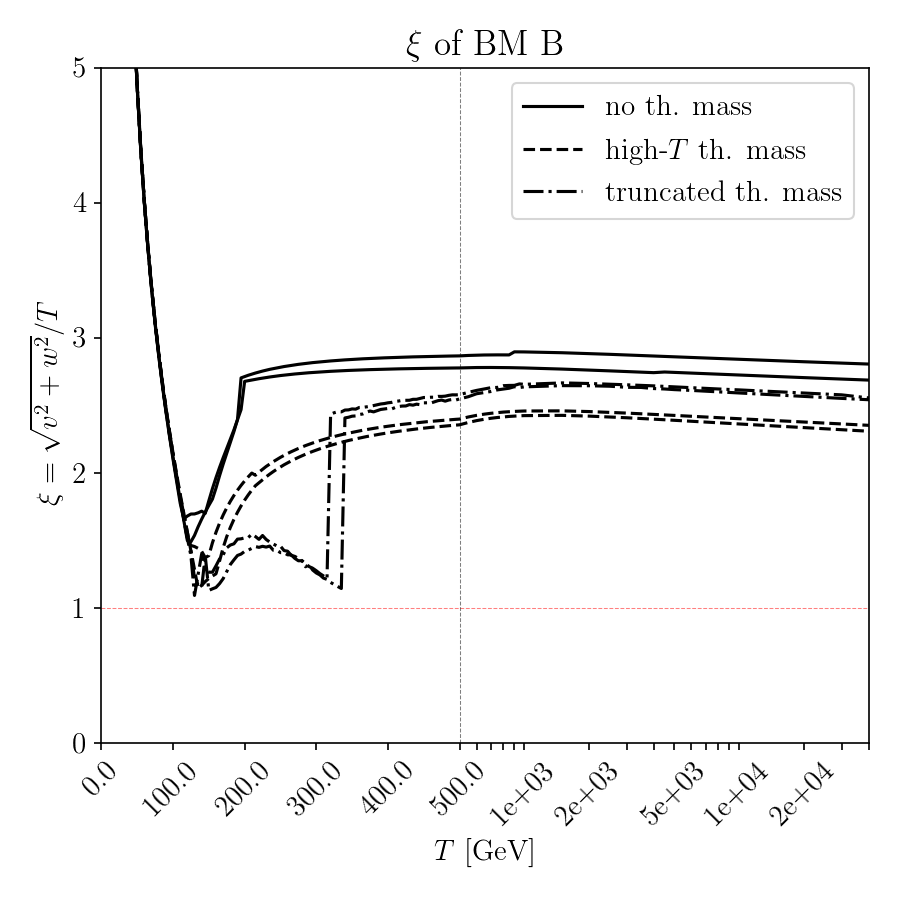}
\caption{Phase structure (upper panel) and EW vev-temperature ratio (lower panel) of BM B. Model parameters of the BM are given in~\autoref{table:BM}. The singlets never obtain a non-zero vev in the entire temperature range. The black and red dotted lines are as in~\autoref{fig:BMlight}.}
\label{fig:BMheavy}
\end{center}
\end{figure}

In~\autoref{fig:BMlight} and~\autoref{fig:BMheavy}, we show the phase structure (upper panel) and EW vev-temperature ratio (lower panel) for BMs A and B, respectively, and for different implementations of the finite temperature effective potential as introduced in~\autoref{sec:pert}. In the phase structure plot, we are showing as a function of the temperature the field values of the SM Higgs (red), inert Higgs (blue), and singlet (green) \footnote{ We assume all singlets have the same vev --- it's either all or none. Given that $\tlambda_{\chi}=0$, which we chose at tree level and is protected against RGE, we have the $SO(N)$ symmetry that we can use to rotate in that form.} at the global minimum. In the vev-temperature ratio plot, we show the value of $\xi(T)$, as defined in eq.~\eqref{eq:c2MFdef}, as a function of the temperature. To showcase the uncertainties associated with different finite temperature implementations, we show results obtained with no daisy resummation (solid lines), daisy resummation with high-T thermal masses, as in eq.~\eqref{eq:daisyhT}, (dashed lines), and daisy resummation with truncated thermal masses, as in eq.~\eqref{eq:daisytrunc}, (dashed-dotted lines). In addition, we have included the RG improvement for all calculations, and consider the uncertainties related to the CW potential, which takes care of multi-scale issues beyond the RG improvement. In the figures, we use the same type of lines to represent a given finite temperature approximation with or without the CW contribution. Hence the space in between the lines shows the uncertainty related to the CW effects. It is apparent from the figures that this accounts for a small effect, and we will not discuss it any further.

In~\autoref{fig:BMlight}, for BM A, one observes that the major uncertainty is caused by the effects of daisy resummation and the impact of different thermal mass treatments within the daisy resummation. However, the most important feature of these results is that the qualitative behavior of the phase structure, in~\autoref{fig:BMlight} upper panel, and the EW vev-temperature ratio affecting the sphaleron rate, in~\autoref{fig:BMlight} lower panel, is not significantly modified by the different finite temperature treatments. In fact, BM A exhibits both the feature of EW non-restoration until high energies and $\xi(T) > 1$.
The plots of BM A are shown up to the temperature of $10^5$ GeV, after which the potential becomes unbounded from below. Observe that there is a kink below/about $200$ GeV, which is due to the phase transition pattern from the P$_{\Phi}$ phase to the P$_{H\Phi}$ phase, and it is a physical effect.
In addition, there is a spike at $T \sim 110$ GeV for the daisy resummation with truncated thermal masses, which is, however, a defect of this finite temperature implementation. We expect this effect to be smoothed out when implementing an improved treatment of the thermal masses \footnote{This spike is a defect associated with the truncated thermal mass calculation, where the second derivative of the thermal potential diverges when its argument, $M^2/T^2$, is close to $0$. Indeed, such an effect does not happen for implementation with high-T thermal masses, as the divergence does not exist for the thermal potential within this approximation. Using the full gap equation, where the IR divergence is cured by including the thermal correction prior to performing the derivative, we expect the spike artifact shown in~\autoref{fig:BMlight} lower panel to disappear.}.

In~\autoref{fig:BMheavy}, we show similar results as for~\autoref{fig:BMlight}, but for a heavier inert Higgs boson mass of the order of $m_h/2$ that will allow for different phenomenology. Same as BM A, BM B exhibits both the feature of EW non-restoration until high energies and $\xi(T) > 1$. The plots are shown up to the temperature of $T=4\cdot 10^{4}$~GeV, after which the potential becomes unbounded from below.
For the BM B, there is a kink above/about $100$ GeV, which is due to the phase transition from the P$_{H\Phi}$ phase to the P$_{H}$ phase.
In addition, analogs to BM A, there is a spike at around $300 - 400$ GeV for the daisy resummation with truncated thermal masses, which we understand is the same type of artifact as discussed above and will be cured by implementing an improved treatment of the thermal masses.

As described above, using the gap equation, eq.~\eqref{eq:daisygap}, to derive the thermal masses is the most robust procedure. 
%DONOT DELETE THIS COMMENT: author Zhen Liu does not agree with the above statement.
However, solving the gap equation at every step in the minimization of the potential is computationally extremely expensive and is beyond the scope of this work. However, in order to secure that the non-restoration behavior at high temperatures survives the most precise treatment of the thermal masses through the gap equation, we checked for several high-temperature values all the way down close to the EW scale, that the non-restoration behavior and $\xi(T) > 1$ survive for both BM scenarios.

\section{Baryogenesis and Sphaleron Rate Supression}
\label{sec:sph}

In this section, we briefly discuss the possibilities of high-scale baryogenesis scenarios based on 1) EW-symmetry non-restoration up to scales as high as the GUT/Planck scale or 2) electroweak symmetry restoration around a UV scale of the order of validity of our model at which a new UV theory is in place. In the latter case, we expect to build a UV theory that allows for EWBG. Hence, in this case, we would like to explore in more detail the sphaleron washout constraints in our BM scenarios to preserve the created asymmetry down to zero temperature.

If the EW~symmetry, through a specific UV completion, were to remain broken well above the scale of validity of our current model \footnote{Our study is only including one-loop RGEs, but in analogy to the SM, we expect the validity of our model to be extended to higher energies by considering higher-order loop RGEs.}, possibly up to the GUT or Planck scale, this would enable baryogenesis mechanisms with little dependence on how the EWSB is triggered. In such case, the baryon asymmetry can be generated by a mechanism that creates a source of B-L$\ne$ 0, such as, for example, GUT-genesis, leptogenesis, or Afflect-Dine baryogenesis (\cite{Dine:2003ax} and references therein). Recall that, sphaleron processes preserve B-L, and hence an asymmetry will subsist once generated. However, they tend to wash out B+L as long as they remain active, thereby enabling conversion of Baryon (anti-Baryon) number into anti-Lepton (Lepton) number. For any specific B-L$\ne$ 0 mechanism, there will be additional model-building considerations for successful baryogenesis, including specifics of the new sources of CP violation and out of equilibrium conditions. It is important to notice that the two BMs we consider in this work imply that the sphaleron rate is suppressed during the broken-electroweak symmetry epoch, hence a mechanism such as Leptogenesis, that requires active sphalerons to convert Leptons into antiBaryons will not work. Other BMs could be studied that allow for sphalerons to become active at some intermediate energy scale during the temporary restoration of the electroweak symmetry, as in the lower right panel of~\autoref{fig:NRglobal}. Exploring these new ideas for baryogenesis will be the subject of future work.

In the case of a UV completion that induces a restoration of the electroweak symmetry at high energy scales of the order of the validity of our model, one can also require that such UV theory induces a strong first order phase transition and enables EWBG. Although building such UV theory will remain a topic of future work, let us briefly comment on the various ways that we can picture such a scenario.

In our minimal model, the restoration can occur through the RGE of the quartic couplings. Under the high-temperature expansion, one can visualize this possibility through the thermal coefficient $c_{\varphi}$ given in eq.~\eqref{eq:cphi}. If $c_{\varphi}$, which at lower temperatures has a negative value, were to become positive at a given high scale through the RGEs, this will render EW~restoration at high temperatures. For simplicity, let's consider the limit where in the IR the mixing quartics $\lambda_{H\Phi}$, $\widetilde \lambda_{H\Phi}$, and $\lambda_{H\chi}$ are zero, and neglect the leading log impact of these mixing quartics. The running of the thermal coefficient is then determined by the running of the linear combination of $\lambda_\Phi/2+(3g^2+g^{\prime\ 2})/16+N\lambda_{\Phi \chi}/24$. In our model the inert doublet self-coupling $\lambda_\Phi$ generically becomes larger at higher scales, while the mixing quartic $\lambda_{\Phi \chi}$, whose initial value is negative, could also increase, depending on the specific region of parameter space. However, we checked that the latter is not fulfilled for our BMs, hence, additional effects will be needed to restore the electroweak symmetry in these cases. There are indeed different ways to change the running behavior of $c_{\varphi}$, to allow for EW~restoration. For instance, one can consider that the inert doublet is charged under some new spontaneously broken U(1) gauge group with coupling $g^{\prime\prime}$. This will affect $c_{\varphi}$ directly by adding a $g^{\prime\prime\ 2}/16$ term after crossing the scale where the new U(1) is restored, rendering its gauge boson massless such that it starts contributing to the thermal mass of the inert doublet. Similarly, one can also introduce some heavy vector-like fermions (under SM gauge groups) that have Yukawa couplings to the inert doublet. When above the heavy fermion mass scale, this will add new positive contributions to the thermal coefficient $c_{\varphi}$ by $ y_F^2 N_F/12$, where $N_F$ is the color-factor or the specifies of new heavy fermions.

Beyond directly changing the thermal coefficient $c_{\varphi}$ above some mass threshold scale, one can also modify the running of the couplings contributing to $c_{\varphi}$, by adding new gauge and/or matter content. The minimal and simplest implementation would be to charge the $\chi$ field under some new $SU(N)$ gauge group. It directly contributes positively to the beta function of $\lambda_{\Phi\chi}$, which is the only source of negative quantities in the thermal coefficient $c_{\varphi}$, helping restore the EW~symmetry at a higher scale. On the contrary, matter fields interacting with the inert Higgs field seem to contribute negatively to the beta functions of the quartics contributing to $c_{\varphi}$, although, as discussed below, they may be required to secure a strong first-order phase transition.
An additional source of symmetry restoration could be to add scalar fields that directly couple to the inert field and acquire masses at high energies at which restoration would take place~\cite{Baldes:2018nel}.

An important additional issue related to the high energy EWBG~mechanism in the framework of delayed electroweak symmetry restoration, is that one needs to secure that a strong first-order phase transition takes place at the time of electroweak symmetry breaking. This is required by the out-of-equilibrium condition of Sakharov. Here, it is possible to exploit the existence of an inert fermion sector that suppresses the strength of the inert self-coupling and thereby enhances the strength of the phase transition.

\renewcommand{\arraystretch}{1.2}
\begin{table*}[t]
\caption{Dilution factors $f_{w.o.} =1 - n_B(t_{\text{now}}) / n_B(t_{\text{high}})$ for our benchmark models as defined by the integral in eq.~\eqref{eq:sph.8}. The upper limit of the integration, $T_{\rm high}$, is taken as the highest temperatures in~\autoref{fig:BMlight} and~\autoref{fig:BMheavy} respectively. The three entries per cell correspond to uncertainty choices of $\kappa$ as $0.01\kappa \;/\; ${\boldmath $\kappa$}$\;/\; 100\kappa $. Top and bottom row per BM refer to using the CW contribution (top) or not (bottom). }
\label{tab:washout}
\begin{tabular}{c|c|c|c|}
&no th. mass & high-$T$ th. mass& truncated th. mass\\
\hline
\multirow{2}{*}{BM A} &
$<10^{-16} \;/\; ${\boldmath $10^{-16}$}$\;/\; 10^{-14} $ &
$10^{-11} \;/\; ${\boldmath $10^{-9}$}$\;/\; 10^{-7} $&
$8\cdot 10^{-11} \;/\; ${\boldmath $8 \cdot 10^{-9}$}$\;/\; 8 \cdot 10^{-7} $\\
&$<10^{-16} \;/\; ${\boldmath $4\cdot 10^{-15}$}$\;/\; 4\cdot 10^{-13} $ &
$2\cdot 10^{-11} \;/\; ${\boldmath $2\cdot 10^{-9}$}$\;/\; 2\cdot 10^{-7} $&
$10^{-12} \;/\; ${\boldmath $10^{-10}$}$\;/\; 10^{-8} $\\
\hline
\multirow{2}{*}{BM B} &$9\cdot 10^{-10} \;/\; ${\boldmath $9\cdot 10^{-8}$}$\;/\; 9\cdot 10^{-6} $ &
$4\cdot 10^{-5} \;/\; ${\boldmath $4\cdot 10^{-3}$}$\;/\; 0.296 $&
$7\cdot 10^{-5} \;/\; ${\boldmath $7 \cdot 10^{-3}$}$\;/\; 0.498 $\\
& $4\cdot 10^{-12} \;/\; ${\boldmath $4\cdot 10^{-10}$}$\;/\; 4\cdot 10^{-8} $ &
$2\cdot 10^{-8} \;/\; ${\boldmath $2\cdot 10^{-6}$}$\;/\; 2\cdot 10^{-4} $&
$10^{-4} \;/\; ${\boldmath $0.012$}$\;/\; 0.694 $\\
\hline
\end{tabular}
\end{table*}
\renewcommand{\arraystretch}{1}

As it is clear from the above discussion, a successful UV model of high-temperature EWBG will demand detailed model building, which we leave for future publication. In the following, we will concentrate on the EW non-restoration case at hand, where the SM Higgs sector is minimally perturbed, to discuss details of the sphaleron rate.

Once the UV completion allows for the creation of the baryon asymmetry through an EWGB mechanism at high temperatures, one needs to evaluate the sphaleron washout factor to preserve the asymmetry all the way down to zero temperatures.
Our model generically predicts a slowly varying $\xi=v_{\rm EW}(T)/T$ up to high temperature, as well as a low scale phase transition between the inert doublet and the SM Higgs doublet phases near the weak scale.
Following a high-scale SFOPT triggered by a UV completion of the model, the sphaleron will become inactive quite fast after that transition, but there will be some dependence on its rate on the model parameters. To properly compute the washout (dilution) of the baryon number density, one should integrate the effects of the sphaleron rate
over a large range of temperatures (a large period of time), instead of the usual assumption that the washout factor is dominated near the vicinity of the phase transition and is treated as a constant.

Specifically, the amount of sphaleron induced washout is determined by two quantities: the product of prefactors entering in the sphaleron rate and the energy of the sphaleron that appears in one of the exponentials. The latter is straightforward to compute and largely depends on the gauge structure of the theory. We provide the necessary steps to get the sphaleron energy~\cite{Moreno:1996zm,Grant:2001at} in detail in appendix~\autoref{sec:ba}. The computation of the prefactors of the sphaleron rate are more model dependent. There are two different sources of deviations from the SM results. First, we have an extended scalar sector and the additional particles might contribute through indirect effects in the prefactors. Second, we focus on the inert doublet, and its quartic coupling is different than the quartic of the SM Higgs. We therefore discuss the specifications of the prefactors from the SM values~\cite{Carson:1989rf,Carson:1990jm} in detail in appendix~\autoref{sec:ba}. \\

The sphaleron rate can be written as
\begin{equation}
\label{eq:sph.2}
\frac{\Gamma}{V} = 4\pi \omega_{-} \mathcal{N}_{tr} \mathcal{N}_{rot} T^{3} \left(\frac{v_{\rm EW}(T)}{T}\right)^{6} \kappa \exp{\left[-E_{sph}(T)/T\right],
}
\end{equation}
where the evaluation of the prefactors $\omega_{-}, \mathcal{N}_{tr}, \mathcal{N}_{rot}, $ and $\kappa$ are explained in detail in~appendix~\autoref{sec:ba} and in~\autoref{fig:prefactor.inputs}.

The survival rate of the baryon number density at any given time $t$, after the onset of the transition at $t=0$ is~\cite{Mottola:1990bz,Patel:2011th,Quiros:1999jp}
\begin{equation}
\label{eq:sph.4}
\frac{n_{B}(t_{now})}{n_{B}(0)} = \exp{\left[- \frac{13 n_{f}}{2} \int_{0}^{t_{now}}dt\, \frac{\Gamma(T(t))}{V T^{3}(t)}\right]},
\end{equation}
where we consider present time, $t = t_{now}$, and with $n_{f}$ the number of fermion families.

In a radiation dominated Universe, changing the integration variable from time to temperature, the above equation reads
\begin{equation}
\label{eq:sph.8}
\frac{n_{B}(t_{\text{now}})}{n_{B}(t_{\text{high}})} = \exp{\left[- \frac{13 n_{f}}{2} \int_{0}^{T_{\text{high}}}dT\, \frac{\Gamma(T)}{V T^{6}} M_{Pl} \sqrt{\frac{90}{8 \pi^{3}g^{*}}}\right]},
\end{equation}
where $M_{Pl}$ is the Planck mass and $g^{*}$ is the number of relativistic degrees of freedom. In our case, it is $g^{*} = 106.75 + 4 + N$ for the range of temperatures under consideration.

Based on the calculation presented above, we can define the washout or dilution factor as $ f_{w.o.} = 1 - \frac{n_{B}(t_{\text{now}})}{n_{B}(t_{\text{high}})} $. In~\autoref{tab:washout} we show the values of $ f_{w.o.}$ for our two benchmarks. We see that BM A has a negligible washout factor for all choices of parameters, even for the most aggressive assumption for the fluctuation determinant $\kappa$, which is a factor 100 larger than the value suggested in~\cite{Baacke:1994ix}. BM B shows sub-percent or even negligible washout for the majority of approximations. Only for the most aggressive choice of $\kappa$, we observe values that can be as high as 70\%, which can be compensated by producing an initial asymmetry about three times larger than the asymmetry we observe now. If we consider the central value for $\kappa$, we see a washout of at most $1.2\%$.
We also note that the washout factor governed by eq.~\eqref{eq:sph.8} is much less sensitive to $T_{\rm high}$ than to effects at temperatures close to the EW~scale. This is the case since at higher temperatures the double exponential in eq.~\eqref{eq:sph.8} is larger than at EW temperatures. Indeed, at temperatures around the EW scale, there is an enhancement from the inverse of the Hubble expansion rate, as well as from the exponent proportional to $ \exp\left[ - \xi(T)\right] $, where $\xi(T)$ has its lowest values. That makes the double exponent in eq.~\eqref{eq:sph.8} to take its smallest values for temperatures close to the EW scale. Hence at such temperatures is when the main effect of the exponential washout takes place.
In other words, the relevant contribution to the washout factor is only at scales between the EW scale and around $500$~GeV, while at high temperatures the exponential washout remains negligible. This  holds as long as $\xi(T)$ does not fall fast below $1$ at high temperatures, which is the case for our BMs. This ensures that high-temperature EWBG could build in through a proper UV completion of our model.

\section{Phenomenological implications}
\label{sec:pheno}
In this section, we discuss the general particle physics phenomenology considerations for our model framework, including Higgs and Z boson invisible decays, disappearing tracks, Higgs global coupling shifts, as well as Higgs diphoton coupling shifts. We note that our benchmark choices in the previous section are explicitly set to satisfy these constraints. Still, the content in this section provides an estimation of current physics constraints and future perspectives for this model. The constraints shall be understood as applied to the parameters defined at the weak scale, which are essentially the bare parameters we used in defining the potential.

There are several phenomenological implications for our benchmark scenarios. 
At the zero temperature EW vacuum, there exists an additional discrete $\mathbb{Z}_2$ symmetry under which the new scalar fields $\Phi$ and $\chi_i$ are odd and the SM fields are even.  This renders the $\chi_i$ and the neutral components of the inert doublet scalar $\Phi$ stable and invisible once produced.

The possible existence of light scalars, $\Phi$ and $\chi_i$ will open the possibilities of the SM Higgs decaying into invisible particles, via the generic portal couplings 
\begin{equation}
\mathcal{L}\supset \lambda_{H\Phi}(H^\dagger H)(\Phi^\dagger \Phi)+\widetilde \lambda_{H\Phi}(H^\dagger \Phi)(\Phi^\dagger H)+\lambda_{H\chi} \chi_i^2 (H^\dagger H).
\end{equation}

The generic Higgs decay width into new scalars via this portal coupling is (per scalar degree of freedom):
\begin{equation}
\Gamma(h\rightarrow ss) = \frac {\lambda_{Hs}^2 v_0^2} {32\pi m_h} \sqrt{1-\frac {4m_s^2} {m_h^2}},
\end{equation}
where the coupling $\lambda_{Hs}$ can be one of the above quartics, $\lambda_{H\Phi}$, $\widetilde\lambda_{H\Phi}$ or $\lambda_{H\chi}$, and $m_s$ can be the mass of the $\Phi$ or $\chi$ states, respectively. 

The current LHC 95\% confidence level (C.L.) limit on Higgs invisible decays is 11\%~\cite{ATLAS:2020kdi} and the HL-LHC projection is 5.6\%~\cite{CidVidal:2018eel}.
When $m_s\ll m_h$, the phase space suppression is negligible and this translates into an upper limit for the SM-new scalars mixing quartics. The current and future limits on the mixing quartics read
\begin{equation}
\sqrt{N\lambda_{H \chi}^2+2(\lambda_{H\Phi}+\widetilde \lambda_{H\Phi})^2+2\lambda_{H\Phi}^2} \leqslant 0.010~(0.007)
\end{equation}
for LHC (HL-LHC).
In the above, by including $2\lambda_{H\Phi}^2$, we also include the Higgs decays into a pair of the charged states from the inert doublet.

In the absence of other mass splitting generating interactions, e.g. $\widetilde \lambda_{H\Phi}$ being zero, one-loop SM effects generate mass splittings between the charged and neutral eigenstate of the inert doublet of about 360~MeV~\cite{Cirelli:2005uq}. The charged state will decay back to the neutral state via a soft charged pion, or via the three-body decay mediated by an off-shell $W$ boson. The typical lifetime is independent of the inert doublet mass and is a few mm. Hence, this charged state can also be treated as invisible at colliders. In fact, precision $Z$ boson measurements of its invisible decays exclude all inert masses below 45~GeV, and hence we shall only consider inert masses beyond the 45~GeV value~\cite{Lundstrom:2008ai}.

Still, one can attempt to look for signals beyond the missing energy at colliders. At high energy colliders, such as the LHC, although challenging, one can look for the disappearing track signatures from the charged eigenstate of the inert doublet. However, it is well-known that this channel is difficult for Higgsinos, to which our inert doublet model signature resembles most. The current sensitivity from LHC disappearing track searches can exclude pure Higgsinos up to 78~GeV~\cite{Egana-Ugrinovic:2018roi}. The inert doublet production rate from the Drell-Yan process is roughly a factor of four lower than that of Higgsino production, due to the inert charged Higgs being a scalar rather than a fermion. Furthermore, for small mixing quartics such as $\widetilde \lambda_{H\Phi}$, one can arrange additional contributions to the mass splitting between the neutral and charged inert doublet states. This will make the charged state decay promptly and therefore the disappearing track searches will no longer apply. Given the above, we are entitled to ignore the disappearing track search limits and only comply with the LEP $Z$ invisible bounds for our benchmark scenarios. Future tests on disappearing tracks could still shed light on our model.

Summarizing, considering direct search constraints for our electroweak symmetry non-restoring model, we observe that the mixing quartics $\lambda_{H\Phi}$ and $\lambda_{H\chi}$ are bounded by constraints on invisible SM Higgs decay rates. This can give a strong handle for testing possible benchmarks, but at the same time there are models, like our BMs, in which they happen to have neglibible values. In this sense, the more direct and inevitable probe for our model at colliders are through the invisible $Z$ decays, relying only on the gauge coupling structure. Disappearing charged track searches open a new window of opportunity, if not undermined by parameter choices of the various mixing quartic couplings. 

There are additional U(1) global symmetries in the inert sector $\Phi$ as well as $\mathbb{Z}_{2}$ symmetries under which the singlet fields $\chi_i$ are odd, that prevent direct mixings between these states with our SM Higgs doublet. There are, however, loop-induced corrections to the SM that can be probed through precision observables.
The leading contribution to the electroweak precision observables (EWPO) is from the custodial symmetry breaking term $\widetilde \lambda_{H\Phi}$, inducing an operator contributing to the T-parameter \cite{Henning:2014wua}
\begin{eqnarray}
\mathcal{O}_T=&\frac 1 2 (H^\dagger \overset\leftrightarrow{D}_\mu H)^2,\ c_T=& \frac {\tlambda_{H\Phi}^2} {192\pi^2 \mu_\Phi^2}.
\end{eqnarray}
For an inert doublet mass scale $\mu_\Phi$ around half the Higgs mass, the EW precision measurement constrains the T-parameter with uncertainty 0.07~\cite{Baak:2012kk,Baak:2014ora}, constraining $|\widetilde \lambda_{H\Phi}|<0.36$ at 95\% C.L. Although this estimation is subject to sizable corrections due to the fact that $\mu_\Phi$ is of the order the Higgs mass, this gives an estimate of the bounds on $\tlambda_{H\Phi}$ not being very stringent coming from one-loop suppressed effects. For our benchmarks, we simply set $\widetilde \lambda_{H\Phi}$ to zero at tree level.

The next set of constraints comes from the Higgs boson coupling precision measurements, through the coefficient of the operator,
\begin{eqnarray} \label{eq:hc}
\mathcal{O}_H&=&\frac 1 2 (\partial_\mu |H|^2)^2,\\
c_H&=&\frac {4\lambda_{H\Phi}^2+4\lambda_{H\Phi}\widetilde \lambda_{H\Phi}+\widetilde \lambda_{H\Phi}^2+ N \lambda_{H\chi}^2 \mu_\Phi^2/\mu_\chi^2} {192\pi^2 \mu_\Phi^2}.\nonumber
\end{eqnarray}
This results in an overall reduction of the Higgs couplings by $1/2 c_H v_0^2$. We note here that this EFT matching is subject to large corrections and higher-order terms since the scales $\mu_\Phi^2$ and $\mu_\chi^2$ are not far from the Higgs mass squared. 
On the other hand, our non-restoration mechanism has limited dependence on these parameters. In particular, we  have set $\widetilde \lambda_{H\Phi}$ and $\lambda_{H\chi}$ to be zero in our BM scenarios, leaving only a shift of the Higgs couplings of about $-1/2 c_H=-{\lambda_{H\Phi}^2 v_0^2} /(96\pi^2 \mu_\Phi^2) $. For an inert doublet with $\mu_\Phi$ around half the Higgs mass, it yields a global shift in the Higgs couplings of around $-\lambda_{H\Phi}^2/(6\pi^2)$, bounding $|\lambda_{H\Phi}|<1.1$ (at 95\% C.L.) if we were to achieve
1\% Higgs coupling precision at the HL-LHC~\cite{Cepeda:2019klc}. This constraint is much weaker when we compare it to bounds from direct invisible Higgs decay searches discussed earlier in this section. It could however be relevant for scenarios with heavy inert masses, since the Higgs invisible decay bound no longer applies. However, in such case, as we shall see next, the precision measurements on Higgs to diphoton coupling provide a stronger constraint than the one derived from eq.~\eqref{eq:hc}.

The EW charged inert doublet also radiatively modifies Higgs couplings to EW~gauge bosons, through
\begin{eqnarray}
&\mathcal{O}_{BB}=g^{\prime 2}|H|^2 B_{\mu\nu}B^{\mu\nu},\ &c_{BB}=\frac {2\lambda_{H\Phi}+\widetilde \lambda_{H\Phi}} {768 \pi^2 \mu_\Phi^2},\nonumber\\
&\mathcal{O}_{WW}=g^{2}|H|^2 W_{\mu\nu}W^{\mu\nu},\ &c_{WW}=\frac {2\lambda_{H\Phi}+\widetilde \lambda_{H\Phi}} {768\pi^2 \mu_\Phi^2},\nonumber\\
&\mathcal{O}_{WB}=2g g^\prime H^\dagger \tau^a H W^a_{\mu\nu}B^{\mu\nu},\ &c_{WB}=\frac {\widetilde \lambda_{H\Phi}} {384\pi^2 \mu_\Phi^2}.
\end{eqnarray}
Here $\tau^a$ are the $SU(2)$ generators. Consequently, the Higgs diphoton coupling is modified by
\begin{equation}
1-\kappa_{\gamma\gamma}\simeq {10\pi^2 v_0^2} \left(c_{BB}+c_{WW}-c_{WB}\right),
\end{equation}
where $\kappa_{\gamma\gamma}\equiv g_{h\gamma\gamma}/g_{h\gamma\gamma}^{\rm SM}$.
Due to the fact that the SM Higgs to diphoton coupling is loop-induced, this provides a strong constraint on $|\lambda_{H\Phi}|$ to be smaller than 0.04 (at 95\% C.L.) for a 1.9\% precision~\cite{Cepeda:2019klc} on the Higgs to diphoton coupling at HL-LHC. The current Higgs precision uncertainty of 17\%~\cite{Sirunyan:2018ouh} translates to a constraint on $|\lambda_{H\Phi}|<0.4$ (at 95\% C.L.). Again, in deriving this limit, we assume that $\widetilde \lambda_{H\Phi}=0$, $\mu_\Phi$ being half the Higgs mass, and ignore the deviation of the form factor from unity from the inert doublet running in the loop.

Beyond the above, the model also generate less constraining effects on EWPO (W and Y parameter) and Higgs self-coupling~\cite{Henning:2014wua,Gu:2017ckc}, whose current and future perspective sensitivities can be found in Refs.~\cite{Gu:2017ckc,DiVita:2017vrr,deBlas:2019wgy}. This may provide, in the future, further complementary information about the model.

\section{Conclusion}
\label{sec:concl}

The exploration of electroweak phase transition patterns leading to electroweak symmetry breaking allows us to envision plausible paths for EWBG, as well as details of the cosmological history of our universe. In particular, the possibility of electroweak symmetry non-restoration up to high energy scales, conceivably up to the GUT or Planck scale, or the opportunity for delayed electroweak symmetry restoration up to scales of the order of $100$ TeVs, opens new windows for baryogenesis mechanisms. In this paper, we propose a novel approach to realize new thermal histories, by enabling the agent of EWSB to be an inert doublet that yields electroweak symmetry non-restoration up to high temperatures. These possibilities allow for diverse thermal histories with multi phase transition patterns, involving the SM Higgs, the inert Higgs and the SM-inert Higgs mixing phases at finite temperatures.

Our new approach for electroweak symmetry non-restoration at high energies has interesting computational requirements. Since the thermal history of our model, as defined in~\autoref{sec:model}, spans over large scale separations from the EW~scale to high temperatures, in our study we carefully implement the effects of RGE and thermal resummation, as detailed in \autoref{sec:pert}. 
When considering daisy resummation, we compute 
thermal masses with different approximations and observe that they lead to similar quantitive results. In~\autoref{sec:pt} we perform an analytical study at leading order in the high-temperature (mean field) approximation that helps us zoom in into the promising region of parameter space for our numerical study. In~\autoref{sec:num}, we report our numerical calculations for two benchmark points, and show that our results are robust under various treatments of thermal resummation while including RGE effects. Most importantly, the non-restoration patterns can hold at least up to high scales of the order of~$10^5$~GeV, within the one loop RG resumed effective potential. An UV completion of our model could take place at higher energy scales. In \autoref{sec:sph}, we present a detailed study of the sphaleron washout effects over a broad range of temperatures, and show that for our two benchmark scenarios, the washout rates are such that high temperature EWBG could be realized after a proper UV completion. 
Observe that the crucial ingredient of our BM scenarios is that the EW symmetry is non-restored from high temperatures all the way down to the EW scale.

Most importantly, our mechanism for transmitting broken electroweak symmetry from the SM sector to an inert sector has a specific interesting feature: It can work {\it even if} one decouples the two Higgs sectors in the tree level scalar potential, implying that the effect of the new doublet enters our zero-temperature particle physics tests at the electroweak-loop level. This enables the existence of large model parameter space compatible with experimental constraints and at the same time calls for new precision tests of the SM. As discussed in~\autoref{sec:pheno}, our model will find scrutiny at the HL-LHC through electroweak and Higgs precision tests, invisible decays and searches for disappearing tracks.

At high temperatures, our model opens up to possible UV completions that would enable various baryogenesis mechanisms. If we go through EWBG, where a strong first-order electroweak phase transition is necessary, it would give rise to gravitational wave signals. The peak frequency, instead of populating around the LISA band (mHZ), will increase to higher frequencies, at reach of facilities~\cite{Moore:2014lga,Breitbach:2018ddu} such as BBO, DECIGO, and even aLIGO. Moreover, the additional singlets $\chi$ in our study can themselves go through phase transitions, further enriching the possible thermal histories of our universe. Beyond all the above, one can also explore such relay of the EW-broken phase between the SM Higgs and scalars under other EW representations.

\section*{Acknowledgments}
We thank N. Blinov, C. Hill, A. Long, M. Perelstein, and C. Wagner for helpful discussions at various stages of this project. CK acknowledges the support of the Alexander von Humboldt Foundation and the grant DE-SC0010008 from the US Department of Energy. YW acknowledges the support of the Fermilab/UChicago Graduate Student Collaborative Research Award. 
This manuscript has been authored by Fermi Research Alliance, LLC under Contract No. DE-AC02-07CH11359 with the U.S. Department of Energy, Office of Science, Office of High Energy Physics. MC, CK and ZL would like to thank the Aspen Center for Physics which
is supported by National Science Foundation grant PHY-1607611, where part of the
study was performed. ZL was supported in part by the National Science Foundation under Grant Number PHY-1914731 at University of Maryland and  by the Maryland Center for Fundamental Physics.

In this work, we used {\tt Mathematica}~\cite{Mathematica} and the {\tt NumPy}~\cite{harris2020array}, {\tt SciPy}~\cite{2020SciPy-NMeth}, {\tt Matplotlib}~\cite{4160265}, {\tt pandas}~\cite{reback2020pandas} and \ct~\cite{Wainwright:2011kj} software packages. 
We wrote the main code in {\tt python}, which is available at {\tt https://gitlab.com/claudius-krause/ew\_nr}.

{\it Note added:} During the completion of this work,~\cite{Biekotter:2021ysx} appeared and considered a specific realization of symmetry non-restoration in a scenario with a 2HDM and one singlet scalar. We note that the main cause for us to require more singlet scalars is to strictly forbid the EW restoration at low temperatures, and thereby avoid the situation as depicted in the bottom right panel of \autoref{fig:NRglobal}. Furthermore, additional number of scalars are needed for the theory to obey tree-level perturbative unitarity up to the high temperature scales of non-restoration. In particular, our BM scenarios satisfy unitarity up to $10^{14}$ GeV. To the best of our understanding, in~\cite{Biekotter:2021ysx}, only benchmark E1, plus gray points in Fig.~7, survive our requirement of no temporary restoration at low temperatures, but due to perturbativity, the validity of the model appears to be limited to scales not far above the EW scale \footnote{We thank the authors of Ref.~\cite{Biekotter:2021ysx} for clarification in private communication.}.

\bibliography{EWNR}

\begin{thebibliography}{100}

\bibitem{Sakharov:1967dj}
A.~D. Sakharov.
\newblock {Violation of CP Invariance, C asymmetry, and baryon asymmetry of the
  universe}.
\newblock {\em Sov. Phys. Usp.}, 34(5):392--393, 1991.
\newblock \href {http://dx.doi.org/10.1070/PU1991v034n05ABEH002497}
  {\path{doi:10.1070/PU1991v034n05ABEH002497}}.

\bibitem{Morrissey:2012db}
David~E. Morrissey and Michael~J. Ramsey-Musolf.
\newblock {Electroweak baryogenesis}.
\newblock {\em New J. Phys.}, 14:125003, 2012.
\newblock \href {http://arxiv.org/abs/1206.2942} {\path{arXiv:1206.2942}},
  \href {http://dx.doi.org/10.1088/1367-2630/14/12/125003}
  {\path{doi:10.1088/1367-2630/14/12/125003}}.

\bibitem{Gavela:1994dt}
M.~B. Gavela, P.~Hernandez, J.~Orloff, O.~Pene, and C.~Quimbay.
\newblock {Standard model CP violation and baryon asymmetry. Part 2: Finite
  temperature}.
\newblock {\em Nucl. Phys. B}, 430:382--426, 1994.
\newblock \href {http://arxiv.org/abs/hep-ph/9406289}
  {\path{arXiv:hep-ph/9406289}}, \href
  {http://dx.doi.org/10.1016/0550-3213(94)00410-2}
  {\path{doi:10.1016/0550-3213(94)00410-2}}.

\bibitem{Kuzmin:1985mm}
V.~A. Kuzmin, V.~A. Rubakov, and M.~E. Shaposhnikov.
\newblock {On the Anomalous Electroweak Baryon Number Nonconservation in the
  Early Universe}.
\newblock {\em Phys. Lett. B}, 155:36, 1985.
\newblock \href {http://dx.doi.org/10.1016/0370-2693(85)91028-7}
  {\path{doi:10.1016/0370-2693(85)91028-7}}.

\bibitem{Carena:1996wj}
Marcela Carena, M.~Quiros, and C.~E.~M. Wagner.
\newblock {Opening the window for electroweak baryogenesis}.
\newblock {\em Phys. Lett. B}, 380:81--91, 1996.
\newblock \href {http://arxiv.org/abs/hep-ph/9603420}
  {\path{arXiv:hep-ph/9603420}}, \href
  {http://dx.doi.org/10.1016/0370-2693(96)00475-3}
  {\path{doi:10.1016/0370-2693(96)00475-3}}.

\bibitem{Delepine:1996vn}
D.~Delepine, J.~M. Gerard, R.~Gonzalez~Felipe, and J.~Weyers.
\newblock {A Light stop and electroweak baryogenesis}.
\newblock {\em Phys. Lett. B}, 386:183--188, 1996.
\newblock \href {http://arxiv.org/abs/hep-ph/9604440}
  {\path{arXiv:hep-ph/9604440}}, \href
  {http://dx.doi.org/10.1016/0370-2693(96)00921-5}
  {\path{doi:10.1016/0370-2693(96)00921-5}}.

\bibitem{Laine:1998qk}
M.~Laine and K.~Rummukainen.
\newblock {The MSSM electroweak phase transition on the lattice}.
\newblock {\em Nucl. Phys. B}, 535:423--457, 1998.
\newblock \href {http://arxiv.org/abs/hep-lat/9804019}
  {\path{arXiv:hep-lat/9804019}}, \href
  {http://dx.doi.org/10.1016/S0550-3213(98)00530-6}
  {\path{doi:10.1016/S0550-3213(98)00530-6}}.

\bibitem{Cline:1998hy}
James~M. Cline and Guy~D. Moore.
\newblock {Supersymmetric electroweak phase transition: Baryogenesis versus
  experimental constraints}.
\newblock {\em Phys. Rev. Lett.}, 81:3315--3318, 1998.
\newblock \href {http://arxiv.org/abs/hep-ph/9806354}
  {\path{arXiv:hep-ph/9806354}}, \href
  {http://dx.doi.org/10.1103/PhysRevLett.81.3315}
  {\path{doi:10.1103/PhysRevLett.81.3315}}.

\bibitem{Balazs:2004ae}
C.~Balazs, Marcela Carena, A.~Menon, D.~E. Morrissey, and C.~E.~M. Wagner.
\newblock {The Supersymmetric origin of matter}.
\newblock {\em Phys. Rev. D}, 71:075002, 2005.
\newblock \href {http://arxiv.org/abs/hep-ph/0412264}
  {\path{arXiv:hep-ph/0412264}}, \href
  {http://dx.doi.org/10.1103/PhysRevD.71.075002}
  {\path{doi:10.1103/PhysRevD.71.075002}}.

\bibitem{Lee:2004we}
Christopher Lee, Vincenzo Cirigliano, and Michael~J. Ramsey-Musolf.
\newblock {Resonant relaxation in electroweak baryogenesis}.
\newblock {\em Phys. Rev. D}, 71:075010, 2005.
\newblock \href {http://arxiv.org/abs/hep-ph/0412354}
  {\path{arXiv:hep-ph/0412354}}, \href
  {http://dx.doi.org/10.1103/PhysRevD.71.075010}
  {\path{doi:10.1103/PhysRevD.71.075010}}.

\bibitem{Carena:2008vj}
M.~Carena, Germano Nardini, M.~Quiros, and C.~E.~M. Wagner.
\newblock {The Baryogenesis Window in the MSSM}.
\newblock {\em Nucl. Phys. B}, 812:243--263, 2009.
\newblock \href {http://arxiv.org/abs/0809.3760} {\path{arXiv:0809.3760}},
  \href {http://dx.doi.org/10.1016/j.nuclphysb.2008.12.014}
  {\path{doi:10.1016/j.nuclphysb.2008.12.014}}.

\bibitem{Weinberg:1974hy}
Steven Weinberg.
\newblock {Gauge and Global Symmetries at High Temperature}.
\newblock {\em Phys. Rev. D}, 9:3357--3378, 1974.
\newblock \href {http://dx.doi.org/10.1103/PhysRevD.9.3357}
  {\path{doi:10.1103/PhysRevD.9.3357}}.

\bibitem{Mohapatra:1979qt}
Rabindra~N. Mohapatra and Goran Senjanovic.
\newblock {Soft CP Violation at High Temperature}.
\newblock {\em Phys. Rev. Lett.}, 42:1651, 1979.
\newblock \href {http://dx.doi.org/10.1103/PhysRevLett.42.1651}
  {\path{doi:10.1103/PhysRevLett.42.1651}}.

\bibitem{Mohapatra:1979vr}
Rabindra~N. Mohapatra and Goran Senjanovic.
\newblock {Broken Symmetries at High Temperature}.
\newblock {\em Phys. Rev. D}, 20:3390--3398, 1979.
\newblock \href {http://dx.doi.org/10.1103/PhysRevD.20.3390}
  {\path{doi:10.1103/PhysRevD.20.3390}}.

\bibitem{Mohapatra:1979bc}
Rabindra~N. Mohapatra and Goran Senjanovic.
\newblock {Broken symmetries at high temperatures and the problem of baryon
  excess of the universe}.
\newblock In {\em {1979 EPS High-Energy Physics Conference}}, Geneva,
  Switzerland, 6 1979. CERN.

\bibitem{Dvali:1995cc}
G.~R. Dvali and Goran Senjanovic.
\newblock {Is there a domain wall problem?}
\newblock {\em Phys. Rev. Lett.}, 74:5178--5181, 1995.
\newblock \href {http://arxiv.org/abs/hep-ph/9501387}
  {\path{arXiv:hep-ph/9501387}}, \href
  {http://dx.doi.org/10.1103/PhysRevLett.74.5178}
  {\path{doi:10.1103/PhysRevLett.74.5178}}.

\bibitem{Dvali:1995cj}
G.~R. Dvali, Alejandra Melfo, and Goran Senjanovic.
\newblock {Is There a monopole problem?}
\newblock {\em Phys. Rev. Lett.}, 75:4559--4562, 1995.
\newblock \href {http://arxiv.org/abs/hep-ph/9507230}
  {\path{arXiv:hep-ph/9507230}}, \href
  {http://dx.doi.org/10.1103/PhysRevLett.75.4559}
  {\path{doi:10.1103/PhysRevLett.75.4559}}.

\bibitem{Bajc:1999cn}
Borut Bajc.
\newblock {High temperature symmetry nonrestoration}.
\newblock In {\em {3rd International Conference on Particle Physics and the
  Early Universe}}, pages 247--253, 2000.
\newblock \href {http://arxiv.org/abs/hep-ph/0002187}
  {\path{arXiv:hep-ph/0002187}}, \href
  {http://dx.doi.org/10.1142/9789812792129_0039}
  {\path{doi:10.1142/9789812792129_0039}}.

\bibitem{Patel:2013zla}
Hiren~H. Patel, Michael~J. Ramsey-Musolf, and Mark~B. Wise.
\newblock {Color Breaking in the Early Universe}.
\newblock {\em Phys. Rev. D}, 88(1):015003, 2013.
\newblock \href {http://arxiv.org/abs/1303.1140} {\path{arXiv:1303.1140}},
  \href {http://dx.doi.org/10.1103/PhysRevD.88.015003}
  {\path{doi:10.1103/PhysRevD.88.015003}}.

\bibitem{Kilic:2015joa}
Can Kilic and Sivaramakrishnan Swaminathan.
\newblock {Can A Pseudo-Nambu-Goldstone Higgs Lead To Symmetry
  Non-Restoration?}
\newblock {\em JHEP}, 01:002, 2016.
\newblock \href {http://arxiv.org/abs/1508.05121} {\path{arXiv:1508.05121}},
  \href {http://dx.doi.org/10.1007/JHEP01(2016)002}
  {\path{doi:10.1007/JHEP01(2016)002}}.

\bibitem{Ramsey-Musolf:2017tgh}
Michael~J. Ramsey-Musolf, Peter Winslow, and Graham White.
\newblock {Color Breaking Baryogenesis}.
\newblock {\em Phys. Rev. D}, 97(12):123509, 2018.
\newblock \href {http://arxiv.org/abs/1708.07511} {\path{arXiv:1708.07511}},
  \href {http://dx.doi.org/10.1103/PhysRevD.97.123509}
  {\path{doi:10.1103/PhysRevD.97.123509}}.

\bibitem{Meade:2018saz}
Patrick Meade and Harikrishnan Ramani.
\newblock {Unrestored Electroweak Symmetry}.
\newblock {\em Phys. Rev. Lett.}, 122(4):041802, 2019.
\newblock \href {http://arxiv.org/abs/1807.07578} {\path{arXiv:1807.07578}},
  \href {http://dx.doi.org/10.1103/PhysRevLett.122.041802}
  {\path{doi:10.1103/PhysRevLett.122.041802}}.

\bibitem{Baldes:2018nel}
Iason Baldes and Géraldine Servant.
\newblock {High scale electroweak phase transition: baryogenesis \& symmetry
  non-restoration}.
\newblock {\em JHEP}, 10:053, 2018.
\newblock \href {http://arxiv.org/abs/1807.08770} {\path{arXiv:1807.08770}},
  \href {http://dx.doi.org/10.1007/JHEP10(2018)053}
  {\path{doi:10.1007/JHEP10(2018)053}}.

\bibitem{Glioti:2018roy}
Alfredo Glioti, Riccardo Rattazzi, and Luca Vecchi.
\newblock {Electroweak Baryogenesis above the Electroweak Scale}.
\newblock {\em JHEP}, 04:027, 2019.
\newblock \href {http://arxiv.org/abs/1811.11740} {\path{arXiv:1811.11740}},
  \href {http://dx.doi.org/10.1007/JHEP04(2019)027}
  {\path{doi:10.1007/JHEP04(2019)027}}.

\bibitem{Carena:2019une}
Marcela Carena, Zhen Liu, and Yikun Wang.
\newblock {Electroweak phase transition with spontaneous Z$_{2}$-breaking}.
\newblock {\em JHEP}, 08:107, 2020.
\newblock \href {http://arxiv.org/abs/1911.10206} {\path{arXiv:1911.10206}},
  \href {http://dx.doi.org/10.1007/JHEP08(2020)107}
  {\path{doi:10.1007/JHEP08(2020)107}}.

\bibitem{Matsedonskyi:2020mlz}
Oleksii Matsedonskyi and Geraldine Servant.
\newblock {High-Temperature Electroweak Symmetry Non-Restoration from New
  Fermions and Implications for Baryogenesis}.
\newblock {\em JHEP}, 09:012, 2020.
\newblock \href {http://arxiv.org/abs/2002.05174} {\path{arXiv:2002.05174}},
  \href {http://dx.doi.org/10.1007/JHEP09(2020)012}
  {\path{doi:10.1007/JHEP09(2020)012}}.

\bibitem{Bai:2021hfb}
Yang Bai, Seung~J. Lee, Minho Son, and Fang Ye.
\newblock {Global Electroweak Symmetric Vacuum}.
\newblock 3 2021.
\newblock \href {http://arxiv.org/abs/2103.09819} {\path{arXiv:2103.09819}}.

\bibitem{Branco:2011iw}
G.~C. Branco, P.~M. Ferreira, L.~Lavoura, M.~N. Rebelo, Marc Sher, and Joao~P.
  Silva.
\newblock {Theory and phenomenology of two-Higgs-doublet models}.
\newblock {\em Phys. Rept.}, 516:1--102, 2012.
\newblock \href {http://arxiv.org/abs/1106.0034} {\path{arXiv:1106.0034}},
  \href {http://dx.doi.org/10.1016/j.physrep.2012.02.002}
  {\path{doi:10.1016/j.physrep.2012.02.002}}.

\bibitem{Gustafsson:2010zz}
Michael Gustafsson.
\newblock {The Inert Doublet Model and Its Phenomenology}.
\newblock {\em PoS}, CHARGED2010:030, 2010.
\newblock \href {http://arxiv.org/abs/1106.1719} {\path{arXiv:1106.1719}},
  \href {http://dx.doi.org/10.22323/1.114.0030}
  {\path{doi:10.22323/1.114.0030}}.

\bibitem{Pomarol:1993mu}
Alex Pomarol and Roberto Vega.
\newblock {Constraints on CP violation in the Higgs sector from the rho
  parameter}.
\newblock {\em Nucl. Phys.}, B413:3--15, 1994.
\newblock \href {http://arxiv.org/abs/hep-ph/9305272}
  {\path{arXiv:hep-ph/9305272}}, \href
  {http://dx.doi.org/10.1016/0550-3213(94)90611-4}
  {\path{doi:10.1016/0550-3213(94)90611-4}}.

\bibitem{Ferreira:2004yd}
P.M. Ferreira, R.~Santos, and A.~Barroso.
\newblock {Stability of the tree-level vacuum in two Higgs doublet models
  against charge or CP spontaneous violation}.
\newblock {\em Phys. Lett. B}, 603:219--229, 2004.
\newblock [Erratum: Phys.Lett.B 629, 114--114 (2005)].
\newblock \href {http://arxiv.org/abs/hep-ph/0406231}
  {\path{arXiv:hep-ph/0406231}}, \href
  {http://dx.doi.org/10.1016/j.physletb.2004.10.022}
  {\path{doi:10.1016/j.physletb.2004.10.022}}.

\bibitem{Barroso:2005sm}
A.~Barroso, P.~M. Ferreira, and R.~Santos.
\newblock {Charge and CP symmetry breaking in two Higgs doublet models}.
\newblock {\em Phys. Lett.}, B632:684--687, 2006.
\newblock \href {http://arxiv.org/abs/hep-ph/0507224}
  {\path{arXiv:hep-ph/0507224}}, \href
  {http://dx.doi.org/10.1016/j.physletb.2005.11.031}
  {\path{doi:10.1016/j.physletb.2005.11.031}}.

\bibitem{Coleman:1973jx}
Sidney~R. Coleman and Erick~J. Weinberg.
\newblock {Radiative Corrections as the Origin of Spontaneous Symmetry
  Breaking}.
\newblock {\em Phys. Rev.}, D7:1888--1910, 1973.
\newblock \href {http://dx.doi.org/10.1103/PhysRevD.7.1888}
  {\path{doi:10.1103/PhysRevD.7.1888}}.

\bibitem{Coleman:1985rnk}
Sidney Coleman.
\newblock {\em {Aspects of Symmetry}}.
\newblock Cambridge University Press, Cambridge, U.K., 1985.
\newblock \href {http://dx.doi.org/10.1017/CBO9780511565045}
  {\path{doi:10.1017/CBO9780511565045}}.

\bibitem{Jackiw:1974cv}
R.~Jackiw.
\newblock {Functional evaluation of the effective potential}.
\newblock {\em Phys. Rev. D}, 9:1686, 1974.
\newblock \href {http://dx.doi.org/10.1103/PhysRevD.9.1686}
  {\path{doi:10.1103/PhysRevD.9.1686}}.

\bibitem{Kang:1974yj}
J.~S. Kang.
\newblock {Gauge Invariance of the Scalar-Vector Mass Ratio in the
  Coleman-Weinberg Model}.
\newblock {\em Phys. Rev. D}, 10:3455, 1974.
\newblock \href {http://dx.doi.org/10.1103/PhysRevD.10.3455}
  {\path{doi:10.1103/PhysRevD.10.3455}}.

\bibitem{Dolan:1974gu}
L.~Dolan and R.~Jackiw.
\newblock {Gauge Invariant Signal for Gauge Symmetry Breaking}.
\newblock {\em Phys. Rev. D}, 9:2904, 1974.
\newblock \href {http://dx.doi.org/10.1103/PhysRevD.9.2904}
  {\path{doi:10.1103/PhysRevD.9.2904}}.

\bibitem{Fukuda:1975di}
Reijiro Fukuda and Taichiro Kugo.
\newblock {Gauge Invariance in the Effective Action and Potential}.
\newblock {\em Phys. Rev. D}, 13:3469, 1976.
\newblock \href {http://dx.doi.org/10.1103/PhysRevD.13.3469}
  {\path{doi:10.1103/PhysRevD.13.3469}}.

\bibitem{Aitchison:1983ns}
I.~J.~R. Aitchison and C.~M. Fraser.
\newblock {Gauge Invariance and the Effective Potential}.
\newblock {\em Annals Phys.}, 156:1, 1984.
\newblock \href {http://dx.doi.org/10.1016/0003-4916(84)90209-4}
  {\path{doi:10.1016/0003-4916(84)90209-4}}.

\bibitem{Patel:2011th}
Hiren~H. Patel and Michael~J. Ramsey-Musolf.
\newblock {Baryon Washout, Electroweak Phase Transition, and Perturbation
  Theory}.
\newblock {\em JHEP}, 07:029, 2011.
\newblock \href {http://arxiv.org/abs/1101.4665} {\path{arXiv:1101.4665}},
  \href {http://dx.doi.org/10.1007/JHEP07(2011)029}
  {\path{doi:10.1007/JHEP07(2011)029}}.

\bibitem{Garny:2012cg}
Mathias Garny and Thomas Konstandin.
\newblock {On the gauge dependence of vacuum transitions at finite
  temperature}.
\newblock {\em JHEP}, 07:189, 2012.
\newblock \href {http://arxiv.org/abs/1205.3392} {\path{arXiv:1205.3392}},
  \href {http://dx.doi.org/10.1007/JHEP07(2012)189}
  {\path{doi:10.1007/JHEP07(2012)189}}.

\bibitem{Andreassen:2014eha}
Anders Andreassen, William Frost, and Matthew~D. Schwartz.
\newblock {Consistent Use of Effective Potentials}.
\newblock {\em Phys. Rev. D}, 91(1):016009, 2015.
\newblock \href {http://arxiv.org/abs/1408.0287} {\path{arXiv:1408.0287}},
  \href {http://dx.doi.org/10.1103/PhysRevD.91.016009}
  {\path{doi:10.1103/PhysRevD.91.016009}}.

\bibitem{Andreassen:2014gha}
Anders Andreassen, William Frost, and Matthew~D. Schwartz.
\newblock {Consistent Use of the Standard Model Effective Potential}.
\newblock {\em Phys. Rev. Lett.}, 113(24):241801, 2014.
\newblock \href {http://arxiv.org/abs/1408.0292} {\path{arXiv:1408.0292}},
  \href {http://dx.doi.org/10.1103/PhysRevLett.113.241801}
  {\path{doi:10.1103/PhysRevLett.113.241801}}.

\bibitem{Quiros:1999jp}
Mariano Quiros.
\newblock {Finite temperature field theory and phase transitions}.
\newblock In {\em {Proceedings, Summer School in High-energy physics and
  cosmology: Trieste, Italy, June 29-July 17, 1998}}, pages 187--259, 1999.
\newblock \href {http://arxiv.org/abs/hep-ph/9901312}
  {\path{arXiv:hep-ph/9901312}}.

\bibitem{FENDLEY1987175}
Paul Fendley.
\newblock The effective potential and the coupling constant at high
  temperature.
\newblock {\em Physics Letters B}, 196(2):175 -- 180, 1987.
\newblock URL:
  \url{http://www.sciencedirect.com/science/article/pii/0370269387905995},
  \href {http://dx.doi.org/https://doi.org/10.1016/0370-2693(87)90599-5}
  {\path{doi:https://doi.org/10.1016/0370-2693(87)90599-5}}.

\bibitem{PhysRevD.9.3320}
L.~Dolan and R.~Jackiw.
\newblock Symmetry behavior at finite temperature.
\newblock {\em Phys. Rev. D}, 9:3320--3341, Jun 1974.
\newblock URL: \url{https://link.aps.org/doi/10.1103/PhysRevD.9.3320}, \href
  {http://dx.doi.org/10.1103/PhysRevD.9.3320}
  {\path{doi:10.1103/PhysRevD.9.3320}}.

\bibitem{Kirzhnits:1974as}
D.A. Kirzhnits and Andrei~D. Linde.
\newblock {A Relativistic phase transition}.
\newblock {\em Sov. Phys. JETP}, 40:628, 1975.

\bibitem{PhysRevD.45.4695}
Rajesh~R. Parwani.
\newblock Resummation in a hot scalar field theory.
\newblock {\em Phys. Rev. D}, 45:4695--4705, Jun 1992.
\newblock URL: \url{https://link.aps.org/doi/10.1103/PhysRevD.45.4695}, \href
  {http://dx.doi.org/10.1103/PhysRevD.45.4695}
  {\path{doi:10.1103/PhysRevD.45.4695}}.

\bibitem{Nakkagawa:1998xc}
Hisao Nakkagawa and Hiroshi Yokota.
\newblock {Phase structure of the massive scalar phi**4 model at finite
  temperature: Resummation procedure a la RG improvement}.
\newblock In {\em {Summer School on Introduction to Thermal Field Theory (TFT
  98)}}, 9 1998.
\newblock \href {http://arxiv.org/abs/hep-ph/9809317}
  {\path{arXiv:hep-ph/9809317}}.

\bibitem{Arnold:1992rz}
Peter~Brockway Arnold and Olivier Espinosa.
\newblock {The Effective potential and first order phase transitions: Beyond
  leading-order}.
\newblock {\em Phys. Rev. D}, 47:3546, 1993.
\newblock [Erratum: Phys.Rev.D 50, 6662 (1994)].
\newblock \href {http://arxiv.org/abs/hep-ph/9212235}
  {\path{arXiv:hep-ph/9212235}}, \href
  {http://dx.doi.org/10.1103/PhysRevD.47.3546}
  {\path{doi:10.1103/PhysRevD.47.3546}}.

\bibitem{Curtin:2016urg}
David Curtin, Patrick Meade, and Harikrishnan Ramani.
\newblock {Thermal Resummation and Phase Transitions}.
\newblock {\em Eur. Phys. J.}, C78(9):787, 2018.
\newblock \href {http://arxiv.org/abs/1612.00466} {\path{arXiv:1612.00466}},
  \href {http://dx.doi.org/10.1140/epjc/s10052-018-6268-0}
  {\path{doi:10.1140/epjc/s10052-018-6268-0}}.

\bibitem{Croon:2020cgk}
Djuna Croon, Oliver Gould, Philipp Schicho, Tuomas V.~I. Tenkanen, and Graham
  White.
\newblock {Theoretical uncertainties for cosmological first-order phase
  transitions}.
\newblock {\em JHEP}, 04:055, 2021.
\newblock \href {http://arxiv.org/abs/2009.10080} {\path{arXiv:2009.10080}},
  \href {http://dx.doi.org/10.1007/JHEP04(2021)055}
  {\path{doi:10.1007/JHEP04(2021)055}}.

\bibitem{Laine:2017hdk}
M.~Laine, M.~Meyer, and G.~Nardini.
\newblock {Thermal phase transition with full 2-loop effective potential}.
\newblock {\em Nucl. Phys. B}, 920:565--600, 2017.
\newblock \href {http://arxiv.org/abs/1702.07479} {\path{arXiv:1702.07479}},
  \href {http://dx.doi.org/10.1016/j.nuclphysb.2017.04.023}
  {\path{doi:10.1016/j.nuclphysb.2017.04.023}}.

\bibitem{Bando:1992np}
Masako Bando, Taichiro Kugo, Nobuhiro Maekawa, and Hiroaki Nakano.
\newblock {Improving the effective potential}.
\newblock {\em Phys. Lett.}, B301:83--89, 1993.
\newblock \href {http://arxiv.org/abs/hep-ph/9210228}
  {\path{arXiv:hep-ph/9210228}}, \href
  {http://dx.doi.org/10.1016/0370-2693(93)90725-W}
  {\path{doi:10.1016/0370-2693(93)90725-W}}.

\bibitem{Bando:1992wy}
Masako Bando, Taichiro Kugo, Nobuhiro Maekawa, and Hiroaki Nakano.
\newblock {Improving the effective potential: Multimass scale case}.
\newblock {\em Prog. Theor. Phys.}, 90:405--418, 1993.
\newblock \href {http://arxiv.org/abs/hep-ph/9210229}
  {\path{arXiv:hep-ph/9210229}}, \href {http://dx.doi.org/10.1143/PTP.90.405,
  10.1143/ptp/90.2.405} {\path{doi:10.1143/PTP.90.405, 10.1143/ptp/90.2.405}}.

\bibitem{Casas:1998cf}
J.~A. Casas, V.~Di~Clemente, and M.~Quiros.
\newblock {The Effective potential in the presence of several mass scales}.
\newblock {\em Nucl. Phys.}, B553:511--530, 1999.
\newblock \href {http://arxiv.org/abs/hep-ph/9809275}
  {\path{arXiv:hep-ph/9809275}}, \href
  {http://dx.doi.org/10.1016/S0550-3213(99)00262-X}
  {\path{doi:10.1016/S0550-3213(99)00262-X}}.

\bibitem{Liao:1995gt}
Sen-Ben Liao and Michael Strickland.
\newblock {Renormalization group approach to field theory at finite
  temperature}.
\newblock {\em Phys. Rev. D}, 52:3653--3671, 1995.
\newblock \href {http://arxiv.org/abs/hep-th/9501137}
  {\path{arXiv:hep-th/9501137}}, \href
  {http://dx.doi.org/10.1103/PhysRevD.52.3653}
  {\path{doi:10.1103/PhysRevD.52.3653}}.

\bibitem{Nakkagawa:1996ju}
H.~Nakkagawa and H.~Yokota.
\newblock {RG improvement of the effective potential at finite temperature}.
\newblock {\em Mod. Phys. Lett. A}, 11:2259--2269, 1996.
\newblock \href {http://dx.doi.org/10.1142/S0217732396002253}
  {\path{doi:10.1142/S0217732396002253}}.

\bibitem{Nakkagawa:1997hg}
Hisao Nakkagawa and Hiroshi Yokota.
\newblock {Effective potential at finite temperature: RG improvement versus
  high temperature expansion}.
\newblock {\em Prog. Theor. Phys. Suppl.}, 129:209--214, 1997.
\newblock \href {http://arxiv.org/abs/hep-ph/9709323}
  {\path{arXiv:hep-ph/9709323}}, \href {http://dx.doi.org/10.1143/PTPS.129.209}
  {\path{doi:10.1143/PTPS.129.209}}.

\bibitem{Affleck:1984fy}
Ian Affleck and Michael Dine.
\newblock {A New Mechanism for Baryogenesis}.
\newblock {\em Nucl. Phys. B}, 249:361--380, 1985.
\newblock \href {http://dx.doi.org/10.1016/0550-3213(85)90021-5}
  {\path{doi:10.1016/0550-3213(85)90021-5}}.

\bibitem{Dine:2003ax}
Michael Dine and Alexander Kusenko.
\newblock {The Origin of the matter - antimatter asymmetry}.
\newblock {\em Rev. Mod. Phys.}, 76:1, 2003.
\newblock \href {http://arxiv.org/abs/hep-ph/0303065}
  {\path{arXiv:hep-ph/0303065}}, \href
  {http://dx.doi.org/10.1103/RevModPhys.76.1}
  {\path{doi:10.1103/RevModPhys.76.1}}.

\bibitem{Moreno:1996zm}
J.~M. Moreno, D.~H. Oaknin, and M.~Quiros.
\newblock {Sphalerons in the MSSM}.
\newblock {\em Nucl. Phys.}, B483:267--290, 1997.
\newblock \href {http://arxiv.org/abs/hep-ph/9605387}
  {\path{arXiv:hep-ph/9605387}}, \href
  {http://dx.doi.org/10.1016/S0550-3213(96)00562-7}
  {\path{doi:10.1016/S0550-3213(96)00562-7}}.

\bibitem{Grant:2001at}
Jackie Grant and Mark Hindmarsh.
\newblock {Sphalerons in two Higgs doublet theories}.
\newblock {\em Phys. Rev.}, D64:016002, 2001.
\newblock \href {http://arxiv.org/abs/hep-ph/0101120}
  {\path{arXiv:hep-ph/0101120}}, \href
  {http://dx.doi.org/10.1103/PhysRevD.64.016002}
  {\path{doi:10.1103/PhysRevD.64.016002}}.

\bibitem{Carson:1989rf}
Larry Carson and Larry~D. McLerran.
\newblock {Approximate Computation of the Small Fluctuation Determinant Around
  a Sphaleron}.
\newblock {\em Phys. Rev.}, D41:647, 1990.
\newblock \href {http://dx.doi.org/10.1103/PhysRevD.41.647}
  {\path{doi:10.1103/PhysRevD.41.647}}.

\bibitem{Carson:1990jm}
Larry Carson, Xu~Li, Larry~D. McLerran, and Rong-Tai Wang.
\newblock {Exact Computation of the Small Fluctuation Determinant Around a
  Sphaleron}.
\newblock {\em Phys. Rev.}, D42:2127--2143, 1990.
\newblock \href {http://dx.doi.org/10.1103/PhysRevD.42.2127}
  {\path{doi:10.1103/PhysRevD.42.2127}}.

\bibitem{Mottola:1990bz}
Emil Mottola and Stuart Raby.
\newblock {Baryon number dissipation at finite temperature in the standard
  model}.
\newblock {\em Phys. Rev.}, D42:4202--4208, 1990.
\newblock \href {http://dx.doi.org/10.1103/PhysRevD.42.4202}
  {\path{doi:10.1103/PhysRevD.42.4202}}.

\bibitem{Baacke:1994ix}
J.~Baacke and S.~Junker.
\newblock {Quantum fluctuations of the electroweak sphaleron: Erratum and
  addendum}.
\newblock {\em Phys. Rev.}, D50:4227--4228, 1994.
\newblock \href {http://arxiv.org/abs/hep-th/9402078}
  {\path{arXiv:hep-th/9402078}}, \href
  {http://dx.doi.org/10.1103/PhysRevD.50.4227}
  {\path{doi:10.1103/PhysRevD.50.4227}}.

\bibitem{ATLAS:2020kdi}
{Combination of searches for invisible Higgs boson decays with the ATLAS
  experiment}.
\newblock 10 2020.

\bibitem{CidVidal:2018eel}
Xabier Cid~Vidal et~al.
\newblock {Report from Working Group 3}: {Beyond the Standard Model physics at
  the HL-LHC and HE-LHC}.
\newblock {\em CERN Yellow Rep. Monogr.}, 7:585--865, 2019.
\newblock \href {http://arxiv.org/abs/1812.07831} {\path{arXiv:1812.07831}},
  \href {http://dx.doi.org/10.23731/CYRM-2019-007.585}
  {\path{doi:10.23731/CYRM-2019-007.585}}.

\bibitem{Cirelli:2005uq}
Marco Cirelli, Nicolao Fornengo, and Alessandro Strumia.
\newblock {Minimal dark matter}.
\newblock {\em Nucl. Phys. B}, 753:178--194, 2006.
\newblock \href {http://arxiv.org/abs/hep-ph/0512090}
  {\path{arXiv:hep-ph/0512090}}, \href
  {http://dx.doi.org/10.1016/j.nuclphysb.2006.07.012}
  {\path{doi:10.1016/j.nuclphysb.2006.07.012}}.

\bibitem{Lundstrom:2008ai}
Erik Lundstrom, Michael Gustafsson, and Joakim Edsjo.
\newblock {The Inert Doublet Model and LEP II Limits}.
\newblock {\em Phys. Rev. D}, 79:035013, 2009.
\newblock \href {http://arxiv.org/abs/0810.3924} {\path{arXiv:0810.3924}},
  \href {http://dx.doi.org/10.1103/PhysRevD.79.035013}
  {\path{doi:10.1103/PhysRevD.79.035013}}.

\bibitem{Egana-Ugrinovic:2018roi}
Daniel Egana-Ugrinovic, Matthew Low, and Joshua~T. Ruderman.
\newblock {Charged Fermions Below 100 GeV}.
\newblock {\em JHEP}, 05:012, 2018.
\newblock \href {http://arxiv.org/abs/1801.05432} {\path{arXiv:1801.05432}},
  \href {http://dx.doi.org/10.1007/JHEP05(2018)012}
  {\path{doi:10.1007/JHEP05(2018)012}}.

\bibitem{Henning:2014wua}
Brian Henning, Xiaochuan Lu, and Hitoshi Murayama.
\newblock {How to use the Standard Model effective field theory}.
\newblock {\em JHEP}, 01:023, 2016.
\newblock \href {http://arxiv.org/abs/1412.1837} {\path{arXiv:1412.1837}},
  \href {http://dx.doi.org/10.1007/JHEP01(2016)023}
  {\path{doi:10.1007/JHEP01(2016)023}}.

\bibitem{Baak:2012kk}
M.~Baak, M.~Goebel, J.~Haller, A.~Hoecker, D.~Kennedy, R.~Kogler, K.~Moenig,
  M.~Schott, and J.~Stelzer.
\newblock {The Electroweak Fit of the Standard Model after the Discovery of a
  New Boson at the LHC}.
\newblock {\em Eur. Phys. J. C}, 72:2205, 2012.
\newblock \href {http://arxiv.org/abs/1209.2716} {\path{arXiv:1209.2716}},
  \href {http://dx.doi.org/10.1140/epjc/s10052-012-2205-9}
  {\path{doi:10.1140/epjc/s10052-012-2205-9}}.

\bibitem{Baak:2014ora}
M.~Baak, J.~C\'uth, J.~Haller, A.~Hoecker, R.~Kogler, K.~M\"onig, M.~Schott,
  and J.~Stelzer.
\newblock {The global electroweak fit at NNLO and prospects for the LHC and
  ILC}.
\newblock {\em Eur. Phys. J. C}, 74:3046, 2014.
\newblock \href {http://arxiv.org/abs/1407.3792} {\path{arXiv:1407.3792}},
  \href {http://dx.doi.org/10.1140/epjc/s10052-014-3046-5}
  {\path{doi:10.1140/epjc/s10052-014-3046-5}}.

\bibitem{Cepeda:2019klc}
M.~Cepeda et~al.
\newblock {Report from Working Group 2}: {Higgs Physics at the HL-LHC and
  HE-LHC}.
\newblock {\em CERN Yellow Rep. Monogr.}, 7:221--584, 2019.
\newblock \href {http://arxiv.org/abs/1902.00134} {\path{arXiv:1902.00134}},
  \href {http://dx.doi.org/10.23731/CYRM-2019-007.221}
  {\path{doi:10.23731/CYRM-2019-007.221}}.

\bibitem{Sirunyan:2018ouh}
A.~M. Sirunyan et~al.
\newblock {Measurements of Higgs boson properties in the diphoton decay channel
  in proton-proton collisions at $\sqrt{s} =$ 13 TeV}.
\newblock {\em JHEP}, 11:185, 2018.
\newblock \href {http://arxiv.org/abs/1804.02716} {\path{arXiv:1804.02716}},
  \href {http://dx.doi.org/10.1007/JHEP11(2018)185}
  {\path{doi:10.1007/JHEP11(2018)185}}.

\bibitem{Gu:2017ckc}
Jiayin Gu, Honglei Li, Zhen Liu, Shufang Su, and Wei Su.
\newblock {Learning from Higgs Physics at Future Higgs Factories}.
\newblock {\em JHEP}, 12:153, 2017.
\newblock \href {http://arxiv.org/abs/1709.06103} {\path{arXiv:1709.06103}},
  \href {http://dx.doi.org/10.1007/JHEP12(2017)153}
  {\path{doi:10.1007/JHEP12(2017)153}}.

\bibitem{DiVita:2017vrr}
Stefano Di~Vita, Gauthier Durieux, Christophe Grojean, Jiayin Gu, Zhen Liu,
  Giuliano Panico, Marc Riembau, and Thibaud Vantalon.
\newblock {A global view on the Higgs self-coupling at lepton colliders}.
\newblock {\em JHEP}, 02:178, 2018.
\newblock \href {http://arxiv.org/abs/1711.03978} {\path{arXiv:1711.03978}},
  \href {http://dx.doi.org/10.1007/JHEP02(2018)178}
  {\path{doi:10.1007/JHEP02(2018)178}}.

\bibitem{deBlas:2019wgy}
Jorge De~Blas, Gauthier Durieux, Christophe Grojean, Jiayin Gu, and Ayan Paul.
\newblock {On the future of Higgs, electroweak and diboson measurements at
  lepton colliders}.
\newblock {\em JHEP}, 12:117, 2019.
\newblock \href {http://arxiv.org/abs/1907.04311} {\path{arXiv:1907.04311}},
  \href {http://dx.doi.org/10.1007/JHEP12(2019)117}
  {\path{doi:10.1007/JHEP12(2019)117}}.

\bibitem{Moore:2014lga}
C.~J. Moore, R.~H. Cole, and C.~P.~L. Berry.
\newblock {Gravitational-wave sensitivity curves}.
\newblock {\em Class. Quant. Grav.}, 32(1):015014, 2015.
\newblock \href {http://arxiv.org/abs/1408.0740} {\path{arXiv:1408.0740}},
  \href {http://dx.doi.org/10.1088/0264-9381/32/1/015014}
  {\path{doi:10.1088/0264-9381/32/1/015014}}.

\bibitem{Breitbach:2018ddu}
Moritz Breitbach, Joachim Kopp, Eric Madge, Toby Opferkuch, and Pedro
  Schwaller.
\newblock {Dark, Cold, and Noisy: Constraining Secluded Hidden Sectors with
  Gravitational Waves}.
\newblock {\em JCAP}, 07:007, 2019.
\newblock \href {http://arxiv.org/abs/1811.11175} {\path{arXiv:1811.11175}},
  \href {http://dx.doi.org/10.1088/1475-7516/2019/07/007}
  {\path{doi:10.1088/1475-7516/2019/07/007}}.

\bibitem{Mathematica}
Wolfram~Research{,} Inc.
\newblock Mathematica, {V}ersion 12.1.
\newblock Champaign, IL, 2020.
\newblock URL: \url{https://www.wolfram.com/mathematica}.

\bibitem{harris2020array}
Charles~R. Harris, K.~Jarrod Millman, St{'{e}}fan~J. van~der Walt, Ralf
  Gommers, Pauli Virtanen, David Cournapeau, Eric Wieser, Julian Taylor,
  Sebastian Berg, Nathaniel~J. Smith, Robert Kern, Matti Picus, Stephan Hoyer,
  Marten~H. van Kerkwijk, Matthew Brett, Allan Haldane, Jaime~Fern{'{a}}ndez
  del R{'{\i}}o, Mark Wiebe, Pearu Peterson, Pierre G{'{e}}rard-Marchant, Kevin
  Sheppard, Tyler Reddy, Warren Weckesser, Hameer Abbasi, Christoph Gohlke, and
  Travis~E. Oliphant.
\newblock Array programming with {NumPy}.
\newblock {\em Nature}, 585(7825):357--362, September 2020.
\newblock URL: \url{https://doi.org/10.1038/s41586-020-2649-2}, \href
  {http://dx.doi.org/10.1038/s41586-020-2649-2}
  {\path{doi:10.1038/s41586-020-2649-2}}.

\bibitem{2020SciPy-NMeth}
Pauli Virtanen, Ralf Gommers, Travis~E. Oliphant, Matt Haberland, Tyler Reddy,
  David Cournapeau, Evgeni Burovski, Pearu Peterson, Warren Weckesser, Jonathan
  Bright, St{\'e}fan~J. {van der Walt}, Matthew Brett, Joshua Wilson, K.~Jarrod
  Millman, Nikolay Mayorov, Andrew R.~J. Nelson, Eric Jones, Robert Kern, Eric
  Larson, C~J Carey, {\.I}lhan Polat, Yu~Feng, Eric~W. Moore, Jake
  {VanderPlas}, Denis Laxalde, Josef Perktold, Robert Cimrman, Ian Henriksen,
  E.~A. Quintero, Charles~R. Harris, Anne~M. Archibald, Ant{\^o}nio~H. Ribeiro,
  Fabian Pedregosa, Paul {van Mulbregt}, and {SciPy 1.0 Contributors}.
\newblock {{SciPy} 1.0: Fundamental Algorithms for Scientific Computing in
  Python}.
\newblock {\em Nature Methods}, 17:261--272, 2020.
\newblock \href {http://dx.doi.org/10.1038/s41592-019-0686-2}
  {\path{doi:10.1038/s41592-019-0686-2}}.

\bibitem{4160265}
J.~D. {Hunter}.
\newblock Matplotlib: A 2d graphics environment.
\newblock {\em Computing in Science Engineering}, 9(3):90--95, 2007.
\newblock \href {http://dx.doi.org/10.1109/MCSE.2007.55}
  {\path{doi:10.1109/MCSE.2007.55}}.

\bibitem{reback2020pandas}
The pandas~development team.
\newblock pandas-dev/pandas: Pandas, February 2020.
\newblock URL: \url{https://doi.org/10.5281/zenodo.3509134}, \href
  {http://dx.doi.org/10.5281/zenodo.3509134}
  {\path{doi:10.5281/zenodo.3509134}}.

\bibitem{Wainwright:2011kj}
Carroll~L. Wainwright.
\newblock {CosmoTransitions: Computing Cosmological Phase Transition
  Temperatures and Bubble Profiles with Multiple Fields}.
\newblock {\em Comput. Phys. Commun.}, 183:2006--2013, 2012.
\newblock \href {http://arxiv.org/abs/1109.4189} {\path{arXiv:1109.4189}},
  \href {http://dx.doi.org/10.1016/j.cpc.2012.04.004}
  {\path{doi:10.1016/j.cpc.2012.04.004}}.

\bibitem{Biekotter:2021ysx}
Thomas Biek\"otter, Sven Heinemeyer, Jos\'e~Miguel No, Mar\'\i{}a~Olalla Olea,
  and Georg Weiglein.
\newblock {Fate of electroweak symmetry in the early Universe: Non-restoration
  and trapped vacua in the N2HDM}.
\newblock {\em JCAP}, 06:018, 2021.
\newblock \href {http://arxiv.org/abs/2103.12707} {\path{arXiv:2103.12707}},
  \href {http://dx.doi.org/10.1088/1475-7516/2021/06/018}
  {\path{doi:10.1088/1475-7516/2021/06/018}}.

\bibitem{Blinov:2015vma}
Nikita Blinov, Stefano Profumo, and Tim Stefaniak.
\newblock {The Electroweak Phase Transition in the Inert Doublet Model}.
\newblock {\em JCAP}, 1507(07):028, 2015.
\newblock \href {http://arxiv.org/abs/1504.05949} {\path{arXiv:1504.05949}},
  \href {http://dx.doi.org/10.1088/1475-7516/2015/07/028}
  {\path{doi:10.1088/1475-7516/2015/07/028}}.

\bibitem{Niemi:2018asa}
Lauri Niemi, Hiren~H. Patel, Michael~J. Ramsey-Musolf, Tuomas~V.I. Tenkanen,
  and David~J. Weir.
\newblock {Electroweak phase transition in the real triplet extension of the
  SM: Dimensional reduction}.
\newblock {\em Phys. Rev. D}, 100(3):035002, 2019.
\newblock \href {http://arxiv.org/abs/1802.10500} {\path{arXiv:1802.10500}},
  \href {http://dx.doi.org/10.1103/PhysRevD.100.035002}
  {\path{doi:10.1103/PhysRevD.100.035002}}.

\bibitem{Kannike:2012pe}
Kristjan Kannike.
\newblock {Vacuum Stability Conditions From Copositivity Criteria}.
\newblock {\em Eur. Phys. J. C}, 72:2093, 2012.
\newblock \href {http://arxiv.org/abs/1205.3781} {\path{arXiv:1205.3781}},
  \href {http://dx.doi.org/10.1140/epjc/s10052-012-2093-z}
  {\path{doi:10.1140/epjc/s10052-012-2093-z}}.

\bibitem{KAPLAN2000203}
Wilfred Kaplan.
\newblock A test for copositive matrices.
\newblock {\em Linear Algebra and its Applications}, 313(1):203 -- 206, 2000.
\newblock URL:
  \url{http://www.sciencedirect.com/science/article/pii/S0024379500001385},
  \href {http://dx.doi.org/https://doi.org/10.1016/S0024-3795(00)00138-5}
  {\path{doi:https://doi.org/10.1016/S0024-3795(00)00138-5}}.

\bibitem{Buchalla:2017jlu}
G.~Buchalla, O.~Cata, A.~Celis, M.~Knecht, and C.~Krause.
\newblock {Complete One-Loop Renormalization of the Higgs-Electroweak Chiral
  Lagrangian}.
\newblock {\em Nucl. Phys.}, B928:93--106, 2018.
\newblock \href {http://arxiv.org/abs/1710.06412} {\path{arXiv:1710.06412}},
  \href {http://dx.doi.org/10.1016/j.nuclphysb.2018.01.009}
  {\path{doi:10.1016/j.nuclphysb.2018.01.009}}.

\bibitem{Buchalla:2019wsc}
Gerhard Buchalla, Alejandro Celis, Claudius Krause, and Jan-Niklas Toelstede.
\newblock {Master Formula for One-Loop Renormalization of Bosonic SMEFT
  Operators}.
\newblock 2019.
\newblock \href {http://arxiv.org/abs/1904.07840} {\path{arXiv:1904.07840}}.

\bibitem{Einhorn:2007rv}
Martin~B. Einhorn and D.~R.~Timothy Jones.
\newblock {The Effective potential, the renormalisation group and vacuum
  stability}.
\newblock {\em JHEP}, 04:051, 2007.
\newblock \href {http://arxiv.org/abs/hep-ph/0702295}
  {\path{arXiv:hep-ph/0702295}}, \href
  {http://dx.doi.org/10.1088/1126-6708/2007/04/051}
  {\path{doi:10.1088/1126-6708/2007/04/051}}.

\bibitem{Gan:2017mcv}
Xucheng Gan, Andrew~J. Long, and Lian-Tao Wang.
\newblock {Electroweak sphaleron with dimension-six operators}.
\newblock {\em Phys. Rev.}, D96(11):115018, 2017.
\newblock \href {http://arxiv.org/abs/1708.03061} {\path{arXiv:1708.03061}},
  \href {http://dx.doi.org/10.1103/PhysRevD.96.115018}
  {\path{doi:10.1103/PhysRevD.96.115018}}.

\bibitem{Akiba:1989xu}
T.~Akiba, H.~Kikuchi, and T.~Yanagida.
\newblock {The Free Energy of the Sphaleron in the {Weinberg-Salam} Model}.
\newblock {\em Phys. Rev.}, D40:588, 1989.
\newblock \href {http://dx.doi.org/10.1103/PhysRevD.40.588}
  {\path{doi:10.1103/PhysRevD.40.588}}.

\bibitem{Dine:1991ck}
Michael Dine, Patrick Huet, and Robert~L. Singleton, Jr.
\newblock {Baryogenesis at the electroweak scale}.
\newblock {\em Nucl. Phys.}, B375:625--648, 1992.
\newblock \href {http://dx.doi.org/10.1016/0550-3213(92)90113-P}
  {\path{doi:10.1016/0550-3213(92)90113-P}}.

\bibitem{Baacke:1993jr}
J.~Baacke and S.~Junker.
\newblock {Quantum corrections to the electroweak sphaleron transition}.
\newblock {\em Mod. Phys. Lett.}, A8:2869--2874, 1993.
\newblock \href {http://arxiv.org/abs/hep-ph/9306307}
  {\path{arXiv:hep-ph/9306307}}, \href
  {http://dx.doi.org/10.1142/S0217732393003251}
  {\path{doi:10.1142/S0217732393003251}}.

\bibitem{Baacke:1993aj}
J.~Baacke and S.~Junker.
\newblock {Quantum fluctuations around the electroweak sphaleron}.
\newblock {\em Phys. Rev.}, D49:2055--2073, 1994.
\newblock \href {http://arxiv.org/abs/hep-ph/9308310}
  {\path{arXiv:hep-ph/9308310}}, \href
  {http://dx.doi.org/10.1103/PhysRevD.49.2055}
  {\path{doi:10.1103/PhysRevD.49.2055}}.

\bibitem{JMLR:v18:17-468}
Atilim~Gunes Baydin, Barak~A. Pearlmutter, Alexey~Andreyevich Radul, and
  Jeffrey~Mark Siskind.
\newblock Automatic differentiation in machine learning: a survey.
\newblock {\em Journal of Machine Learning Research}, 18(153):1--43, 2018.
\newblock URL: \url{http://jmlr.org/papers/v18/17-468.html}.

\bibitem{Brent:113464}
Richard~P Brent.
\newblock {\em {Algorithms for minimization without derivatives}}.
\newblock Prentice-Hall series in automatic computation. Prentice-Hall,
  Englewood Cliffs, NJ, 1973.
\newblock URL: \url{http://cds.cern.ch/record/113464}.

\end{thebibliography}
\bibliographystyle{unsrturl}

\onecolumngrid

\appendix
\numberwithin{equation}{section}

\section{Effective field-dependent masses}
\label{app:mass}

In this appendix, we list field-dependent masses of all degrees of freedoms in the plasma, which are relevant calculating one-loop effective potentials. The field-dependent scalar mass matrix squared $m^{2}(\hat{\Phi})$ is defined as 
\begin{equation}
  \label{eq:T.QFT.2}
  m^{2}_{ab}(\hat\Phi) \equiv \left.\frac{\delta^{2} V}{(\delta \Phi_{a})(\delta \Phi_{b})}\right|_{\Phi=\hat{\Phi}},
\end{equation}
where we introduced a short-handed notation $\Phi \equiv \{ h, \varphi, \chi_1, \chi_2,\cdots, \chi_N\}$, and a caret is used to indicate background fields. The field-dependent gauge field mass matrix squared is given by\cite{Coleman:1985rnk}
\begin{equation}
  \label{eq:T.QFT.4}
  M^{2}(\hat{\Phi})= g_{a} g_{b} (T_{a}\hat{\Phi})(T_{b}\hat{\Phi}),
\end{equation}
with $g_{a}$ the gauge coupling and $T_{a}$ the generator of the $a$th gauge field\footnote{It might hava to be symmetrized.}. The field-dependent fermion mass matrix squared is $mm^{\dagger}(\hat{\Phi})$ where $m(\hat{\Phi})$ is defined in the Lagrangian as
\begin{equation}
  \label{eq:T.QFT.6}
  \mathcal{L} = \bar{\Psi}^{a}m_{ab}(\hat{\Phi})\Psi^{b}+\dots .
\end{equation}
All contributions to the CW-potential are formally taken as traces of the squared mass matrices, which in practice can be diagonalized and the potential is then evaluated for the eigenvalues. Next, we list all field-dependent mass matrices squared in our model. \\

In the space of $\left(h,\frac{-i}{\sqrt{2}}(G^+-G^-),\frac{1}{\sqrt{2}}(G^++G^-) ,-G_0,\varphi,\frac{-i}{\sqrt{2}}(\phi^+-\phi^-),\frac{1}{\sqrt{2}}(\phi^++\phi^-) ,-\phi_0,\chi_{i}\right)$,
 the symmetric field dependent mass matrix squared of the scalar sector is given by

 \begin{align}
   \begin{aligned}
     \label{eq:fielddepmass.full}
     & m^{2}( \hat{h},\hat{\varphi},\hat{\chi_i}) = \left(\begin{array}{ccc}
                                                            M^2_{H} & M^2_{H\Phi} & M^2_{H\chi} \\
                                                                  M^2_{H\Phi}  & M^2_{\Phi} & M^2_{\Phi\chi} \\
                                                                 M^2_{H\chi}   &  M^2_{\Phi\chi}& M^2_{\chi} 
                                                          \end{array}\right)
                                                      \end{aligned}
 \end{align}
with
\begin{align}
  \begin{aligned}
    \label{eq:fielddepmass.HH}
     M^2_{H} ( \hat{h},\hat{\varphi},\hat{\chi_i})  & = \\
    &\hspace*{-2em}\left(\begin{array}{cccc}
             -\mu_{H}^{2}+3\lambda_{H}\hat{h}^{2} +\tfrac{\lambda_{H\Phi}}{2}\hat{\varphi}^{2} & & &\\
             +\tfrac{\tlambda_{H\Phi}}{2}\hat{\varphi}^{2}  + \tfrac{\lambda_{H\chi}}{2} \sum_{i}\hat{\chi}_i^2  &  &  & \\
                                                                             & -\mu_{H}^2 + \lambda_{H} \hat{h}^{2} & & \\
             & +\tfrac{\lambda_{H\Phi}}{2}\hat{\varphi}^{2}+ \tfrac{\lambda_{H\chi}}{2} \sum_{i}\hat{\chi}_i^2 &  & \\
                                                                             &  & -\mu_{H}^2 + \lambda_{H} \hat{h}^{2} & \\
             &&+\tfrac{\lambda_{H\Phi}}{2}\hat{\varphi}^{2}+ \tfrac{\lambda_{H\chi}}{2} \sum_{i}\hat{\chi}_i^2 & \\
                                                                             &&& -\mu_{H}^2 + \lambda_{H} \hat{h}^{2} +\tfrac{\lambda_{H\Phi}}{2}\hat{\varphi}^{2}\\
             &&&+\tfrac{\tlambda_{H\Phi}}{2}\hat{\varphi}^{2}+ \tfrac{\lambda_{H\chi}}{2} \sum_{i}\hat{\chi}_i^2
           \end{array}\right),
       \end{aligned}
\end{align}
\begin{align}
  \begin{aligned}
    \label{eq:fielddepmass.phiphi}
    M^2_{\Phi} ( \hat{h},\hat{\varphi},\hat{\chi_i})  =&\\
   & \hspace*{-2em} \left(\begin{array}{cccc}
               \mu_{\Phi}^{2}+3\lambda_{\Phi}\hat{\varphi}^{2}+\tfrac{\lambda_{H\Phi}}{2}\hat{h}^{2} &&&\\
               +\tfrac{\tlambda_{H\Phi}}{2}\hat{h}^{2}+\frac{ \lambda_{\Phi\chi}}{2} \sum_{i}\hat{\chi}_i^2   &  &  & \\
                                                                     & \mu_{\Phi}^{2}+\lambda_{\Phi}\hat{\varphi}^{2}&&\\
                                                                     &+\tfrac{\lambda_{H\Phi}}{2}\hat{h}^{2} +\tfrac{ \lambda_{\Phi\chi}}{2} \sum_{i}\hat{\chi}_i^2    & & \\
                                                                                    & &\mu_{\Phi}^{2}+\lambda_{\Phi}\hat{\varphi}^{2}&\\
               &&+\tfrac{\lambda_{H\Phi}}{2}\hat{h}^{2} +\tfrac{ \lambda_{\Phi\chi}}{2} \sum_{i}\hat{\chi}_i^2   & \\
                                                                     &  & & \mu_{\Phi}^{2}+\lambda_{\Phi}\hat{\varphi}^{2}+\tfrac{\lambda_{H\Phi}}{2}\hat{h}^{2}\\
               &&&+\tfrac{\tlambda_{H\Phi}}{2}\hat{h}^{2}+\tfrac{ \lambda_{\Phi\chi}}{2} \sum_{i}\hat{\chi}_i^2 
                                        \end{array}\right),
  \end{aligned}
\end{align}

\begin{align}
  \begin{aligned}
    \label{eq:fielddepmass.chichi}
   M^2_{\chi} ( \hat{h},\hat{\varphi},\hat{\chi_i}) = \left(\begin{array}{cccc}
    \makecell{ 3 \tlambda_{\chi}  \hat{\chi}_1^2 + \lambda_{\chi}(\sum_{i}\hat{\chi}_i^2+2\hat{\chi}_1^2)  \\ +\tfrac{ \lambda_{\Phi\chi}}{2} \hat{\varphi}^2   + \tfrac{\lambda_{H\chi}}{2} \hat{h}^2+  \mu_{\chi}^{2} }
               &\quad 2\lambda_{\chi} \hat{\chi}_1\hat{\chi}_2 
              &  \dots
             &2\lambda_{\chi} \hat{\chi}_1\hat{\chi}_N\\[5pt]
%   +\tfrac{ \lambda_{\Phi\chi}}{2} \hat{\varphi}^2   + \tfrac{\lambda_{H\chi}}{2} \hat{h}^2+  \mu_{\chi}^{2} & & &   \\
  2\lambda_{\chi} \hat{\chi}_2\hat{\chi}_1    & \ddots & \ddots &   \vdots                                                   \\[5pt]
    \vdots   & \ddots&  \ddots &  2\lambda_{\chi} \hat{\chi}_{N-1}\hat{\chi}_N \\[5pt]
   2\lambda_{\chi} \hat{\chi}_N\hat{\chi}_1     & \dots & 2\lambda_{\chi} \hat{\chi}_N\hat{\chi}_{N-1} & \quad \makecell{ 3 \tlambda_{\chi}  \hat{\chi}_N^2 + \lambda_{\chi}(\sum_{i}\hat{\chi}_i^2+2\hat{\chi}_N^2)\\ +\tfrac{ \lambda_{\Phi\chi}}{2} \hat{\varphi}^2   + \tfrac{\lambda_{H\chi}}{2} \hat{h}^2 +  \mu_{\chi}^{2} }\\
%                                                            &      &&+\tfrac{ \lambda_{\Phi\chi}}{2} \hat{\varphi}^2   + \tfrac{\lambda_{H\chi}}{2} \hat{h}^2 +  \mu_{\chi}^{2}
                                  \end{array}\right),
  \end{aligned}
\end{align}

\begin{align}
  \begin{aligned}
    \label{eq:fielddepmass.Hphi}
   &   M^2_{H\phi} ( \hat{h},\hat{\varphi},\hat{\chi_i}) = \left(\begin{array}{cccc}
(\lambda_{H\Phi}+\tlambda_{H\Phi}) \hat{h}\hat{\varphi}  &  &  &\\
       & \tfrac{\tlambda_{H\Phi}}{2} \hat{h}\hat{\varphi}  &  & \\
                                                                   &  & \tfrac{\tlambda_{H\Phi}}{2} \hat{h}\hat{\varphi} & \\
                                                                   &&&0
                                        \end{array}\right),
  \end{aligned}
\end{align}
\begin{align}
  \begin{aligned}
    \label{eq:fielddepmass.Hchi}
   &   M^2_{H\chi} ( \hat{h},\hat{\varphi},\hat{\chi_i}) = \left(\begin{array}{ccc}
 \lambda_{H\chi}\hat{h}\hat{\chi}_{1} &\cdots  & \lambda_{H\chi}\hat{h}\hat{\chi}_{N}   \\
    0   &\cdots  &0  \\
    0   &\cdots  &0  \\
    0   &\cdots  &0                                          \end{array}\right),
  \end{aligned}
\end{align}
\begin{align}
  \begin{aligned}
    \label{eq:fielddepmass.phichi}
   &   M^2_{\phi\chi} ( \hat{h},\hat{\varphi},\hat{\chi_i}) = \left(\begin{array}{ccc}
  \lambda_{\Phi\chi}\hat{\varphi}\hat{\chi}_{1} &\cdots  & \lambda_{\Phi\chi}\hat{\varphi}\hat{\chi}_{N}   \\
    0   &\cdots  &0  \\
    0   &\cdots  &0  \\
    0   &\cdots  &0                                       
                                        \end{array}\right),
  \end{aligned}
\end{align}
where cells left blank are zero, while cells represented by dots are following previous cells' pattern.
In addition, we have~\cite{Blinov:2015vma,Niemi:2018asa}
\begin{equation}
  \label{eq:fielddep.WZ}
  m_{W}^{2} = \frac{g^{2}}{4}(\hat{h}^{2}+\hat{\varphi}^{2}), \quad m_{Z}^{2} = \frac{g^{2}+g^{\prime 2}}{4}(\hat{h}^{2}+\hat{\varphi}^{2}), \quad m_{t}^{2} = \frac{y_{t}^{2}}{2}\hat{h}^{2}.
\end{equation}

\section{Bounded from below conditions and zero temperature vacuum structure}
\label{app:BFB}
We show detailed derivations of bounded form below (BFB) condition for the potential in eq.\eqref{eq:Model.Lag}. A scalar potential, whose quartic part can be written as the form $\lambda_{ab} \varphi_a^2 \varphi_b^2$, is bounded from below if the matrix of quartic couplings $\lambda_{ab}$ is copositive \cite{Kannike:2012pe}. A symmetric matrix is strictly copositive if and only if the associated eigenvalues to non-negative eigenvectors of all principal submatrices are stricly positive~\cite{KAPLAN2000203}. 

Let's work in the basis where
\begin{equation}
  \label{eq:BFB.0}
H^{\dagger} H = \frac{1}{2} h_1^2,\quad \Phi^{\dagger}\Phi =  \frac{1}{2} h_2^2,\quad H^{\dagger} \Phi =  \frac{1}{2} h_1 h_2 \rho e^{i\eta}.
\end{equation}
The parameter $|\rho|  \in [0, 1]$ parametrizes the Cauchy inequality $ 0 \le |H^{\dagger} \Phi | \le |H| |\Phi | $. Note that the potential of eq.~\eqref{eq:Model.Lag} is independent of $ \eta $. The matrix of quartic couplings takes the form
\begin{equation}
  \label{eq:BFB.1}
  M_{\text{quartic}} = \begin{pmatrix}\lambda_{H} & \Lambda_{H\Phi}/2 & \lambda_{H\chi}/2 &\dots&\dots& \lambda_{H\chi}/2\\
     \Lambda_{H\Phi}/2&\lambda_{\Phi}&\lambda_{\Phi\chi}/2 &\dots &\dots&\lambda_{\Phi\chi}/2\\
    \lambda_{H\chi}/2&\lambda_{\Phi\chi}/2& \tlambda_{\chi} + \lambda_{\chi} & \lambda_{\chi} &\dots &\lambda_{\chi} \\
    \vdots & \vdots &  \lambda_{\chi}  &\ddots & \ddots&\vdots \\
    \vdots & \vdots & \vdots &\ddots & \ddots& \lambda_{\chi}  \\
    \lambda_{H\chi}/2&\lambda_{\Phi\chi}/2& \lambda_{\chi} &\dots & \lambda_{\chi}  & \tlambda_{\chi}+ \lambda_{\chi} 
  \end{pmatrix},
\end{equation}
where for simplicity we have defined a shorthanded notation
 \begin{equation}
  \label{eq:}
\Lambda_{H\Phi} \equiv \lambda_{H\phi} + \tlambda_{H\Phi} \rho^2.
\end{equation}

For copositivity, the principal submatrices are:\\
Order 1:
\begin{equation}
  \label{eq:BFB.2}
  \lambda_{H} >0 \qquad \lambda_{\Phi}>0 \qquad \tlambda_{\chi} + \lambda_{\chi} >0
\end{equation}
Order 2:
\begin{equation}
  \label{eq:BFB.3}
  \begin{pmatrix} \lambda_{H} & \Lambda_{H\Phi}/2\\ \Lambda_{H\Phi}/2& \lambda_{\Phi}\end{pmatrix} \quad \Rightarrow \quad \Lambda_{H\Phi}> - \sqrt{4 \lambda_{H} \lambda_{\Phi}}
\end{equation}
Notice that if $\tlambda_{H\Phi} \ge 0$, the condition should be $\lambda_{H\phi} > - \sqrt{4 \lambda_{H} \lambda_{\Phi}}$ while if $\tlambda_{H\Phi} < 0$, the condition should be $\lambda_{H\phi} +  \tlambda_{H\Phi} > - \sqrt{4 \lambda_{H} \lambda_{\Phi}}$ .
\begin{equation}
  \label{eq:BFB.4}
  \begin{pmatrix} \lambda_{\Phi} &\lambda_{\Phi\chi}/2 \\ \lambda_{\Phi\chi}/2 & \tlambda_{\chi}+ \lambda_{\chi}\end{pmatrix} \quad \Rightarrow \quad \lambda_{\Phi\chi}> - \sqrt{4  \lambda_{\Phi} (\tlambda_{\chi} + \lambda_{\chi}) }
\end{equation}
\begin{equation}
  \label{eq:BFB.5}
  \begin{pmatrix} \lambda_{H} & \lambda_{H\chi}/2 \\ \lambda_{H\chi}/2 & \tlambda_{\chi} + \lambda_{\chi} \end{pmatrix} \quad \Rightarrow \quad \lambda_{H\chi}> - \sqrt{4  \lambda_{H} (\tlambda_{\chi} + \lambda_{\chi}) }
\end{equation}
\begin{equation}
  \label{eq:BFB.6}
  \begin{pmatrix} \tlambda_{\chi}+\lambda_{\chi} & \lambda_{\chi} \\ \lambda_{\chi} & \tlambda_{\chi} + \lambda_{\chi} \end{pmatrix} \quad \Rightarrow \quad \tlambda_{\chi} + 2\lambda_{\chi} >0 
\end{equation}

Higher order:
\begin{equation}
  \label{eq:BFB.7}
  \begin{pmatrix} 
  \tlambda_{\chi} + \lambda_{\chi} &\lambda_{\chi}& \dots&\lambda_{\chi}  \\ 
  \lambda_{\chi} & \tlambda_{\chi} + \lambda_{\chi}&\lambda_{\chi}&\lambda_{\chi} \\
    \vdots&\lambda_{\chi}&\ddots&\lambda_{\chi}\\
    \lambda_{\chi} & \lambda_{\chi}&\lambda_{\chi}&\tlambda_{\chi} + \lambda_{\chi}\\
  \end{pmatrix}_{n \times n} \quad \Rightarrow \quad \tlambda_{\chi} + n \lambda_{\chi} >0\quad {\rm with}\quad n = 3, \cdots N
\end{equation}
\begin{equation}
  \label{eq:BFB.8}
  \hspace*{-2em}
  \begin{pmatrix} \lambda_{\Phi} &\lambda_{\Phi\chi}/2& \dots&\lambda_{\Phi\chi}/2 \\ \lambda_{\Phi\chi}/2 & \tlambda_{\chi} + \lambda_{\chi}&\lambda_{\chi}&\lambda_{\chi} \\
    \vdots&\lambda_{\chi}&\ddots&\lambda_{\chi}\\
    \lambda_{\Phi\chi}/2 & \lambda_{\chi}&\lambda_{\chi}&\tlambda_{\chi} + \lambda_{\chi}\\
  \end{pmatrix}_{(1+n) \times (1+n)} \Rightarrow \quad \lambda_{\Phi\chi}> - \sqrt{4  \lambda_{\Phi} \left( \frac{\tlambda_{\chi}}{n} +  \lambda_{\chi}\right)} \quad {\rm with}\quad n = 2, \cdots N
\end{equation}
\begin{equation}
  \label{eq:BFB.9}
  \hspace*{-2em}
  \begin{pmatrix} \lambda_{H} &\lambda_{H\chi}/2& \dots& \lambda_{H\chi}/2 \\ 
  \lambda_{H\chi}/2 & \tlambda_{\chi} + \lambda_{\chi}&\lambda_{\chi}&\lambda_{\chi} \\
    \vdots&\lambda_{\chi}&\ddots&\lambda_{\chi}\\
    \lambda_{H\chi}/2 & \lambda_{\chi}&\lambda_{\chi}&\tlambda_{\chi} + \lambda_{\chi}\\
  \end{pmatrix}_{(1+n) \times (1+n)}  \Rightarrow \quad  \lambda_{H\chi}> - \sqrt{4 \lambda_{H} \left( \frac{\tlambda_{\chi}}{n} +  \lambda_{\chi}\right)}\quad {\rm with}\quad n = 2, \cdots N
\end{equation}
Eq.~\eqref{eq:BFB.7} is derived when the eigenvalues of the matrix being
\begin{equation}
  \label{eq:}
 \{ \tlambda_{\chi}, \cdots, \tlambda_{\chi}, \tlambda_{\chi}+ n \lambda_{\chi} \}
 \end{equation}
with the corresponding eigenvectors
\begin{equation}
  \label{eq:}
 \{ \{-1, 1, 0,\dots, 0\}, \cdots, \{-1, 0, \dots,0, 1 \} , \{ 1, \cdots, 1\} \}.
 \end{equation}
Only the last eigenvalue $\tlambda_{\chi} + n \lambda_{\chi}$ corresponds to a positive eigenvector, and accordingly such an eigenvalue must be positive, yielding $\tlambda_{\chi} + n \lambda_{\chi} > 0$. Equation~\eqref{eq:BFB.8} is derived when the eigenvalues of the matrix being
\begin{equation}
  \label{eq:}
 \{ \tlambda_{\chi}, \cdots, \tlambda_{\chi},e_1,e_2 \}
 \end{equation}
 where 
 \begin{equation}
  \label{eq:}
e_{1,2} =  \frac{1}{2} \left[ \lambda_{\Phi} + \tlambda_{\chi} + n \lambda_{\chi} \pm \sqrt{(\lambda_{\Phi} - \tlambda_{\chi} - n \lambda_{\chi})^2 + n \lambda_{\Phi\chi}^2} \right]
   \end{equation}
with the corresponding eigenvectors
\begin{equation}
  \begin{aligned}
  \label{eq:}
& \left\{ \{0, -1, 1, 0,\dots, 0\}, \cdots, \{0, -1, 0, \dots,0, 1 \} ,\right.\\
 & \qquad\left.\{\frac{e_1 - (\tlambda_{\chi} + n \lambda_{\chi})}{\lambda_{\Phi\chi}/2} ,1, \cdots, 1\},\{ \frac{e_2 - (\tlambda_{\chi} + n \lambda_{\chi})}{\lambda_{\Phi\chi}/2},1, \cdots, 1\} \right\}.
   \end{aligned}
 \end{equation}
The first $n-1$ eigenvectors are not positive, hence they do not induce conditions. According to conditions \eqref{eq:BFB.2} and \eqref{eq:BFB.7}, the eigenvalue $e_1$ is positive definite, thus it does not induce new conditions either. Notice that for the eigenvalue $e_2$, the corresponding eigenvector is only positive if $\lambda_{\Phi\chi} < 0$, in which case, $e_2$ needs to be positive leading to $n \lambda_{\Phi\chi}^2 < 4 \lambda_{\Phi} (\tlambda_{\chi} + n \lambda_{\chi}) $. Combining this with the allowed range of the case $\lambda_{\Phi\chi}  > 0$, one arrives at the condition \eqref{eq:BFB.8}. The condition \eqref{eq:BFB.9} can be derived similarly.

Lastly, the principle submatrices in the form of 
\begin{align}
  \begin{aligned}
  \label{eq:BFB.10}
\hspace*{-3cm}  \begin{pmatrix}\lambda_{H} & \Lambda_{H\Phi}/2& \lambda_{H\chi}/2&\dots&\dots& \lambda_{H\chi}/2\\
    \Lambda_{H\Phi}/2&\lambda_{\Phi}&\lambda_{\Phi\chi}/2 &\dots &\dots&\lambda_{\Phi\chi}/2\\
    \lambda_{H\chi}/2&\lambda_{\Phi\chi}/2& \tlambda_{\chi} + \lambda_{\chi}&\lambda_{\chi}&\dots &\lambda_{\chi}\\
    \vdots & \vdots & \lambda_{\chi} &\ddots & \ddots&\vdots \\
    \vdots & \vdots & \vdots &\ddots & \ddots&\lambda_{\chi} \\
    \lambda_{H\chi}/2&\lambda_{\Phi\chi}/2&\lambda_{\chi}&\dots &\lambda_{\chi} & \tlambda_{\chi}+\lambda_{\chi}\end{pmatrix}_{(n+2) \times (n+2)} \quad {\rm with} \quad n = 1,\cdots, N   \end{aligned}
\end{align}
need to be copositive. The eigenvalues of such matrices are
\begin{equation}
  \label{eq:}
 \{ \tlambda_{\chi}, \cdots, \tlambda_{\chi},e_1,e_2, e_3 \}
 \end{equation}
where $e_{1,2,3}$ are roots of the cubic polynomial
\begin{align}
  \label{eq:BFB.11}
- e^3 + A e^2 -B  e + C
\end{align}
with
\begin{align}
  \label{eq:BFB.12}
&A = \lambda_{H} + \lambda_{\Phi} + \tlambda_{\chi} + n \lambda_{\chi} \\
&B =  \frac{1}{4}\left[4 \lambda_{H} \lambda_{\Phi} -  \Lambda_{H\Phi}^2 \right] + \frac{1}{4}\left[ 4\lambda_{\Phi} (\tlambda_{\chi} +n\lambda_{\chi}) - n \lambda_{\Phi\chi}^2 \right] +  \frac{1}{4}\left[ 4 \lambda_{H} (\tlambda_{\chi} + n\lambda_{\chi}) -  n \lambda_{H\chi}^2 \right]\\
& C = \lambda_{H} \lambda_{\Phi} (\tlambda_{\chi}+n\lambda_{\chi}) + \frac{1}{4}n \Lambda_{H\Phi} \lambda_{H\chi} \lambda_{\Phi\chi}  -\frac{1}{4} \Lambda_{H\Phi}^2 (\tlambda_{\chi} +n\lambda_{\chi})- \frac{1}{4} n \lambda_{H} \lambda_{\Phi\chi}^2-   \frac{1}{4}n \lambda_{\Phi} \lambda_{H\chi}^2 .
\end{align}
The corresponding eigenvectors read
\begin{equation}
  \begin{aligned}
  \label{eq:}
& \left\{ \{0,0, -1, 1, 0,\dots, 0\}, \cdots, \{0,0, -1, 0, \dots,0, 1 \} ,\right.\\
 & \qquad\left.\{x_1, y_1,1, \cdots, 1\},\{ x_2, y_2,1, \cdots, 1\},\{ x_3, y_3,1, \cdots, 1\} \right\},
   \end{aligned}
 \end{equation}
 where $x_i$ and $ y_i$ are given by the solution of
 \begin{equation}
  \begin{aligned}
  \label{eq:BFB.14}
&( \lambda_{H} - e_i) x_i +  \frac{\Lambda_{H\Phi}}{2} y_i + n\lambda_{H\chi}/2 = 0\\
&  \frac{\lambda_{H\chi}}{2} x +\frac{\lambda_{\Phi\chi}}{2} y + \tlambda_{\chi} + n \lambda_{\chi} - e_i= 0.
   \end{aligned}
 \end{equation}
Notice that the eigenvectors of the $n-1$ eigenvalues $\tlambda_{\chi}$ are non-positive, and accordingly they do not give any constraints on the copositivity of the matrix. For the last three eigenvalues $e_i$ with $i =1,2,3$, the last entries of their eigenvectors is unity, and accordingly, the positivity of the eigenvectors is determined by the sign of $x_i$ and $y_i$. Thus, such $(n+2)\times(n+2)$ matrices are copositive if and only if there is no $e_i < 0$ with corresponding $x_i, y_i >0$.

Now let's use the fact that the roots $\{e_1, e_2, e_3\}$ are also eigenvalues of a $3 \times 3$ matrix 
\begin{align}
  \begin{aligned}
  \label{eq:BFB.15}
\hspace*{-3cm}  \begin{pmatrix}\lambda_{H} &\Lambda_{H\Phi}/2& \sqrt{n} \lambda_{H\chi}/2 \\
   \Lambda_{H\Phi}/2 &\lambda_{\Phi}& \sqrt{n}\lambda_{\Phi\chi}/2 \\
    \sqrt{n}\lambda_{H\chi}/2&  \sqrt{n}\lambda_{\Phi\chi}/2& \tlambda_{\chi} +n \lambda_{\chi}\end{pmatrix}\quad {\rm for} \quad n = 1,\cdots, N   \end{aligned}
\end{align}
with the corresponding eigenvectors being
\begin{equation}
  \begin{aligned}
  \label{eq:}
  \left\{ \{x_1, y_1,\sqrt{n}\},\{ x_2, y_2,\sqrt{n}\},\{ x_3, y_3,\sqrt{n}\} \right\},
   \end{aligned}
 \end{equation}
where $x_i,y_i$ are solved by the same conditions given in eq.~\eqref{eq:BFB.14}. Since the last entry of the above eigenvectors is positive, one can immediately see that the condition of the copositivity of such a $3\times3$ matrix is identical to the conditions for the above $(n+2)\times(n+2)$ matrices: there is no $e_i < 0$ with corresponding $x_i, y_i >0$. The copositivity of a $3 \times 3$ matrix in terms of its entries has been discussed in the literature. The conditions are \cite{Kannike:2012pe}:
\begin{equation}
  \begin{aligned}
  \label{eq:BFB.16}
&\lambda_{H} >0,\quad \lambda_{\Phi} >0,\quad \tlambda_{\chi} + n \lambda_{\chi} > 0 \\
&\Lambda_{H\Phi} > - \sqrt{4 \lambda_{H} \lambda_{\Phi}},\quad \lambda_{\Phi\chi} > - \sqrt{4 \lambda_{\Phi} (\tlambda_{\chi}/n + \lambda_{\chi})},\quad \lambda_{H\chi} > -  \sqrt{ 4 \lambda_{H} (\tlambda_{\chi}/n + \lambda_{\chi})} \\
&\sqrt{4 \lambda_{H} \lambda_{\Phi} (\tlambda_{\chi}/n + \lambda_{\chi})} + \Lambda_{H\Phi} \sqrt{\tlambda_{\chi}/n + \lambda_{\chi}} + \lambda_{\Phi\chi} \sqrt{\lambda_{H}} + \lambda_{H\chi} \sqrt{\lambda_{\Phi}} \\
&\quad + \sqrt{\left( \Lambda_{H\Phi} + \sqrt{4 \lambda_{H} \lambda_{\Phi}}\right) \left( \lambda_{\Phi\chi} + \sqrt{4 \lambda_{\Phi} (\tlambda_{\chi}/n + \lambda_{\chi})}\right)\left( \lambda_{H\chi} + \sqrt{ 4 \lambda_{H} (\tlambda_{\chi}/n + \lambda_{\chi})}\right) } > 0.\\
   \end{aligned}
 \end{equation}
% Notice that the above conditions conclude all the conditions we have talked about for other types of matrices.

Now, we have derived the copositivity conditions for all type of principle submatrices of the quartic coupling matrix of our two doublets + $N$ singlets tree-level potential. Notice that we work in a generic basis including all CP even, CP odd and charged components of the doublets. The potential will be bounded from below if all the conditions are satisfied. We define for convenience:
 \begin{align}
  \label{eq:BFB.abbrev}
\Lambda_{\chi,n} \equiv \frac{1}{n}\tlambda_{\chi} +  \lambda_{\chi},\quad \Lambda_{H\Phi} \equiv  \lambda_{H\Phi}+ \tlambda_{H\Phi} \rho^2
\end{align}
with which the BFB conditions can be written as
 \begin{align}
  \label{eq:}
\lambda_{H} >0,\quad \lambda_{\Phi} >0,\quad \Lambda_{\chi,n} > 0
\end{align}
 \begin{align}
  \label{eq:}
\Lambda_{H\Phi} > - \sqrt{4 \lambda_{H} \lambda_{\Phi}},\quad \lambda_{\Phi\chi} > - \sqrt{4 \lambda_{\Phi} \Lambda_{\chi,n}},\quad \lambda_{H\chi} > -  \sqrt{ 4 \lambda_{H} \Lambda_{\chi,n}}
\end{align}
 \begin{align}
  \label{eq:}
&\sqrt{4 \lambda_{H} \lambda_{\Phi} \Lambda_{\chi,n}} + \Lambda_{H\Phi} \sqrt{ \Lambda_{\chi,n}} + \lambda_{\Phi\chi} \sqrt{\lambda_{H}} + \lambda_{H\chi} \sqrt{\lambda_{\Phi}} \\
&\quad + \sqrt{\left( \Lambda_{H\Phi} + \sqrt{4 \lambda_{H} \lambda_{\Phi}}\right) \left( \lambda_{\Phi\chi} + \sqrt{4 \lambda_{\Phi}  \Lambda_{\chi,n}}\right)\left( \lambda_{H\chi} + \sqrt{ 4 \lambda_{H}  \Lambda_{\chi,n}}\right) } > 0.
 \end{align}
For the potential to be bounded from below, these conditions have to hold for all $n \in \{1, \dots, N\}$ and $|\rho|  \in [0, 1]$. In practice, we check the conditions above for the boundary values only, as these give the smallest/largest values of $\Lambda_{\chi,n}$ and $\Lambda_{H\Phi}$.

%%%%% END BFB %%%%%%%%%%%%%%

\renewcommand{\arraystretch}{2.}
\begin{table}[] 
\hspace*{-5cm}
\caption{Tree level potential extrema structure: CP even neutral components. The vevs of scalar fields, the discriminant of the quadratic function $\Delta$, and the potential value at the respective extremum $V$ are listed in the table.
}
\label{table:minima1}
\begin{tabular}{c|c|c|c|c|c|c|c}
\hline
 & 
 $(0,0,0)$ & $(0,0,\chi_i)$ & $(h,0,0)$  & $(0,\varphi,0)$   & $(h,0, \chi_i)$ & 
  $(0, \varphi,\chi_i)$ & $(h, \varphi,0)$    
 %\Tstrut\Bstrut 
 \\ \hline

 $\langle h\rangle^2$  
 & $0$ & $0$ & $\frac{\mu_{H}^2}{\lambda_{H}}$ & $0$ & $\frac{\lambda_{H\chi} \mu_{\chi}^2 + 2\Lambda_{\chi,n} \mu_{H}^2}{2\Delta} $ & $0$  
 & $\frac{\Lambda_{H\Phi} \mu_{\Phi}^2 + 2 \lambda_{\Phi} \mu_{H}^2}{ 2\Delta }$ 
 \\
 
  $\langle \varphi \rangle^2$ 
  & $0$& $0$ & $0$& $- \frac{\mu_{\Phi}^2}{\lambda_{\Phi}}$ & $0$ & $\frac{ \lambda_{\Phi\chi} \mu_{\chi}^2 - 2\Lambda_{\chi,n} \mu_{\Phi}^2}{2 \Delta }$  & $\frac{- \Lambda_{H\Phi} \mu_{H}^2 -2 \lambda_{H} \mu_{\Phi}^2}{2\Delta }$  
   \\ 
  
 $\langle \chi_i\rangle^2$ 
 & $0$ & $-\frac{\mu_{\chi}^2}{n\Lambda_{\chi,n}}$ & $0$ & $0$  & $\frac{- \lambda_{H\chi} \mu_{H}^2 -2 \lambda_{H} \mu_{\chi}^2}{2 n \Delta }$ &  $\frac{ \lambda_{\Phi\chi} \mu_{\Phi}^2 - 2\lambda_{\Phi} \mu_{\chi}^2}{2n \Delta }$ & $0$
\\ \hline
 
     $\Delta$ 
    & $-$ & $-$ & $-$ & $-$  & $\lambda_{H} \Lambda_{\chi,n} - \frac{1}{4} \lambda_{H\chi}^2 $  & $\lambda_{\Phi} \Lambda_{\chi,n} - \frac{1}{4}  \lambda_{\Phi\chi}^2  $& $\lambda_{H} \lambda_{\Phi}  - \frac{1}{4} \Lambda_{H\Phi}^2  $
  \\ \hline

  $V$ & 
  $0$ & $-  \frac{\mu_{\chi}^4}{4\Lambda_{\chi,n}}$ & $- \frac{\mu_{H}^4}{4\lambda_{H}}$& $- \frac{\mu_{\Phi}^4}{4\lambda_{\Phi}}$    & $- \frac{\lambda_{H} \mu_{\chi}^4  + \Lambda_{\chi,n} \mu_{H}^4 + \lambda_{H\chi} \mu_{\chi}^2 \mu_{H}^2 }{4 \Delta}$ & $ - \frac{\lambda_{\Phi} \mu_{\chi}^4  + \Lambda_{\chi,n} \mu_{\Phi}^4 -  \lambda_{\Phi\chi} \mu_{\chi}^2 \mu_{\Phi}^2 }{4\Delta}$
  & $- \frac{\lambda_{H} \mu_{\Phi}^4  + \lambda_{\Phi} \mu_{H}^4 + \Lambda_{H\Phi} \mu_{\Phi}^2 \mu_{H}^2 }{4 \Delta}$
  \\ \hline
\end{tabular}
\end{table}

\begin{table}[] 
\caption{Tree level potential extrema structure: CP even neutral components - continued 
}
\label{table:minima3}
\begin{tabular}{c|c}
\hline
 & 
   $(h, \varphi,\chi_i)$    
 \\ \hline

 $\langle h\rangle^2$
 &  $\frac{\mu_{H}^2 (4\lambda_{\Phi} \Lambda_{\chi,n} - \lambda_{\Phi\chi}^2) + \mu_{\Phi}^2 (2 \Lambda_{\chi,n}\Lambda_{H\Phi}- \lambda_{H\chi} \lambda_{\Phi\chi}  ) + \mu_{\chi}^2 (2 \lambda_{\Phi} \lambda_{H\chi} -\Lambda_{H\Phi}\lambda_{\Phi\chi}) }{4\Delta}$ 
 \\
 
  $\langle \varphi \rangle^2$ 
&    $\frac{- \mu_{\Phi}^2 (4\lambda_{H} \Lambda_{\chi,n} -\lambda_{H\chi}^2) - \mu_{H}^2 (2 \Lambda_{\chi,n}\Lambda_{H\Phi}- \lambda_{H\chi} \lambda_{\Phi\chi}  ) + \mu_{\chi}^2 (2 \lambda_{H} \lambda_{\Phi\chi} - \Lambda_{H\Phi}\lambda_{H\chi}) }{4\Delta}$ 
   \\ 
  
 $\langle \chi_i\rangle^2$ 
 &     $\frac{- \mu_{\chi}^2 (4\lambda_{H} \lambda_{\Phi} - \Lambda_{H\Phi}^2) - \mu_{H}^2 (2 \lambda_{\Phi}  \lambda_{H\chi}  -\Lambda_{H\Phi}\lambda_{\Phi\chi}  ) + \mu_{\Phi}^2 (2 \lambda_{H} \lambda_{\Phi\chi} - \Lambda_{H\Phi} \lambda_{H\chi} ) }{4n\Delta}$ 
 \\ \hline
 
     $\Delta$ 
    &     $\lambda_{H} \lambda_{\Phi} \Lambda_{\chi,n} +\frac{1}{4} \Lambda_{H\Phi}\lambda_{\Phi\chi} \lambda_{H\chi}  - \frac{1}{4}\Lambda_{\chi,n} \Lambda_{H\Phi}^2  - \frac{1}{4} \lambda_{H} \lambda_{\Phi\chi}^2 - \frac{1}{4} \lambda_{\Phi}  \lambda_{H\chi} ^2$ 
  \\ \hline

  $V$ 
 & \makecell{ $- \frac{1}{16 \Delta} \left[ \mu_{H}^4 (4 \lambda_{\Phi} \Lambda_{\chi,n} - \lambda_{\Phi\chi}^2) + \mu_{\Phi}^4 (4\lambda_{H} \Lambda_{\chi,n} - \lambda_{H\chi}^2 ) + \mu_{\chi}^4 (4 \lambda_{H} \lambda_{\Phi} - \Lambda_{H\Phi}^2)  \right. $ \\
$ \qquad \left.  + 2 \mu_{H}^2 \mu_{\Phi}^2 (2 \Lambda_{\chi,n}\Lambda_{H\Phi}-  \lambda_{\Phi\chi} \lambda_{H\chi}) + 2 \mu_{H}^2 \mu_{\chi}^2 (2 \lambda_{\Phi} \lambda_{H\chi} -\Lambda_{H\Phi}\lambda_{\Phi\chi}) \right. $ \\ $  \left. -2 \mu_{\Phi}^2 \mu_{\chi}^2 (2 \lambda_{H}\lambda_{\Phi\chi} - \Lambda_{H\Phi}\lambda_{H\chi}) \right] $ }
  \\ \hline
\end{tabular}
\end{table}
\renewcommand{\arraystretch}{1}

In~\autoref{table:minima1} and~\autoref{table:minima3}, we show all possible extrema of the zero temperature tree level potential for our a model. For the extrema to be realized by the potential, the squared field values $\langle h \rangle^{2}$, $\langle \varphi \rangle^{2}$, and $\langle \chi_{i} \rangle^{2}$ need to be positive. For them to be a minimum, the Hessian needs to be positive definite. We require the EW~minimum $(v_0,0,0)$ to be the deepest. In practice, especially when we include the Coleman-Weinberg contribution, we check that the EW~vacuum is the deepest by numerical minimization.
\section{Leading order daisy coefficients and details on improved daisy resummation treatments}
\label{sec:odaisy}

As stated in the main text, at high temperatures, there will be sizable higher loop thermal contributions which may break the perturbative validity at some field values. In order to resum such contributions, a naive treatment is to include a thermal mass contribution on top of the tree level effective mass. Formally, the thermal mass should be calculated using the gap equation. However, if one truncates the thermal potential at leading order in the expansion of the thermal mass, as well as in the leading order in the high-temperature expansion, one would obtain analytical leading order thermal mass contributions to each degree of freedom. Here we quote such leading order contributions of our model. 
\begin{align}
  \label{eq:thermal.scalar.1}
  \Pi_{0,h}&=\Pi_{0,G}= c_h T^2= \left( \frac{\lambda_{H}}{2} + \frac{\lambda_{H\Phi}}{6} + \frac{3g^{2}+g^{\prime 2}}{16} + \frac{y_{t}^{2}}{4}   + \frac{\tlambda_{H\Phi}}{12}+ \frac{N}{24} \lambda_{H\chi}  \right) T^2\\   \label{eq:thermal.scalar.2}
  \Pi_{0,\varphi}&=\Pi_{0,\phi}= c_{\varphi} T^2=\left(\frac{\lambda_{\Phi}}{2} + \frac{\lambda_{H\Phi}}{6}+ \frac{3g^{2}+g^{\prime 2}}{16} + \frac{\tlambda_{H\Phi}}{12} + \frac{N}{24}\lambda_{\Phi\chi} \right) T^2\\   \label{eq:thermal.scalar.3}
  \Pi_{0,\chi_{i}}&= c_{\chi} T^2=\left(\frac{\lambda_{\chi}}{12}(N+2) + \frac{\lambda_{\Phi\chi}}{6}  + \frac{\lambda_{H\chi}}{6} +\frac{\tlambda_{\chi}}{4} \right) T^2.
\end{align}
For later convenience, we define constants $c_i =\left. \Pi_i^2/T^2 \right|_{\hat{h}=\hat{\varphi}=0}$.
The thermal masses of $W$ and $Z$ are as given in~\cite{Blinov:2015vma,Niemi:2018asa}, they only contribute to the longitudinal components:
\begin{align}
  \label{eq:thermal.WZ}
  \Pi_{0,W_{L}} &=  2 g^{2}T^{2}\\
  \Pi_{0,Z_{L},A_{L}}&= - \frac{g^{2}+g^{\prime 2}}{8}(\hat{h}^{2}+\hat{\varphi}^{2}) + (g^{2}+g^{\prime 2})T^{2} \pm \Delta\\
  \Delta^{2} &=  \left(\frac{g^{2}+g^{\prime 2}}{8}\right)^2\left(\hat{h}^{2}+\hat{\varphi}^{2} +8T^{2}\right)^{2}-g^{2}g^{\prime 2}T^{2}\left(\hat{h}^{2}+\hat{\varphi}^{2}+4T^{2}\right).
\end{align}

\section{RGEs}
\label{app:rge}
RG improvement is necessary to resum large log contributions at large field values. To compute the RGEs, we follow the steps discussed in~\cite{Buchalla:2017jlu}, using the real representation of the $SU(2)$-doublets discussed in~\cite{Buchalla:2019wsc}. This approach utilizes the background-field method and super-heat-kernel expansion. Our results have been checked in the SM limit~\cite{Einhorn:2007rv} and the pure inert 2HDM limit presented in~\cite{Blinov:2015vma}, as well as two independent computations. Note that the wavefunction renormalizations are gauge dependent (therefore there is a difference compared to~\cite{Einhorn:2007rv}). This is another manifestation of the gauge-dependence of $v_{\rm EW}(T)$ that was discussed in~\cite{Jackiw:1974cv,Kang:1974yj,Dolan:1974gu,Fukuda:1975di,Aitchison:1983ns,Patel:2011th,Garny:2012cg,Andreassen:2014eha,Andreassen:2014gha}. Given the Lagrangian of eq.~\eqref{eq:Model.Lag}, we find the $\beta$-functions, defined as $ \beta(c)\equiv 16\pi^{2} \frac{d }{ d \log{\mu_{R}}}c $, to be:
\begin{align}
  \label{eq:RGEs}
  \beta(g_{s}) &= -7 g_{s}^{3} \\
  \beta(g) &= -3 g^{3} \\ 
  \beta(g^{\prime}) &= 7 g^{\prime 3} \\
  \beta(\mu_{H}^{2}) &= -4\lambda_{H\Phi}\mu_{\Phi}^{2}-2\tlambda_{H\Phi}\mu_{\Phi}^{2}-\mu_{H}^{2}(-12\lambda_{H} + \tfrac{3}{2} (3g^{2}+g^{\prime 2})-6 y_{t}^{2})- N \mu_{\chi}^{2}\lambda_{H\chi}\\
  \beta(\mu_{\Phi}^{2}) &= -4\lambda_{H\Phi}\mu_{H}^{2}-2\tlambda_{H\Phi}\mu_{H}^{2}-\mu_{\Phi}^{2}(-12\lambda_{\Phi} + \tfrac{3}{2} (3g^{2}+g^{\prime 2}))- N \mu_{\chi}^{2}\lambda_{\Phi\chi}\\
  \beta(\mu_{\chi}^{2}) &= 4 \mu_{\Phi}^{2}\lambda_{\Phi\chi} + 6 \tlambda_{\chi}\mu_{\chi}^{2}-4 \mu_{H}^{2}\lambda_{H\chi} + 2(N+2) \mu_{\chi}^{2}\lambda_{\chi} \\
  \beta(\lambda_{H}) &= 2 \lambda_{H\Phi}^{2} + 2\lambda_{H\Phi}\tlambda_{H\Phi}+ \tlambda_{H\Phi}^{2}+24\lambda_{H}^{2}-3\lambda_{H}(3g^{2}+g^{\prime 2}) \nonumber \\
               &+\frac{3}{8}(3g^{4}+2g^{2}g^{\prime 2}+g^{\prime 4}) + 12\lambda_{H} y_{t}^{2}-6 y_{t}^{4}+\tfrac{N}{2}\lambda_{H\chi}^{2}\\
  \beta(\lambda_{\Phi}) &= 2 \lambda_{H\Phi}^{2} + 2\lambda_{H\Phi}\tlambda_{H\Phi}+ \tlambda_{H\Phi}^{2}+24\lambda_{\Phi}^{2}-3\lambda_{\Phi}(3g^{2}+g^{\prime 2}) \nonumber \\
               &+\frac{3}{8}(3g^{4}+2g^{2}g^{\prime 2}+g^{\prime 4}) +\tfrac{N}{2}\lambda_{\Phi\chi}^{2}\\
  \beta(\lambda_{\chi}) &= 2\lambda_{\Phi\chi}^{2} + 2 \lambda_{H\chi}^{2}+16 \lambda_{\chi}^{2} + 12 \tlambda_{\chi}\lambda_{\chi}+ 2N \lambda_{\chi}^{2}\\
  \beta(\lambda_{H\Phi}) &= \frac{3}{4}(3g^{4}-2g^{2}g^{\prime 2}+g^{\prime 4})+4 \lambda_{H\Phi}^{2} + 2\tlambda_{H\Phi}^{2} +4\tlambda_{H\Phi}(\lambda_{H} + \lambda_{\Phi})\nonumber \\
               &+ \lambda_{H\Phi}(12\lambda_{\Phi}+12\lambda_{H} - 3(3g^{2}+g^{\prime 2})) + 6\lambda_{H\Phi}y_{t}^{2}\\
  \beta(\tlambda_{\chi}) &= 18 \tlambda_{\chi}^{2} + 24 \tlambda_{\chi}\lambda_{\chi}\\
  \beta(\lambda_{\Phi\chi}) &= (-\tfrac{3}{2}(3g^{2}+g^{\prime 2})+12\lambda_{\Phi}+6\tlambda_{\chi}+4\lambda_{\Phi\chi}+2N\lambda_{\chi}+4\lambda_{\chi})\lambda_{\Phi\chi} \nonumber \\
               &+4 \lambda_{H\Phi}\lambda_{H\chi} + 2 \tlambda_{H\Phi}\lambda_{H\chi}\\
  \beta(\lambda_{H\chi}) &= (-\tfrac{3}{2}(3g^{2}+g^{\prime 2})+12\lambda_{H}+6\tlambda_{\chi}+4\lambda_{H\chi}+2N\lambda_{\chi}+4\lambda_{\chi}+6y_{t}^{2})\lambda_{H\chi} \nonumber \\
               &+ 4 \lambda_{H\Phi}\lambda_{\Phi\chi} + 2 \tlambda_{H\Phi}\lambda_{\Phi\chi}\\
  \beta(\tlambda_{H\Phi}) &= 3 g^{2}g^{\prime 2} - 3 \tlambda_{H\Phi}(3g^{2}+g^{\prime 2}) + 6 \tlambda_{H\Phi}y_{t}^{2} + 4\tlambda_{H\Phi}(\lambda_{H} + \lambda_{\Phi})\nonumber \\
  &+ 8 \lambda_{H\Phi}\tlambda_{H\Phi}+4\tlambda_{H\Phi}^{2}\\
  \beta(y_{t}) &= -8 y_{t} g_{s}^{2} - \tfrac{9}{4}y_{t}g^{2}-\tfrac{17}{12}y_{t}g^{\prime 2} + \tfrac{9}{2} y_{t}^{3}.
\end{align}
As mentioned in~\autoref{sec:model}, the couplings $\tlambda_{H\Phi}$ and $\lambda_{H\chi}$ are not protected by a symmetry (hypercharge breaks the custodial symmetry of $\tlambda_{H\Phi}$) and will run away from their initial, vanishing value. The wave function renormalization of the scalar fields are
\begin{align}
  \label{eq:WFRs}
  \gamma_{H} &= -3 g^{2} - g^{\prime 2} + 3 y_{t}^{2},\\
  \gamma_{\Phi} &= -3 g^{2} - g^{\prime 2}, \\
  \gamma_{\chi} &= 0.
\end{align}

\section{Details on the sphaleron rate calculation}\label{sec:ba}

\begin{figure}[t]
\centering
\includegraphics[width=0.49\textwidth]{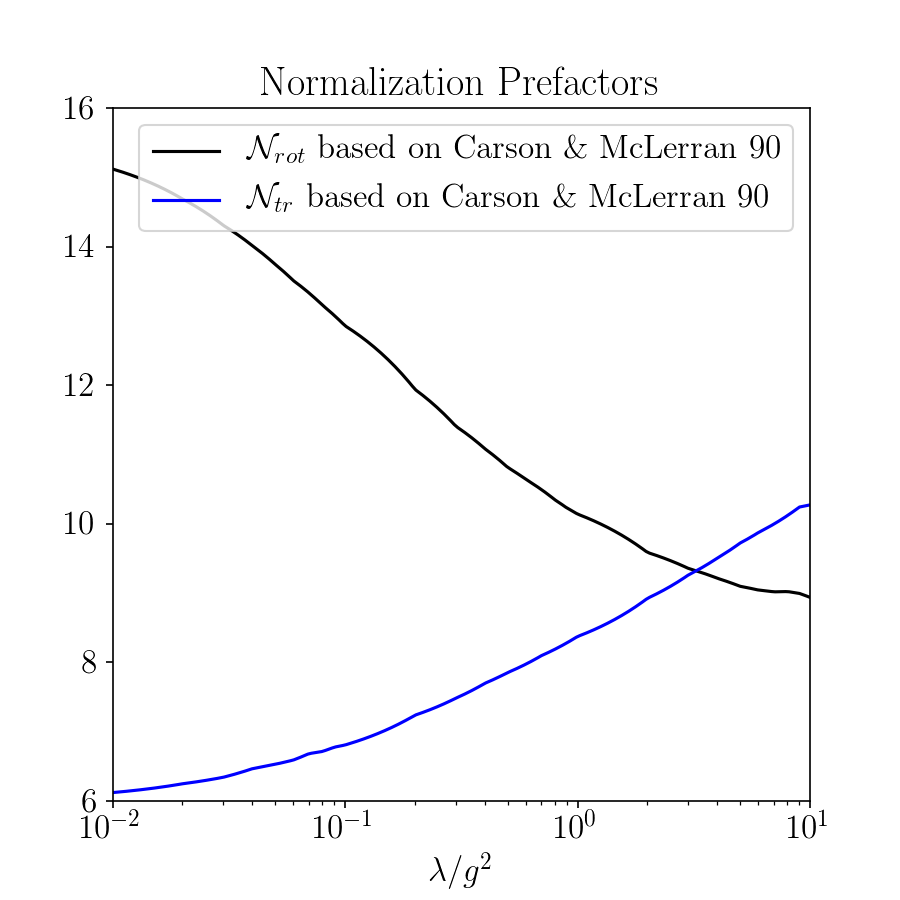}
\includegraphics[width=0.49\textwidth]{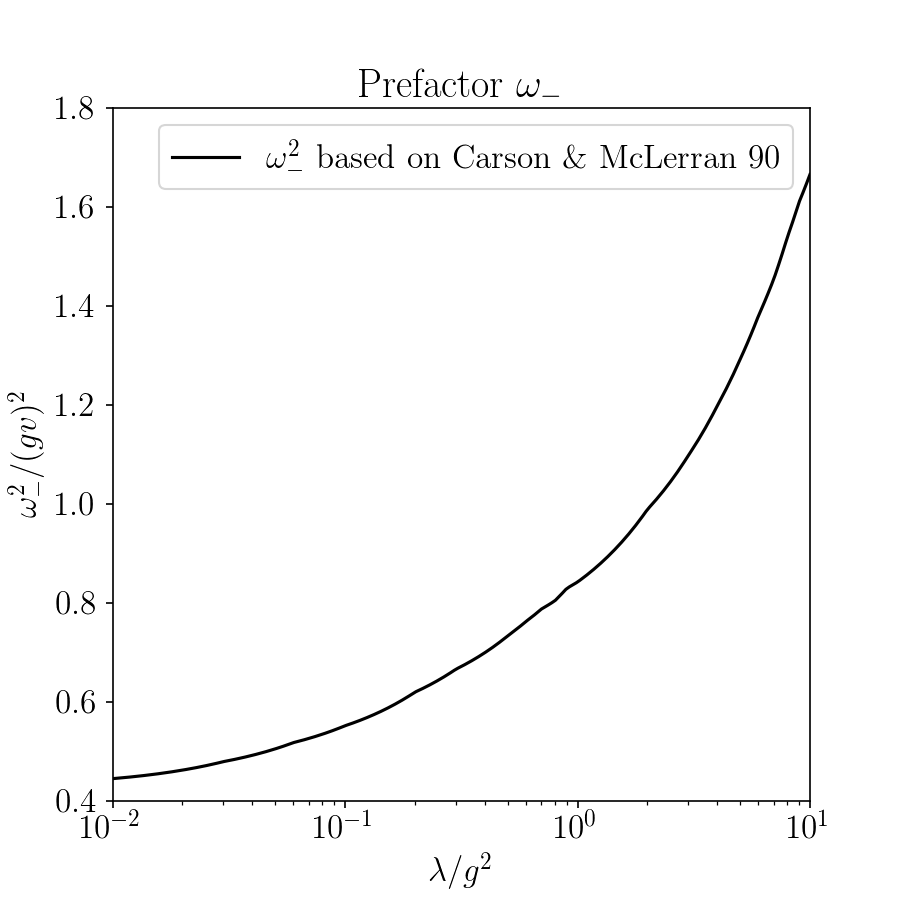}\\
\includegraphics[width=0.49\textwidth]{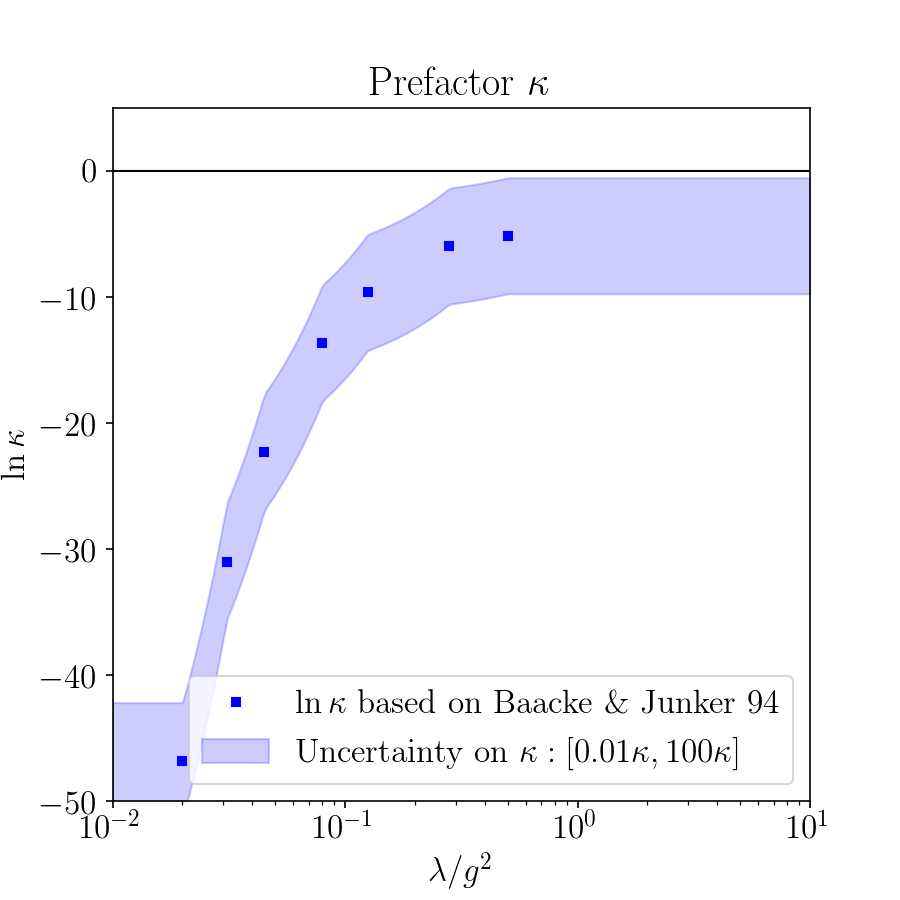}
\includegraphics[width=0.49\textwidth]{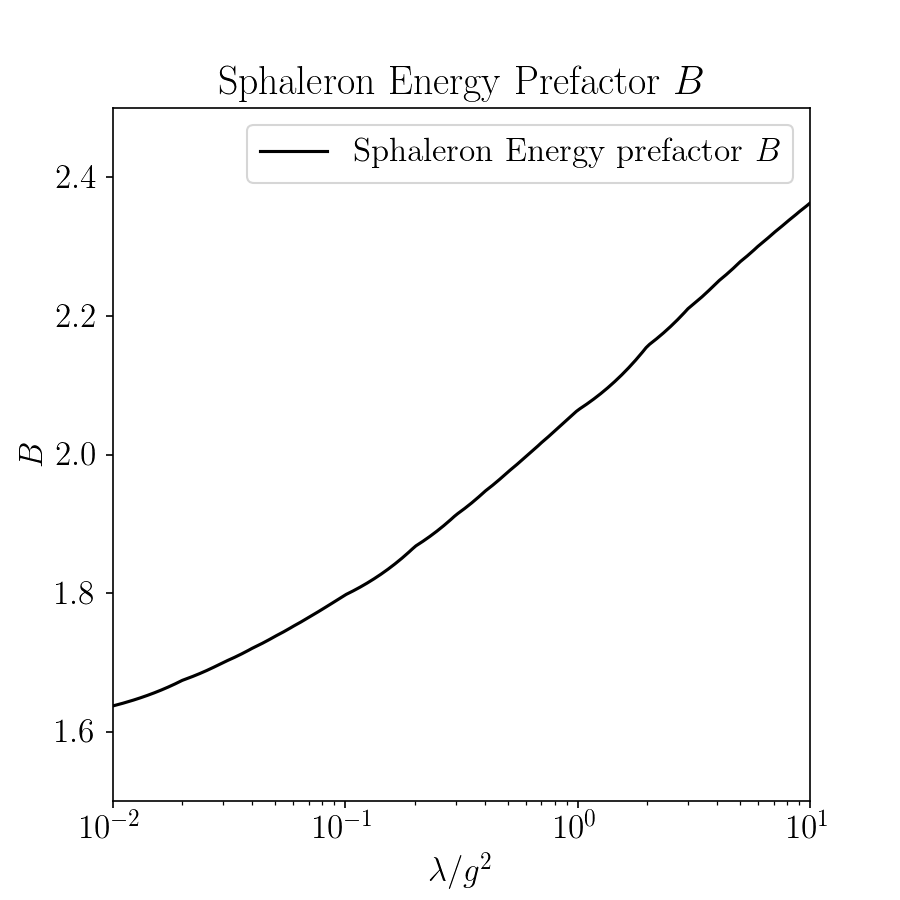}
\caption{Input values for the sphaleron decay rate in eq.~\eqref{eq:sph.2}. $\mathcal{N}_{tr} $ and $\mathcal{N}_{rot}$ are taken from~\cite[Fig. 5]{Carson:1989rf}, $\omega_{-} $ is taken from~\cite[Fig. 6]{Carson:1989rf}, and $\kappa$ is taken from~\cite{Baacke:1994ix}, including uncertainties as explained in the text, and the energy prefactor $B$ is computed also as discussed in the text. All quantities are plot against $\lambda/g^2$, the ratio of the corresponding quartic to the gauge coupling.}
\label{fig:prefactor.inputs}
\end{figure}

In this section, we give some details on calculating the sphaleron rate across a large range of temperatures, which would prove essential in evaluating the baryon asymmetry in a model with UV EWBG. The sphaleron rate per unit volume is~\cite{Carson:1989rf,Carson:1990jm}
\begin{equation}
\label{eq:sph.1}
\frac{\Gamma}{V} = \frac{\omega_{-}}{2\pi} \mathcal{N}_{tr} (\mathcal{N}V)_{rot} \left(\frac{\alpha_{w}T}{4\pi}\right)^{3} \alpha_{3}^{-6}\kappa \exp{\left[-E_{sph}(T)/T\right]} .
\end{equation}
This rate depends on the profile functions of the sphaleron solution that can be obtained by solving the equations of motion for  the SU(2) and U(1) gauge bosons, and Higgs doublets~\cite{Moreno:1996zm,Grant:2001at}. In the limit of neglecting the U(1) gauge coupling, $g' = 0$, a spherically symmetric ansatz gives a system of differential equations that can be numerically solved, for example, using the Newton-Kantorovich method as done in~\cite{Gan:2017mcv}. Given the uncertainties of the thermal potential calculation, we use the values of $\lambda/g^2$ as shown in~\autoref{fig:prefactor.inputs}, where $g= g(\mu_{\rm R})$ is the SU(2) gauge coupling, and we consider $\lambda = \lambda_H(\mu_{\rm R})$ when we are in the phase P$_H$, and $\lambda = \lambda_{\Phi}(\mu_{\rm R})$ when we are in the phases P$_{\Phi}$ or P$_{H\Phi}$. This is justified by the fact that the sphaleron solution depends on the $SU(2)$ structure of the theory, and our model mostly has either the SM Higgs or the inert Higgs taking a vev.

The sphaleron energy is then given by $E_{sph}(T) = E_{sph}(T=0) \tfrac{v_{\rm EW}(T)}{v_{\rm EW}({T=0})} = \tfrac{4 \pi}{g} B\, v_{\rm EW}(T)$, where the energy prefactor $B$ can be obtained by performing the volume integral of the stress-energy tensor using the previously obtained profile functions. Our choice of $B$ as a function of $\lambda/g^2$ is shown in~\autoref{fig:prefactor.inputs}, which is consistent with \cite{Quiros:1999jp}.

$\mathcal{N}_{tr}$ and $\mathcal{N}_{rot}$ are the normalization of the zero-frequency translation and rotation modes~\cite{Carson:1989rf}. They can be computed from small fluctuations around the sphaleron solution. The resulting formula depends again on the profile functions and can therefore be either computed numerically or read off from~\cite[Fig. 5]{Carson:1989rf}. We pursue the latter and show the values we use in~\autoref{fig:prefactor.inputs}.

$\omega_{-}$ is the frequency of the unstable mode~\cite{Carson:1989rf,Akiba:1989xu}. It can be found as a negative eigenvalue of a system of equations that also depends on the profile functions. We use the values of~\cite[Fig. 6]{Carson:1989rf} directly and show them in~\autoref{fig:prefactor.inputs} (Note that this plot shows $\omega_{-}^{2}$ in units of $(gv)^{2}$).

$\kappa$ is the fluctuation determinant. A first numerical evaluation was given in~\cite{Carson:1990jm}, and later improved in~\cite{Dine:1991ck,Baacke:1993jr,Baacke:1993aj,Baacke:1994ix}. We use the values given in~\cite{Baacke:1994ix} and assume a rather large uncertainty of $[0.01 \kappa, 100 \kappa]$ to also partially parametrize uncertainties in the other prefactors~\cite{Gan:2017mcv}.

Finally, $V_{rot} = 8\pi^{2}$ is the volume of the rotation group; $\alpha_{w}= g^{2} / 4\pi^{2}$ is the weak coupling constant; and $\alpha_{3}= \alpha_{w}  / (g \xi(T))$ is the weak coupling in the three-dimensional high-temperature effective theory.

\section{Details on the Numerical Implementation}
\label{app:numerics}
Here we discuss our numerical implementation of the effective potential discussed and thermal calculation in~\autoref{sec:pert}. We wrote the main code in {\tt python}, which is available at {\tt https://gitlab.com/claudius-krause/ew\_nr}. We have a second, independent implementation of the code using Mathematica~\cite{Mathematica}, which we extensively cross-checked against the {\tt python} code.

When including the Coleman-Weinberg potential, we shift the numerical values of $\mu^{2}_{H}$ and $\lambda_{H}$ so that the full potential satisfies
\begin{equation}
\label{eq:T.QFT.7}
\left.\frac{\delta V}{\delta h}\right|_{\hat{h}=v_0} = 0 \qquad \& \qquad \left.\frac{\delta^{2} V}{(\delta h)^{2}}\right|_{\hat{h}=v_0} =m_{h}^{2}
\end{equation}
at $T=0$. These values are then used throughout the computation, including inside the RGEs.

The finite temperature potential (defined in eq.~\eqref{eq:T.QFT.8}) can either be evaluated numerically for each point $y=M^{2}/T^{2}$ (which is slow), or a pre-computed look-up table and subsequent spline interpolation can be used. In the benchmark points discussed in the main text, we use a modified version of the spline implementation of \ct~\cite{Wainwright:2011kj}. Compared to the original implementation, we extended the pre-computed grid of exact evaluations of the $J_{B/F}(y)$-functions to include more points in the negative $y$ direction and re-wrote the exact evaluation of $J_{B/F}(y)$ to reduce numerical noise. The corresponding files are also included in the \gl\ repository.

Daisy corrections beyond the high-$T$ approximation require the second derivative of the thermal potential. We use a numerical, finite-difference derivative based on 9 points chosen symmetrically around the desired functional argument, with a stepsize that increases with large field values or temperatures. We checked that this choice gives a stable value for the derivative for various temperatures and field configurations. Automatic differentiation (AD), as nowadays widely used in the machine learning community~\cite{JMLR:v18:17-468}, would greatly improve the computation of the derivatives. However, implementation of AD in the computation of the effective potential as we do it here would be beyond the scope of this work. In practice, we include the following Daisy approximations in the truncated full dressing scheme of~\cite{Curtin:2016urg}: vanishing thermal masses, i.e., no Daisy correction; leading-order thermal masses in the high-$T$ expansion, i.e. the formulas given in eqs.~\eqref{eq:thermal.scalar.1}--\eqref{eq:thermal.scalar.3}; field-dependent thermal masses in the definition of eq.~\eqref{eq:daisytrunc}; and the thermal masses as defined by the gap equation \eqref{eq:daisygap}. Note that in the latter two approaches we do not include the Coleman-Weinberg contribution, to properly have the limit $\Pi_{i}(T=0) = 0$.

To reduce the dimension of field space that we have to scan, we assume that all $\chi_{i}$ acquire a vev simultaneously. This is justified as long as $\tlambda_{\chi}=0$ because then the $\chi$-sector exhibits an additional $SO(N)$ symmetry that allows us to rotate them freely into each other. Because of this enhanced symmetry, the condition $\tlambda_{\chi}=0$ is also conserved under the RGE.

We use the renormalization scale that we discussed in eq. \eqref{eq:ren.scale}, which is given by the largest square root of the absolute values of the eigenvalues of the bosonic mass matrix including thermal masses in the high-temperature approximation (eqs.~\eqref{eq:thermal.scalar.1}--\eqref{eq:thermal.scalar.3}), or the EW~scale $v_0=246~$GeV, whichever is larger. Since the mass matrix at a given point in field space itself also depends on the renormalization scale via the couplings, we have to solve eq.~\eqref{eq:ren.scale} numerically. We use Brent's method~\cite{Brent:113464}, as implemented in {\tt SciPy}~\cite{2020SciPy-NMeth} for this purpose. Note that we do not include the wavefunction renormalization factors of eq.~\eqref{eq:WFRs} at this point, as this would be numerically more complicated. Instead, we compute these factors at the end, after the minimization of the potential, and rescale the minima positions accordingly. We checked that even at large scales around $100~$ TeV the factors are at most around $1.05$ for the Higgs and at least $0.92$ for the inert scalar, so the feedback effect we neglect is in fact small. 

A given potential is then numerically minimized. To ensure that we found the global minimum instead of a local one, we use 8 different initial guesses in field space, which cover all possible directions in the three-dimensional space spanned by the Higgs, the inert, and the singlets. As field value, we choose 1.5 (2.5) times the current temperature for BM A (B), as we expect the minimum to grow with temperature in the non-restoring phase. At $T=0$, we use the tree-level extrema that we list in appendix~\autoref{app:BFB}. We require that at $T=0$ we are in the EW~minimum with $\langle h \rangle = 246~$GeV and $\langle \varphi \rangle = \langle \chi_{i}\rangle = 0$, otherwise either the BFB (minimum at large field values) or the $T=0$ vacuum structure would not be satisfied. 
\end{document}